\documentclass[aps,physrev,groupedaddress]{revtex4-2}
\usepackage{graphicx}
\usepackage{caption}
\usepackage{subcaption}
\usepackage{amssymb, amsthm, amsmath}
\usepackage{braket}
\usepackage{multirow}
\usepackage{dcolumn}
\usepackage{bm}
\usepackage{bbm}
\usepackage{natbib}
\usepackage{url}
\begin{document}
\title{Engineering of non-Hermitian trajectories and phase structure in an open Bose-Hubbard model via rate
operator transformations}
\author{Jaakko Luomala}
\email{jaakko.s.luomala@utu.fi}
\affiliation{Department of Physics and Astronomy,
University of Turku, FI-20014 Turun yliopisto, Finland}

\author{Kimmo Luoma}
\affiliation{Department of Physics and Astronomy,
University of Turku, FI-20014 Turun yliopisto, Finland}

\author{Iiro Vilja}
\affiliation{Department of Physics and Astronomy,
University of Turku, FI-20014 Turun yliopisto, Finland}

\author{Jyrki Piilo}
\affiliation{Department of Physics and Astronomy,
University of Turku, FI-20014 Turun yliopisto, Finland}

\begin{abstract}

Non-Hermitian evolution can be realized through post-selection on stochastic pure-state trajectories arising in continuously monitored open quantum systems. The rate operator formalism provides a versatile and systematic framework for unraveling a master equation into stochastic pure-state evolutions, offering enhanced control over the resulting non-Hermitian dynamics.
In the present work, we explore the applicability of the rate operator formalism as a tool for engineering non-Hermitian dynamics. Specifically, we apply this approach to the Bose–Hubbard model subject to environmental dephasing, examining its consequences for controlled state manipulation. Our analysis is framed within the broader contexts of quantum state engineering and measurement-induced phase transitions.
We demonstrate that the rate operator formalism enables the construction of effective non-Hermitian Hamiltonians exhibiting a unique steady state—even in regimes where the standard Monte Carlo wavefunction method fails to produce one. Furthermore, we show that this framework facilitates transitions between distinct steady-state phases, governed by tunable parameters such as the interaction strength and a non-Hermiticity control parameter introduced via the rate operator formalism.

\end{abstract}

\maketitle

\section{Introduction}

Non-Hermitian Hamiltonians have been used as an effective description of open quantum systems in many contexts, such as in the decay of a nucleus, scattering of particles, and as a mean-field description of open quantum systems. Non-Hermitian Hamiltonians also arise in stochastic trajectories realizable via continuous measurements on open quantum systems. For a review in non-Hermitian dynamics, see e.g. \cite{El-Ganainy2018,Ashida2020}. An important contribution to the rise of popularity of non-Hermitian dynamics was a discovery that some non-Hermitian Hamiltonians, like parity-time (PT) symmetric Hamiltonians, can have purely real eigenvalues \cite{Bender1998}. Another extraordinary quality that non-Hermitian Hamiltonian can have, and is found in PT-symmetry breaking, is the presence of exceptional points (EP) in its parameter space. At EPs two or more eigenvalues and eigenstates coalesce, leading to spectral and dynamical properties without analogies in Hermitian dynamics \cite{Graefe2008a,Jin2013,Zhang2021,Teixeira2023,Li2023}. Non-Hermitian dynamics have been incorporated to systems such as optics at the mean-field level using gain and loss \cite{El-Ganainy2007-optics,Makris2008-optics,Guo2009-optics}, a transmon qubit \cite{Naghiloo2019} and atomic systems in optical lattices \cite{Zhang2016}. Potential applications for non-Hermitian dynamics include quantum state engineering, \cite{Lee2014-entanglement-spinsqueezing,Arkhipov2024-engineering} and enhanced sensing \cite{Hodaei2017-sensing,Wiersig2020-sensing,Wu2026-sensing}.

One way to realize non-Hermitian quantum dynamics is via its association to stochastic jump methods
used to unravel a given open system master equation. These methods consider an ensemble of stochastically
evolving pure-state trajectories with a non-Hermitian Hamiltonian, where deterministic evolution is interrupted by randomly occurring quantum jumps. Conditioning, or post-selecting, to no-jump evolution realizes 
a given non-Hermitian dynamics. Indeed, there exists a wide variety of jump-methods both in Markovian and non-Markovian 
regimes~\cite{Plenio1998RMP, Molmer1992PRL, Federico-Jyrki-Review, Smirne2020}. Recently developed rate operator formalism displays promising features for both realizing and in particular
for engineering non-Hermitian deterministic dynamics~\cite{Smirne2020, Chruciski2022, Settimo2024, Settimo2026}. This is based on using a transformation leaving the solution of the master equation invariant while simultaneously modifying both the non-Hermitian Hamiltonian -- giving the deterministic evolution -- and the type of jumps that occur. Indeed, this opens the possibility for almost arbitrary engineering of the non-Hermitian deterministic dynamics for a given physical system.

In this paper, we focus on non-Hermitian dynamics in a paradigmatic many-body model, i.e., Bose-Hubbard model~\cite{Gersch1963,Anglin1997,Jaksch2005,Kordas2015}.
Within this context, we consider two different decoherence mechanisms, which we call local and non-local dephasing. The motivation and impact of the work arise from the following aspects. We demonstrate the use and applicability of the recently developed rate-operator formalism -- beyond its original aim of solving complicated master equations --  for quantum control and engineering purposes. This is done both in the dynamical and steady-state sense. For the former, the interest is in displaying the richness and classification of non-Hermitian time evolution itself, via various rate-operator transformations, even if the considered physical system remains the same.  For the latter, our work reveals an interesting steady-state phase structure as a function of the Bose-Hubbard and rate-operator transformation parameters. In other words, in terms of steady state populations, after deterministic non-Hermitian  dynamics, the populations of the state vector display abrupt changes and phase transitions by varying the used control parameters. This opens new possibilities, e.g., for quantum control purposes in the presence of noise within a many body system.

The paper is organized in the following way. In Section \ref{sec-ROQJ-B} we introduce the basics of the rate operator formalism
and in Sec.~\ref{sec-BH} the Bose-Hubbard model as it is used for our purposes. The results for the evolution and engineering of the dynamical non-Hermitian evolution are shown in Sec.~\ref{sec-results} whereas the results for the discovered steady-state structure and transitions  are displayed in Sec.~\ref{sec-results_b}. We conclude in Sec.~\ref{summary}.    

\section{Rate operator unraveling \label{sec-ROQJ}}

The theory of open quantum systems (OQS) investigates the dynamics of a quantum system interacting with its environment \cite{Breuer2007,Vacchini2024open}. The composite system evolves in the Hilbert space $\mathcal{H}_S \otimes \mathcal{H}_E$, where $\mathcal{H}_S$ and $\mathcal{H}_E$ denote the system and environment Hilbert spaces, respectively. While the full system undergoes unitary evolution, the primary interest in OQS theory lies in describing the reduced dynamics of the subsystem. This reduced evolution is inherently non-unitary and is typically formulated in terms of a master equation for the density operator. Several master equations have been derived under different physical approximations (see, e.g., \cite{Breuer2007}), and multiple techniques exist for solving a given master equation.

In the present section, we introduce the tools most relevant for our analysis. Section~\ref{sec-ROQJ-A} addresses the general structure of time-local completely positive and trace-preserving (CPTP) master equations, which represent one of the most widely employed frameworks in the study of OQS dynamics. Section~\ref{sec-ROQJ-B} provides a concise overview of the Monte Carlo wave function (MCWF) method \cite{Mlmer1993} and rate-operator-based unravelings \cite{Smirne2020,Chruciski2022,Settimo2024}, with a particular emphasis on the rate operator formalism. The results presented in this work focus on quantum dynamical semigroups governed by the Gorini–Kossakowski–Sudarshan–Lindblad (GKSL) equation \cite{Gorini1976,Lindblad1976}. Throughout the following, we assume that the system Hilbert space $\mathcal{H}_S$ is finite dimensional and set $\hbar = 1$.

\subsection{Open quantum system dynamics \label{sec-ROQJ-A}}

\indent The general form of a CPTP master equation $d\rho(t)/dt=\mathcal{L}_t[\rho(t)]$ for the open system density matrix $\rho(t)$ where the linear operator $\mathcal{L}_t$ can be written in diagonal form as
\begin{align}
    \mathcal{L}_t[\rho(t)]=-i[H(t),\rho]+\sum_{i=1}^{d^2-1}\left[\gamma_i(t)F_i(t)\rho(t) F_i^\dagger(t)-\frac{1}{2}\gamma_i(t)\left(F_i^\dagger(t) F_i(t)\rho(t) + \rho(t) F_i^\dagger(t) F_i(t)\right)\right], \label{Lindblad-me}
\end{align}
\noindent where $d$ is the dimension of the Hilbert space $\mathcal{H_S}$ of the system, $H(t)$ is the Hamiltonian of the system, and $\gamma_i(t)$ are real functions of time, called decay rates.
The operators $F_i(t)$ are generic Lindblad operators, which code the influence of the environment to the studied system such as decay, absorption, and dephasing.  By defining $\Gamma(t)=\sum_{i}^{d^2-1}\gamma_i(t)F_i^\dagger(t) F_i(t)$, $K(t)=H(t)-\frac{i}{2}\Gamma(t)$ and $\mathcal{J}_t[\rho(t)]=\sum_{i}^{d^2-1}\gamma_i(t)F_i(t)\rho F_i^\dagger(t)$, we can write the master equation (\ref{Lindblad-me}) as
\begin{align}
     \frac{d\rho(t)}{dt}&=-i\left(K(t)\rho(t)-\rho(t)K^\dagger(t)\right) + \mathcal{J}_t[\rho(t)] \nonumber\\
     &=\mathcal{D}_t[\rho(t)] + \mathcal{J}_t[\rho(t)], \label{Lindblad-me2}
\end{align}
\noindent where $K(t)$ is generally non-hermitian effective Hamiltonian. The two parts of (\ref{Lindblad-me2}) are called the driving term ($\mathcal{D}_t[\rho(t)]$) and the jump term ($\mathcal{J}_t[\rho(t)]$). In stochastic unravelings the driving term controls the deterministic evolution, while the jump term controls the quantum jumps. 

In the context of the rate operator formalism an important property of the generator $\mathcal{L}_t$ is its invariance in transformations \cite{Chruciski2022}
\begin{align}
    \mathcal{J}_t'[\rho(t)] &= \mathcal{J}_t[\rho(t)] + \frac{1}{2}\left(C(t)\rho(t)+\rho(t) C^\dagger(t)\right) \label{jumpterm-transformed},\\
    K'(t) &= K(t) -\frac{i}{2}C(t), \label{effective-hamiltonian}
\end{align}
\noindent where the operator $C(t)$ is an arbitrary operator on $\mathcal{H_S}$.

\subsection{Rate operator quantum jumps \label{sec-ROQJ-B}}

\indent There are multiple approaches to unravel the master equation (\ref{Lindblad-me}). Quantum-jump unravelings correspond to piecewise deterministic stochastic processes—referred to as realizations or trajectories—that evolve pure states in the Hilbert space $\mathcal{H}$. Such trajectories consist of deterministic evolution governed by the effective Hamiltonian $K(t)$, interspersed with stochastic quantum jumps occurring at random times. Among the available methods, the most general framework is provided by rate-operator unravelings \cite{Smirne2020,Chruciski2022,Settimo2024}. For clarity, it is useful to contrast this formalism with the widely used Monte Carlo wave function (MCWF) method \cite{Mlmer1993}.
In the MCWF approach, the deterministic evolution over an infinitesimal time interval $dt$, from time $t$ to $t+dt$, is described by
\begin{align}
\ket{\psi(t)}\to\ket{\psi(t+dt)}&=\frac{\left(\mathbbm{1}-iK(t)dt\right)\ket{\psi(t)}}{|\left(\mathbbm{1}-iK(t)dt\right)\ket{\psi(t)}|},\label{Kevol}\\
K(t)&= H(t)-\frac{i}{2}\Gamma(t),
\end{align}
\noindent while the stochastic component consists of discontinuous jumps at time $t$ of the form
\begin{align}
\ket{\psi(t)}\to \frac{F_i(t)\ket{\psi(t)}}{|F_i(t)\ket{\psi(t)}|},
\end{align}
\noindent occurring with probability $\gamma_i(t)|F_i(t)\ket{\psi(t)}|^2dt$. It follows directly from this construction that the decay rates functions $\gamma_i(t)$ must be non-negative for the MCWF formalism to be well-defined.

In the generalized rate-operator quantum-jump formalism, both the deterministic evolution and the stochastic jumps are governed by the transformed effective Hamiltonian (\ref{effective-hamiltonian}) and the transformed jump term (\ref{jumpterm-transformed}). The associated \textit{rate-operator transformation} $C(t)$ may, in general, depend explicitly on the instantaneous state of the trajectory; hence we denote it by $C_\psi(t)$. For a single realization $\ket{\psi(t)}$, the corresponding density operator is $\rho(t)=\ket{\psi(t)}\bra{\psi(t)}$. The transformed jump term $\mathcal{J'}_t[\ket{\psi(t)}\bra{\psi(t)}]$ is referred to as the generalized rate operator, which can be expressed as
\begin{align}
R_\psi &= \mathcal{J}[\ket{\psi}\bra{\psi}] + \frac{1}{2}\left(C_\psi\ket{\psi}\bra{\psi}+\ket{\psi}\bra{\psi} C_\psi^\dagger\right) \nonumber\\
&= \sum_{i=1}^{d^2-1}\gamma_i F_i\ket{\psi}\bra{\psi} F_i^\dagger + \frac{1}{2}\left(C_\psi\ket{\psi}\bra{\psi}+\ket{\psi}\bra{\psi} C_\psi^\dagger\right), \label{rateoperator-transformed}
\end{align}
 where explicit time dependence has been omitted for brevity. Quantum jumps at time $t$ are performed to eigenstates $\ket{\phi_{\psi,t,i}}$ of the generalized rate operator $R_{\psi,t}$, with corresponding probabilities $p_{\psi,t,i}=\lambda_{\psi,t,i}dt$, where each $\lambda_{\psi,t,i}$ is the eigenvalue associated with $\ket{\phi_{\psi,t,i}}$. Since the rate operator is Hermitian, all its eigenvalues are guaranteed to be real.

For the unraveling to be well defined with independent realizations, the eigenvalues $\lambda_i$ must be non-negative, which is equivalent to requiring that the rate operator $R_\psi$ be positive semi-definite. Under this assumption, the trajectories consist of deterministic evolution according to (\ref{Kevol}) with the modified effective Hamiltonian $K'(t)$,
\begin{align}
K'(t) &= H(t)-\frac{i}{2}\Gamma(t) - \frac{i}{2}C(t), \label{general-effective-Hamiltonian}
\end{align}
followed by stochastic discontinuous jumps of the form
\begin{align}
\ket{\psi(t)}\to \ket{\phi_{\psi,t,i}},
\end{align}
\noindent occurring with probability $p_{\psi,t,i}=\lambda_{\psi,t,i}dt$. It is therefore evident that both the deterministic and stochastic components of the unraveling are highly non-unique \cite{Settimo2024}, as they depend on the specific choice of the operator $C_\psi$. We note that unravelings can still be constructed even when some eigenvalues are negative by employing the reverse-jump method \cite{Smirne2020}, although in this case the realizations are no longer independent.

Since the rate operator (\ref{rateoperator-transformed}) is Hermitian, its eigenstates can always be chosen to form an orthogonal set. Consequently, the ROQJ formalism admits a continuous-measurement interpretation in which the system is subjected to a time-dependent orthogonal measurement whose possible outcomes are $\emptyset$ or $\ket{\phi_i}_t ,\, i=1,\ldots,d$,
where the states $\ket{\phi_i}_t$ are the orthogonal eigenstates of the rate operator $R_{\psi,t}$ at time $t$, given that the system is in the state $\ket{\psi}$. The null outcome $\emptyset$ corresponds to the case in which no jump occurs, and the system evolves deterministically according to the evolution equation.
Crucially, this measurement interpretation depends on the specific choice of the transformation $C_{\psi,t}$: different choices of $C_{\psi,t}$ lead to distinct measurement schemes, provided that the resulting rate operator $R_{\psi,t}$ remains positive.

In the present work we focus on time-independent generators, denoted by $\mathcal{L}$, and assume non-negative constant decay rates $\gamma_i$ for all $i$. Under these conditions, the dynamics form a quantum dynamical semigroup characterized by the family of dynamical maps $\Lambda_t=\exp(\mathcal{L}t)$, which satisfy the semigroup property $\Lambda_{s+t}=\Lambda_t\Lambda_s$ for all $t,s\ge 0$. In this setting, the evolution of the density operator is governed by the GKSL equation
\begin{align}
\mathcal{L}[\rho(t)]=-i[H,\rho(t)]+\sum_{i=1}^{d^2-1}\left[\gamma_iF_i\rho(t) F_i^\dagger-\frac{1}{2}\gamma_i\left(F_i^\dagger F_i\rho(t) + \rho(t) F_i^\dagger F_i\right)\right]. \label{GKSL}
\end{align}
The rate operator formalism for both time-dependent Markovian and non-Markovian dynamics has been analyzed extensively in previous works \cite{Smirne2020,Chruciski2022,Settimo2024}.

\section{The Bose-Hubbard model \label{sec-BH}}

Previously, the ROQJ method has been applied primarily to simple systems, such as two-level systems \cite{Chruciski2022,Settimo2024}. In the present study, we implement the ROQJ approach for the first time in a more complex many-body setting, namely the Bose–Hubbard model \cite{Gersch1963} describing interacting bosons on a lattice. This model captures, for example, the physics of interacting bosonic atoms confined in optical lattices \cite{Raithel1997}. Such systems are realized experimentally by cooling bosonic atoms to ultracold temperatures and trapping them in periodic potentials generated through interfering laser fields.
The Bose–Hubbard Hamiltonian has also been employed to model arrays of transmon qubits \cite{Yanay2020} as well as transmon networks beyond the strict two-level approximation \cite{Mansikkamki2022,Busel2023}. These platforms offer a high degree of experimental control, making them well-suited for the investigation of open many-body quantum systems.

The general Hamiltonian of the open-chain Bose–Hubbard model is given by
\begin{align}
H=\sum_{i=1}^{M}\epsilon_i a_i^\dagger a_i
-\sum_{i=1}^{M-1}J_{i}(a_{i+1}^\dagger a_i+a_i^\dagger a_{i+1})
+\frac{1}{2}\sum_{i=1}^{M}U_i a_i^\dagger a_i^\dagger a_i a_i,
\end{align}
\noindent where $M$ is the number of lattice sites, $a_i$ is the annihilation operator on site $i$, $\epsilon_i$ denotes the on-site potential, $U_i$ the on-site interaction strength, and $J_i$ the tunneling amplitude between adjacent sites. In this work, we restrict attention to a homogeneous lattice in which all parameters are site-independent, and to a finite-dimensional Hilbert space $\mathcal{H}$ with a fixed particle number $N\in\mathbb{N}$. Under these conditions, the potential terms proportional to $\epsilon_i$ contributes only a constant energy shift and may be omitted. The resulting Hamiltonian takes the form
\begin{align}
H=-J\sum_{i=1}^{M-1}(a_{i+1}^\dagger a_i+a_i^\dagger a_{i+1})
+\frac{U}{2}\sum_{i=1}^{M}a_i^\dagger a_i^\dagger a_i a_i, \label{BH_Hamiltonian}
\end{align}
which is the model employed throughout this article.

The Bose–Hubbard model, when treated as an open quantum system, can be described using the GKSL equation (\ref{GKSL}). A survey of commonly employed Lindblad operators in this context is provided in \cite{Kordas2015}. Typical examples include the annihilation operators $a_i$, which model particle loss at lattice site $i$. Another widely used choice is the number operators $n_i=a_i^\dagger a_i$, which represent on-site dephasing processes.
In studies of full density-matrix dynamics, the influence of localized particle dissipation, with or without additional dephasing, on the coherence properties of a Bose–Einstein condensate has been extensively investigated \cite{Mansikkamki2022,Busel2023,Witthaut2008,Witthaut2009,Kordas2012,Kordas2013}. Furthermore, dissipation has been shown to enable the dynamical preparation of breather states and dark solitons \cite{Witthaut2011}. The effects of pure dephasing on the evolution of the density matrix have also been analyzed in several works \cite{Pichler2010,Poletti2012,Poletti2013,Sciolla2015}.
Additional Lindblad operators used in the Bose–Hubbard setting include those describing multi-particle loss processes. Two-body losses arising from inelastic collisions are captured by operators of the form $L_i=a_i^2$ \cite{Garca-Ripoll2009}, while three-body losses are modeled via $L_i=a_i^3$ \cite{Daley2009}. Non-local dissipation processes may be represented by operators $L_i = a_i a_{i+b}$, where $b$ specifies the distance between the involved lattice sites. Moreover, engineered dissipation enabling the preparation of dark states has been demonstrated using operators of the form $L_{ij}=(a_i^\dagger+a_j^\dagger)(a_i-a_j)$ \cite{Diehl2008}.
Non-Hermitian extensions of the Bose–Hubbard Hamiltonian have also been studied in \cite{Graefe2008a,Graefe2008b,Graefe2010,Jin2013,Znojil2019}, including investigations within the framework of $\mathcal{PT}$-symmetric quantum mechanics \cite{Graefe2008a,Jin2013,Znojil2019}.

We focus on two types of dephasing Lindblad operators. The first class consists of the on-site number operators $L_i=n_i$, which are Hermitian and give rise to the master equation
\begin{align}
\dot{\rho}=-i[H,\rho]+\sum_{i=1}^{M}\left[\gamma_{i}L_i\rho L_i-\frac{1}{2}\gamma_{i}\left(L_i^2\rho + \rho L_i^2\right)\right]. \label{me-local-dephasing}
\end{align}
In the following, we simplify the notation by omitting explicit time dependence of the density matrix $\rho$ and the state vector $\ket{\psi}$.
The second class of dephasing operators is defined by
$L_{ij}=\ket{i}\bra{i}-\ket{j}\bra{j}=\text{diag}(\cdots,1,\cdots,-1,\cdots),\ 1\le i<j\le d\equiv \dim\mathcal{H}$ \cite{Chruciski2022b}, where the states $\ket{i}$, $i=1,\dots,d$, are the vectors of the number-state basis of $\mathcal{H}$,
$\{\ket{n_1,\ldots,n_M}\,:\,\sum_i n_i = N\}$.
This choice leads to the master equation
\begin{align}
\dot{\rho}=-i[H,\rho]+\sum_{i=1}^{d-1}\sum_{j=i+1}^{d}\left[\kappa_{ij}L_{ij}\rho L_{ij}-\frac{1}{2}\kappa_{ij}\left(L_{ij}^2\rho + \rho L_{ij}^2\right)\right], \label{me-nonlocal-dephasing}
\end{align}
\noindent describing a more general form of dephasing.
The key distinction between these two classes of operators is that the number operators $n_i$ act locally on individual lattice sites, whereas the operators $\ket{i}\bra{i}-\ket{j}\bra{j}$ act non-locally in the many-body basis. For clarity, we refer to these two dephasing mechanisms as the local dephasing (LD) and non-local dephasing (NLD) schemes, respectively.

The non-local dephasing scheme captures more intricate dynamical behavior, as it involves $d-1$ linearly independent Lindblad operators $L_{ij}$, whereas the number of Lindblad operators in the local dephasing scheme scales only with the number of lattice sites. Moreover, the non-local dephasing scheme has not yet been extensively explored in the literature. These considerations form the primary motivation for our choice to use the operators $L_{ij}$ to model dephasing, rather than the more conventional number operators $n_i$.
In the present work, we restrict our analysis to a one-dimensional lattice consisting of three sites and containing between one and three particles. This limitation serves partly to reduce computational complexity, but it also suffices to illustrate the advantages of rate-operator transformations in the context of state engineering. Furthermore, since the Hamiltonian (\ref{BH_Hamiltonian}) and the Lindblad operators $n_i$ and $L_{ij}$ all commute with the total particle number operator $\sum_i n_i$, the particle number is conserved. Consequently, the finite-dimensional $N$-particle subspaces defined by vectors $\ket\psi$ such that $\sum_i n_i \ket{\psi} = N \ket{\psi}$ remain invariant under all relevant operations. The dimension $d$ of the Hilbert space for a system with $M$ sites and $N$ particles is $\dim \mathcal{H} = \binom{N+M-1}{N}$.

\section{Results: Evolution and Engineering \label{sec-results}}

In this section, we investigate several aspects of the dissipative Bose–Hubbard model governed by the GKSL equation. Section~\ref{sec-ME-detev} examines the evolution of the system's density matrix under the master equations (\ref{me-local-dephasing}) and (\ref{me-nonlocal-dephasing}). In Section~\ref{sec-detev}, we study the dynamics generated by the effective Hamiltonian of the form (\ref{general-effective-Hamiltonian}) for both dephasing schemes and compare these trajectories with the unitary evolution dictated by the Bose–Hubbard Hamiltonian (\ref{BH_Hamiltonian}). Section~\ref{sec-asymptotic} focuses on applications of the rate operator formalism in the context of state engineering. Specifically, Section~\ref{sec-asymptotic-A} presents the evolution of the absolute squares of the amplitudes of the system state together with the reduction in state norm induced by the non-Hermitian dynamics. 

In the following analysis, we fix the decay rates to be equal, $\gamma\equiv\kappa_{jk}=\gamma_i$, unless stated otherwise, and give all parameters in units of the decay rate. The number of lattice sites is $M=3$. We further employ the rate-operator transformation $C = k\, n_2$ for various values of the parameter $k$; the maximal admissible value of $k$ is defined as the largest value for which the resulting rate operator $R_\psi$ remains non-negative. The parameters and the ranges explored are summarized in Table~\ref{par-table}.

\subsection{Master equation and density matrix solutions \label{sec-ME-detev}}

\begin{table}[]
\begin{tabular}{|c|c|ccc|}
\hline 
\multirow{2}{*}{\quad$J/\gamma$\quad} & \multirow{2}{*}{\quad$U/\gamma$\quad}  & \multicolumn{3}{c|}{Maximal $k/\gamma$ value}    \\ \cline{3-5} 
                     &                         & \multicolumn{1}{c|}{$N=1$}   & \multicolumn{1}{c|}{$N=2$}    & $N=3$  \\ \hline
20            &  ~ $0-500$~          & \multicolumn{1}{c|}{\begin{tabular}[c]{@{}c@{}}NLD: 6.0\\~~ LD: $(1+\sqrt{3})$~~\end{tabular}} & \multicolumn{1}{c|}{\begin{tabular}[c]{@{}c@{}}NLD: 4.8\\ ~~LD: $2\cdot(1+\sqrt{3})$~~\end{tabular}} & \begin{tabular}[c]{@{}c@{}}NLD: 4.0\\~~ LD: $3\cdot(1+\sqrt{3})$~~\end{tabular} \\ \hline
\end{tabular}
\caption{Table of parameters used in this article. NLD refers to non-local dephasing and LD refers to local dephasing.}
\label{par-table}
\end{table}

\begin{figure}[t]
    \begin{subfigure}[t]{50mm}
        \includegraphics[width=\linewidth]{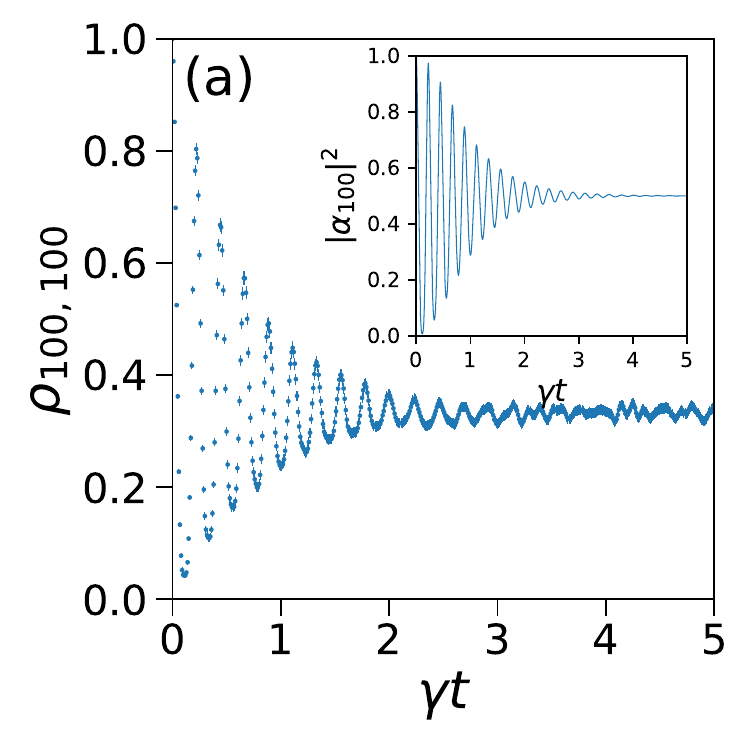}
    \end{subfigure}
    \begin{subfigure}[t]{50mm}
        \includegraphics[width=\linewidth]{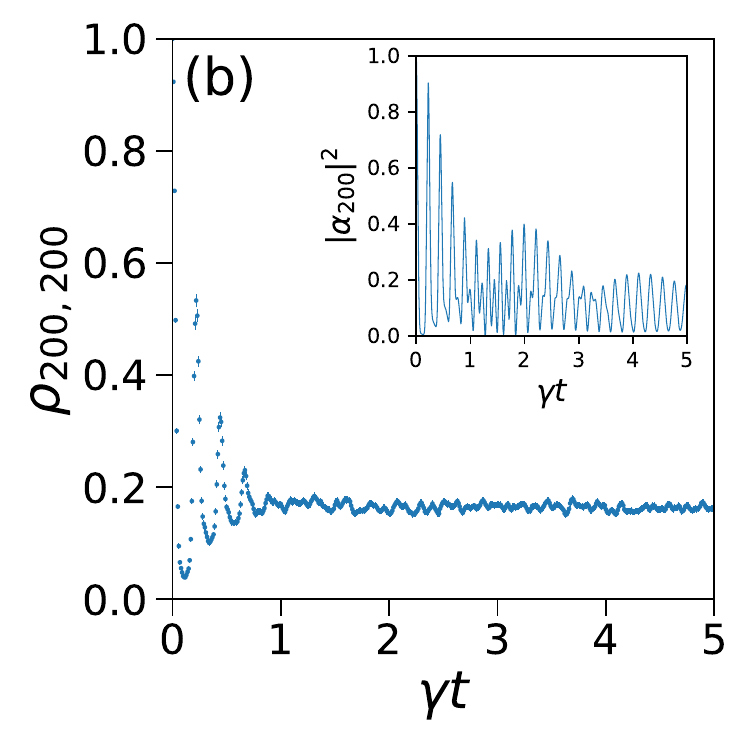}
    \end{subfigure}
    \begin{subfigure}[t]{50mm}
        \includegraphics[width=\linewidth]{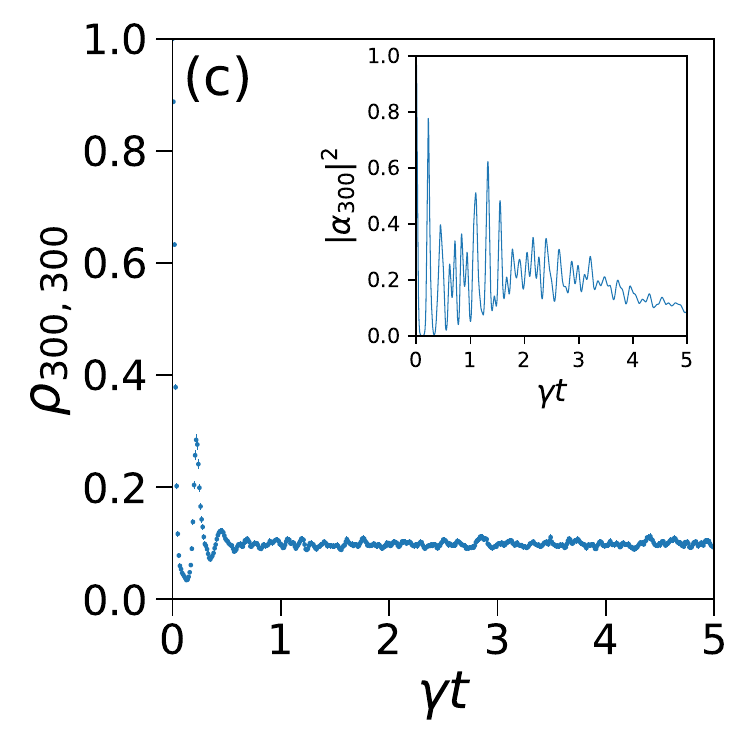}
    \end{subfigure}
    \caption{Solutions to the non-local dephasing master equation (\ref{me-nonlocal-dephasing}) for populations $\rho_{N00,N00}$ solved using the rate-operator formalism with (a): $N=1$, (b): $N=2$ and (c): $N=3$. Parameters are $J/\gamma=20$, $U/\gamma=5$. Insets have a corresponding deterministic evolutions controlled by effective Hamiltonian with rate operator transformations $C=k\,n_2$, with $k/\gamma$ values (a): 6.0, (b): 4.8 and (c): 4.0. The result was calculated using 1000 trajectories. The error bars represent the standard deviation of the populations divided by the square root of the number of trajectories.}
    \label{ME-solutions}
\end{figure}

Before turning to state-engineering study, we first examine the evolution of the density matrix $\rho(t)$. Figure~\ref{ME-solutions} shows the populations $\rho_{N00,N00} \equiv \bra{N,0,0}\rho\ket{N,0,0}$ ($N=1,2,3$) obtained by solving the master equation (\ref{me-nonlocal-dephasing}) via the ROQJ method with non-local dephasing operators $\{L_{ij}\}_{i,j}=\{\ket{i}\bra{i}-\ket{j}\bra{j}\}_{i,j}$, where pair $(i,j)$ runs over all pairs of the number-state basis elements with total particle number $N$. The figure also displays the deterministic trajectories of the absolute squares $|\alpha_{N00}|^2 \equiv |\braket{N,0,0|\psi}|^2$ generated by the effective Hamiltonian $K$, using rate-operator transformations of the form $C = k\, n_2$ with the maximal admissible values of $k$. The remaining populations exhibit asymptotic behavior analogous to that shown in Figure~\ref{ME-solutions}.
From the figure, it is evident that the characteristic time scales of the full density-matrix evolution are shorter than those associated with the deterministic evolution under the effective Hamiltonian $K$. In this case, the dependence on particle number $N$ shows contrasting trends: the time scale of the density-matrix evolution decreases with increasing $N$, whereas the deterministic evolution becomes slower for larger particle numbers.

\begin{figure}[t]
    \begin{subfigure}[t]{65mm}
        \includegraphics[width=\linewidth]{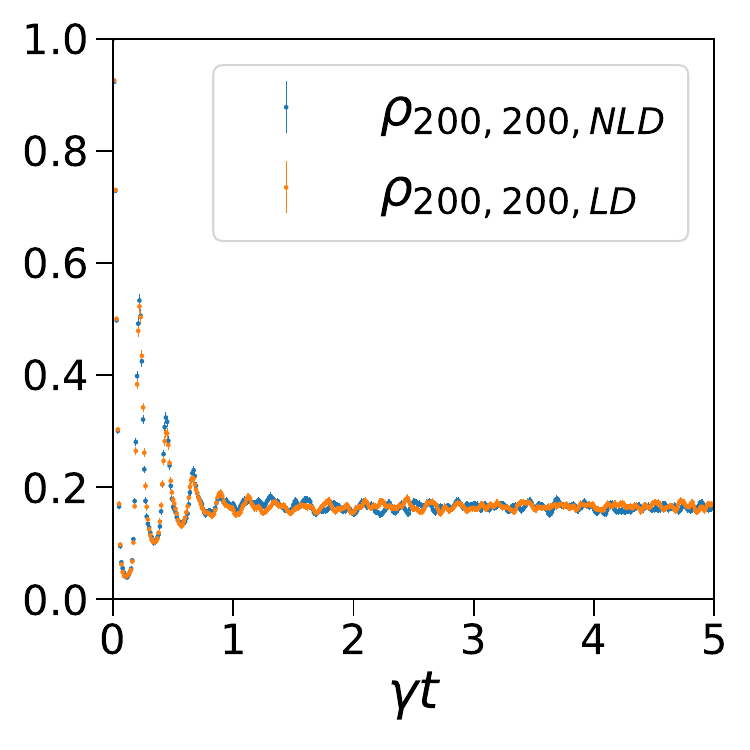}
    \end{subfigure}
    \caption{Master equation solutions for population $\rho_{200,200}$ solved using the rate-operator formalism for two different dephasing schemes for $N=2$. Parameters are $J/\gamma=20$, $U/\gamma=5$. The result was calculated using 1000 trajectories. The error bars represent the standard deviation of the populations divided by the square root of the number of trajectories.}
    \label{ME-LDvNLD-solutions}
\end{figure}

For a comparison of the two dephasing schemes, $L_i = n_i$ and $L_{ij} = \ket{i}\bra{i} - \ket{j}\bra{j}$, Figure~\ref{ME-LDvNLD-solutions} shows the evolution of the population $\rho_{200,200}$. As illustrated, no significant qualitative differences arise between the two approaches: the oscillation frequency is governed exclusively by the tunneling amplitude $J$, and the decay of the oscillation amplitudes proceeds in a similar fashion for both dephasing mechanisms.

For a general GKSL equation, determining the precise conditions under which the Lindblad dynamics possesses a unique non‑equilibrium steady state remains an open problem. Recently, however, Ref.~\cite{Yoshida2024} established a sufficient condition for the existence of a unique positive‑definite steady state in finite-dimensional Hilbert spaces under Markovian dynamics. The result states that a Lindblad master equation admits a unique positive‑definite non‑equilibrium steady state $\rho_\infty$ if the set
\begin{align}
\left \{ H - \frac{i}{2}\sum_{i=1}^{d^2 - 1} F_i^\dagger F_i,\ F_1,\ F_2,\ \ldots,\ F_{d^2-1} \right \} \label{eq-set-for-NESS}
\end{align}
generates the entire operator algebra under addition, multiplication, and multiplication by complex constants.
A corollary further states that when all Lindblad operators $F_i$ are Hermitian, the unique steady state is proportional to the identity, the maximally mixed state
$\rho_\infty = \frac{1}{d}\mathbbm{1}$.
Furthermore, Ref.~\cite{Yoshida2024} shows that in the case the GKSL equation has a strong symmetry, i.e. there is unitary operator $O$ on $\mathcal{H_S}$ such that
\begin{align}
    [H,O]=0 \quad \text{and}\quad [L_i,O]=0 \quad \forall i ,
\end{align}
there is instead a unique positive‑definite non‑equilibrium steady state $\rho_\infty^\alpha$ in each invariant subspace of $\mathcal{L}$ of the form 
$\mathcal{B}_\alpha=\{ \ket{\psi} \bra{\phi} ;  \ket{\psi}, \ket{\phi} \in \mathcal{K}_\alpha \}$,
where $\mathcal{K}_\alpha$ is an eigenspace of $O$, provided that the set~\eqref{eq-set-for-NESS} generates all the operators that commute with $O$ under addition, multiplication, and multiplication by complex constants. A corollary again states that for Hermitian Lindblad operators $F_i$ these non‑equilibrium steady states are
$\rho_\infty^\alpha = \frac{1}{d_\alpha}\mathbbm{1}$.
Our findings are consistent with this result, as the master‑equation solutions considered here appear to converge toward $\rho_\infty$ in a subspace related to an eigenspace of the total particle number operator $N=\sum_i n_i$. Indeed, Ref.~\cite{Yoshida2024} proves that this corollary applies to the tight-binding model on a general connected lattice $(\Lambda, B)$,
where $\Lambda$ denotes the set of lattice sites, and $B$ the set of edges such that the tunneling amplitudes $J_{i,j}= J_{j,i}^* \neq 0$ for all $(i,j) \in B$,
whenever the Hamiltonian is
\begin{align}
H = \sum_{(i,j)\in B} J_{i,j} a_i^\dagger a_j, \label{tight-binding-Hamiltonian}
\end{align}
and Lindblad operators are the number operators. The result holds provided the rates are strictly positive, $\gamma_i > 0$ for all $i \in \Lambda$.

The above result can be directly extended to the full Bose–Hubbard model, in which interaction and on-site potential terms are added to the Hamiltonian (\ref{tight-binding-Hamiltonian}):
\begin{align}
\sum_{i\in\Lambda}\frac{U_i}{2}n_i(n_i-1)+\sum_{i\in\Lambda}\epsilon_i n_i,
\end{align}
each of which depends solely on powers of the number operators $n_i$. Consequently, with Lindblad operators $L_i = n_i$, the Bose–Hubbard Hamiltonian and the tight-binding Hamiltonian (\ref{tight-binding-Hamiltonian}) generate the same operator algebra.
For the non-local dephasing operators $L_{ij} = \ket{i}\bra{i} - \ket{j}\bra{j}$, one can similarly generate all rank-one projectors. In particular, projections of the form 
$\ket{i}\bra{i}$ can be written as $L_{ij} L_{ik}$ for indices $i \neq j \neq k \neq i$, if the number of sites is greater than or equal to three. From these projectors, it is straightforward to generate the full operator space closed under addition, multiplication, and multiplication by complex numbers. Thus, the conditions of Ref.~\cite{Yoshida2024} are satisfied, and the unique steady state of the dynamics in each total particle number sector is the maximally mixed i.e. maximum entropy state of that particle number sector.

\subsection{Engineering deterministic evolution \label{sec-detev}}

The deterministic evolution generated by the effective Hamiltonian $K'$ can be controlled through the decay rates $\kappa_{ij}$ or $\gamma_i$ and through the choice of the transformation $C$. 
Variations in the values of $\kappa_{ij}$, $\gamma_i$, and in the form of the transformation are examined in Appendix~\ref{sec-non-hom-rates}. In the following, we again fix $\gamma \equiv\kappa_{jk} = \gamma_i$ and employ the family of transformations $C=k\, n_2$ with $k \in \mathbb{R}$.
This particular choice—while by no means unique—has the useful property that it yields well-defined steady states for the deterministic evolution. Moreover, the contribution $\frac{i}{2} k\, n_2$ appearing in the effective Hamiltonian due to the transformation is precisely the same term one would obtain by introducing the Lindblad operator $a_2$ directly into the master equation, corresponding to particle dissipation at site 2 with rate $k$. Thus, by appropriately modifying the measurement scheme, one can simulate deterministic evolution driven by particle loss even in systems governed solely by dephasing dynamics.
Furthermore, in the case of local dephasing, $L_i = n_i$, it is possible to choose negative $k$ while still preserving the non-negativity of the rate operator. This corresponds to deterministic evolution induced by non-Markovian particle dissipation, while retaining a consistent continuous-measurement interpretation.

We express a general $N$-particle state $\ket{\psi}$ of the system in the number basis as
\begin{align}
\ket{\psi}=\sum_{i+j+k=N}\alpha_{ijk}\ket{i,j,k}.
\end{align}
Assuming that the effective Hamiltonian $K$ ($K'$) is diagonalizable, the time evolution of the state $\ket{\psi(t)}$ can be written in terms of its eigenvalues $\lambda_i$, eigenstates $\ket{\phi_i}$, and the initial state $\ket{\psi_0}=\sum_i \beta_i(0)\ket{\phi_i}$ as
\begin{align}
\ket{\psi(t)}
= \sum_{i=1}^{d}\beta_i(0)e^{-i\lambda_i t}\ket{\phi_i}
= \sum_{i=1}^{d}\beta_i(0)e^{-i\,\mathrm{Re} \lambda_i t} e^{\mathrm{Im}\lambda_i t}\ket{\phi_i}.
\label{state-evolution}
\end{align}
Equation~(\ref{state-evolution}) provides a convenient representation of the system’s evolution. Although the effective Hamiltonian $K$ is non-Hermitian and therefore not unitarily diagonalizable, in all cases considered it possesses $d$ distinct eigenvalues and, consequently, $d$ linearly independent eigenvectors.
For non-negative decay rates and a non-negative rate operator, all imaginary parts 
$\mathrm{Im}\,\lambda_i$ are guaranteed to be non-positive. The magnitudes $\mathrm{Im}\,\lambda_i$ determine the decay rates at which the corresponding eigenvector components vanish from the unnormalized state. When the state is normalized during time evolution, the relevant decay rates are given by $\mathrm{Im}\,\lambda_1 - \mathrm{Im}\,\lambda_i$, where 
$\mathrm{Im}\,\lambda_1$ denotes the largest imaginary part among the eigenvalues.
Thus, the asymptotic dynamics are governed by the eigenstates associated with the eigenvalues whose imaginary part equals $\mathrm{Im}\,\lambda_1$. If this leading eigenvalue is non-degenerated, the corresponding eigenstate is the steady state of the evolution.

\begin{figure}[t]
    \begin{subfigure}[t]{50mm}
        \includegraphics[width=\linewidth]{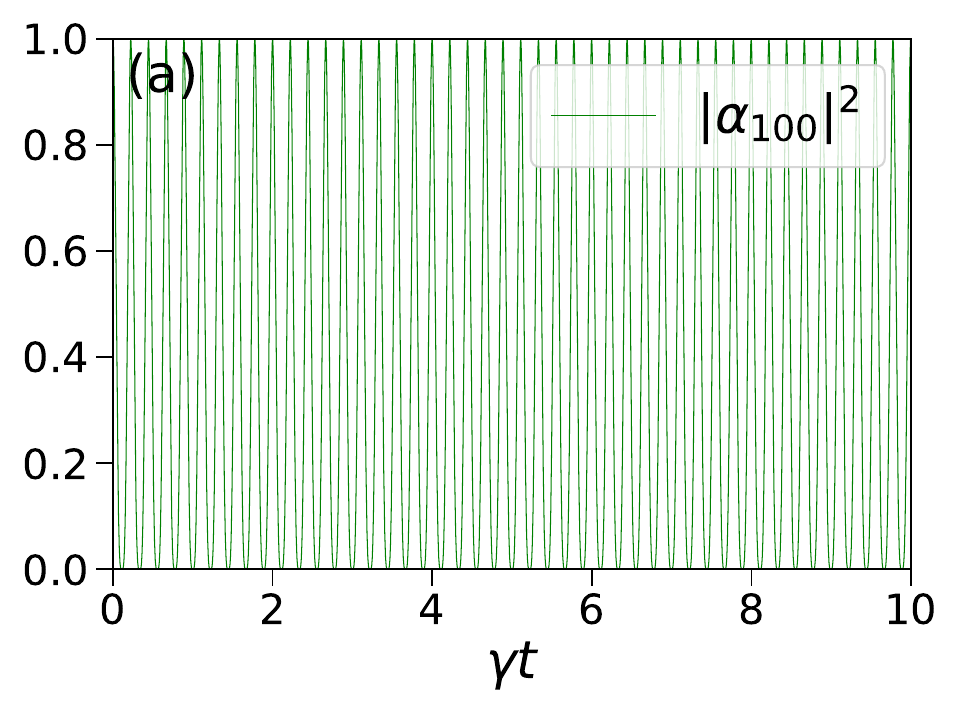}
    \end{subfigure}
    \begin{subfigure}[t]{50mm}
        \includegraphics[width=\linewidth]{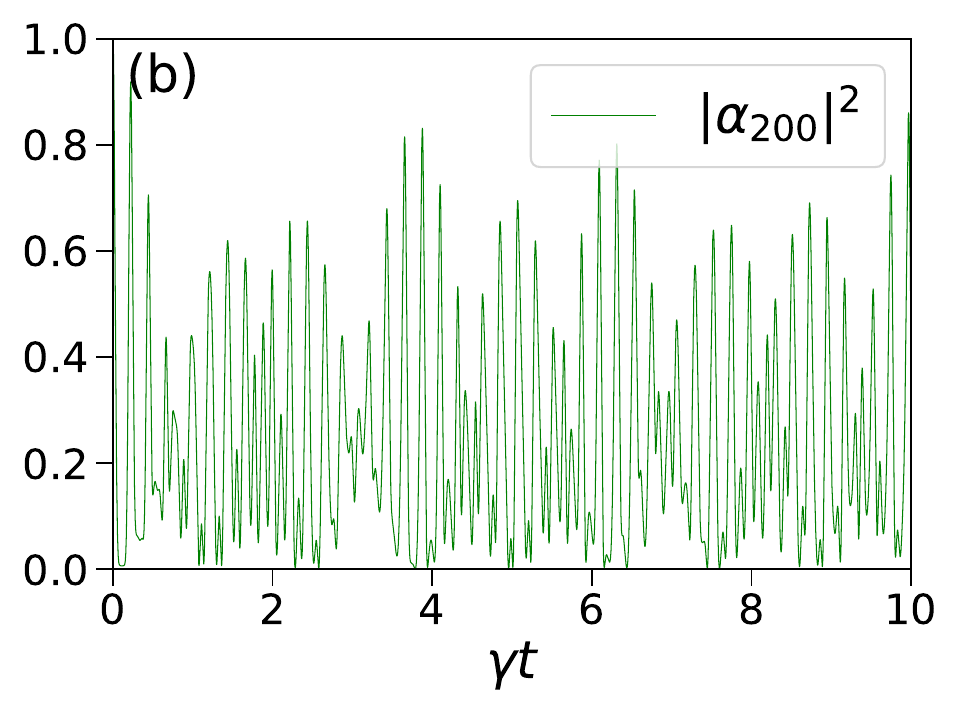}
    \end{subfigure}
    \begin{subfigure}[t]{50mm}
        \includegraphics[width=\linewidth]{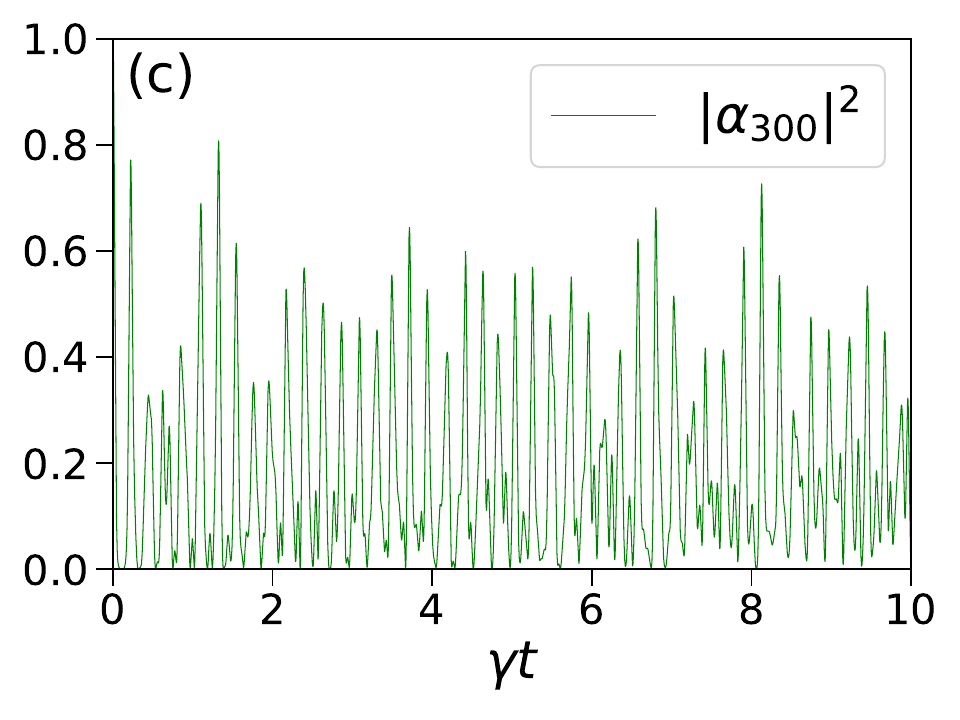}
    \end{subfigure}
    \caption{Examples of unitary evolutions for the particle numbers $N=1$, $N=2$ and $N=3$ with initial states $\ket{N,0,0}$ and parameters are $J/\gamma=20$ and $U/\gamma=5$. Fig. (a): Absolute square of the amplitude $\alpha_{100}$. Fig. (b): Absolute square of the amplitude $\alpha_{200}$. Fig. (c): Absolute square of the amplitude $\alpha_{300}$.}
    \label{uni}
\end{figure}

We begin by examining the unitary evolution of the Bose–Hubbard model governed by the Hamiltonian (\ref{BH_Hamiltonian}). Figure~\ref{uni} presents the unitary dynamics for the cases $N=1, N=2$, and $N=3$, using the parameter values $J/\gamma=20$ and $U/\gamma=5$. Since all eigenvalues of the Hamiltonian are real, no decay of eigenstate amplitudes occurs, and the time evolution remains purely oscillatory in the long-time limit. As the particle number increases, the oscillatory dynamics exhibit multiple characteristic timescales, reflecting the increased complexity of the underlying many-body spectrum.

Figure~\ref{N=1,2,3-U05-kij01-Cn2} presents the non-Hermitian evolution under the non-local dephasing scheme $L_{ij}=\ket{i}\bra{i}-\ket{j}\bra{j}$, using the same parameters $J/\gamma$ and $U/\gamma$ as in the unitary case and the transformation $C = k\, n_2$, where $k$ is chosen to be the largest real value for which the rate operator $R$ remains non-negative. The corresponding effective Hamiltonian is
\begin{align}
K' &= H - \frac{i}{2}\sum_{i<j}\kappa_{ij}\left(\ket{i}\bra{i} + \ket{j}\bra{j}\right) - \frac{i}{2}k\, n_2 \\
&= -J\sum_{i=1}^{2}(a_{i+1}^\dagger a_i + a_i^\dagger a_{i+1})
+ \frac{U}{2}\sum_{i=1}^{3} n_i(n_i - \mathbbm{1})
- i\frac{\gamma (d-1)}{2}\mathbbm{1}
- \frac{i}{2}k\, n_2 .
\label{Eff-Hamiltonian-NLD}
\end{align}
The term $-\tfrac{i}{2}\gamma(d-1)\mathbbm1$ contributes only an overall decay of the state norm and a global phase shift; it does not influence the relative dynamics of the amplitudes. The term $-\tfrac{i}{2} k \, n_2$ is the component that directly modifies the deterministic evolution and determines the asymptotic behavior. For all initial states that are not orthogonal to the steady state, the dynamics converge to that steady state.
The characteristic time scale of the evolution under this transformation is shorter than the longest time scales observed in the corresponding unitary evolution. As noted previously, the deterministic evolution becomes slower as the particle number increases, a trend that is also visible here.

\begin{figure}[h]
    \begin{subfigure}[t]{50mm}
        \includegraphics[width=\linewidth]{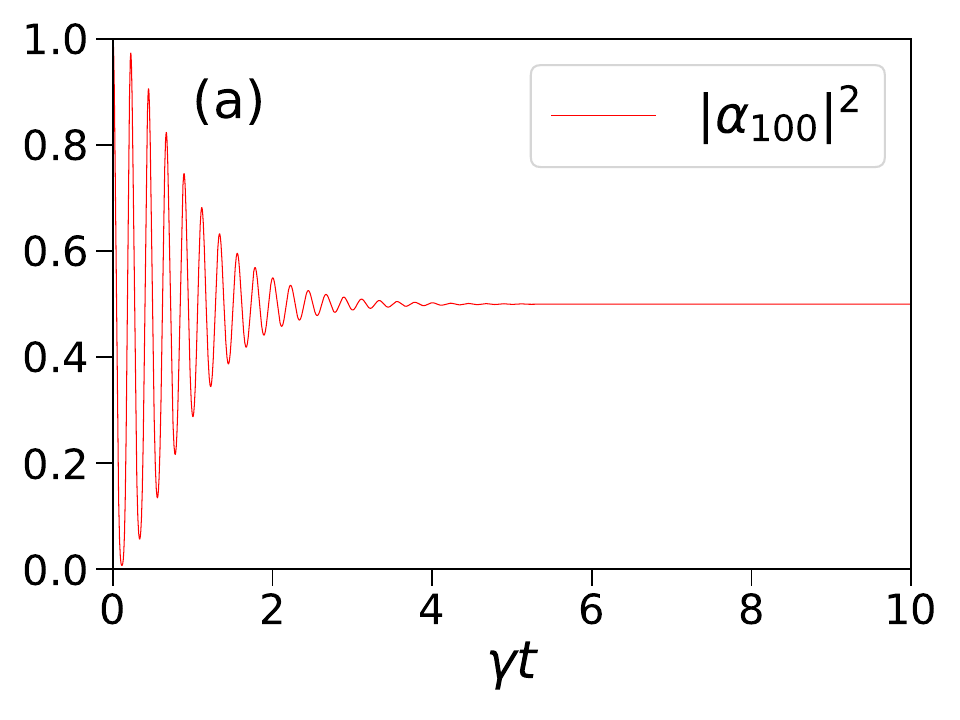}
    \end{subfigure}
    \begin{subfigure}[t]{50mm}
        \includegraphics[width=\linewidth]{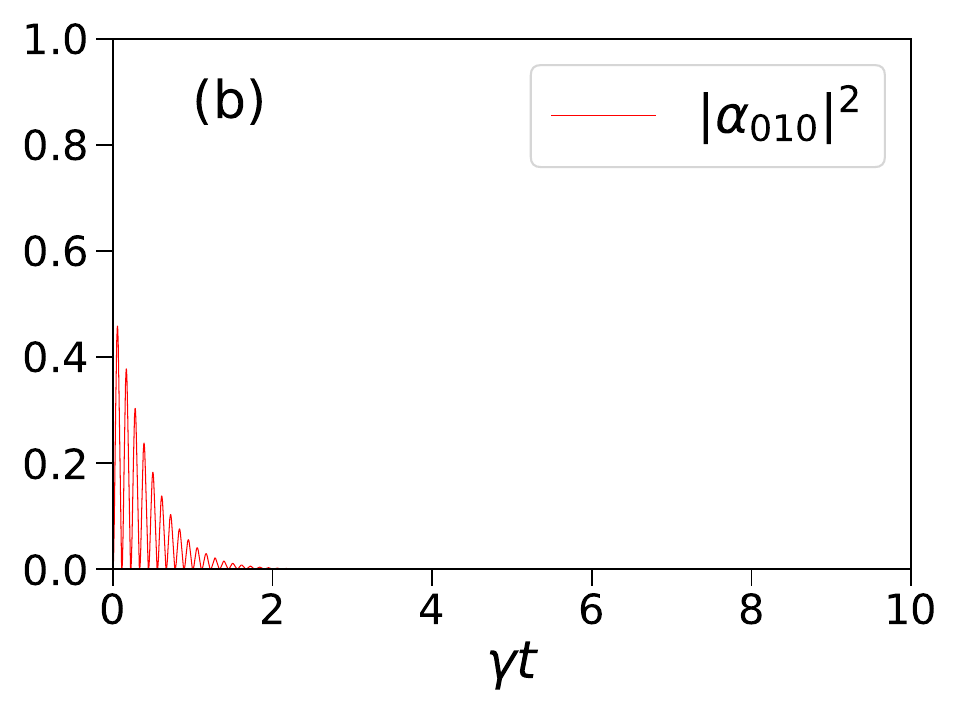}
    \end{subfigure}\\
    \begin{subfigure}[t]{50mm}
        \includegraphics[width=\linewidth]{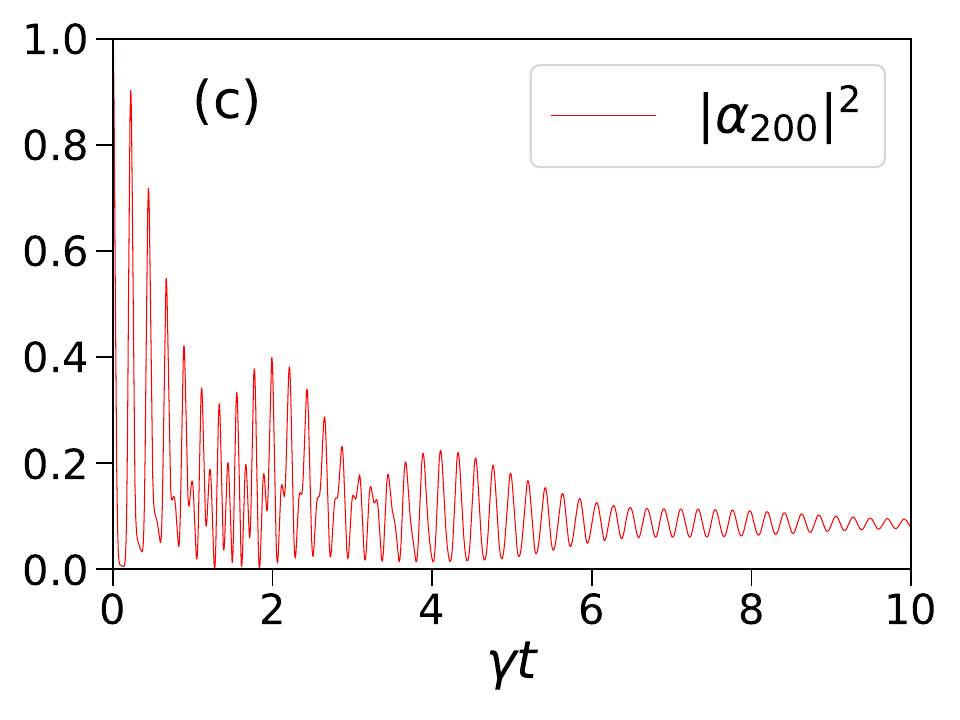}
    \end{subfigure}
    \begin{subfigure}[t]{50mm}
        \includegraphics[width=\linewidth]{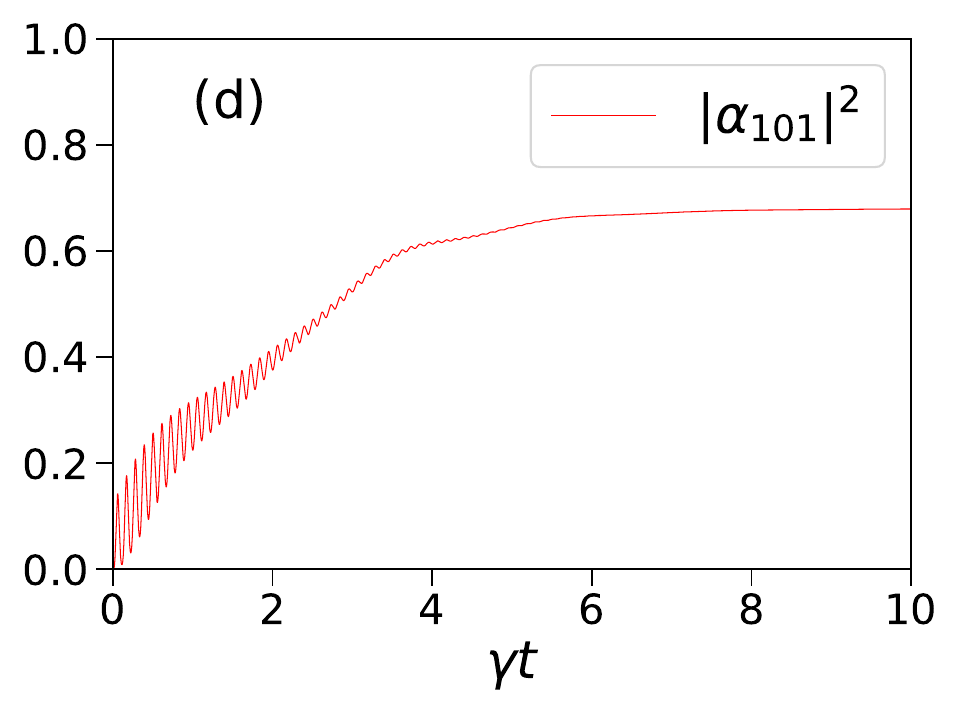}
    \end{subfigure}
    \begin{subfigure}[t]{50mm}
        \includegraphics[width=\linewidth]{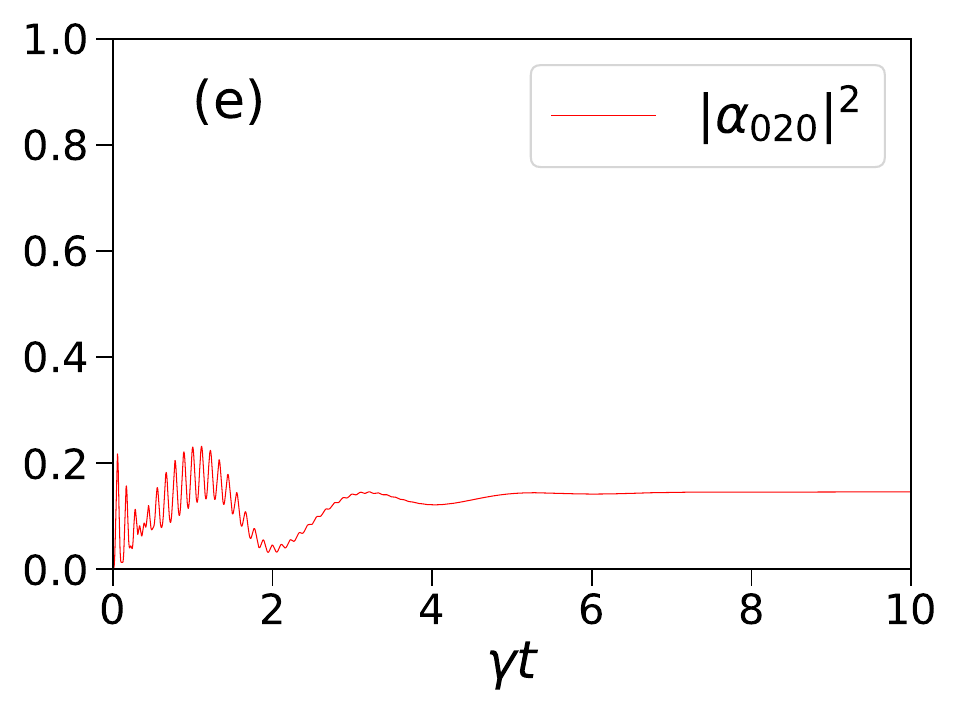}
    \end{subfigure}\\
    \begin{subfigure}[t]{50mm}
        \includegraphics[width=\linewidth]{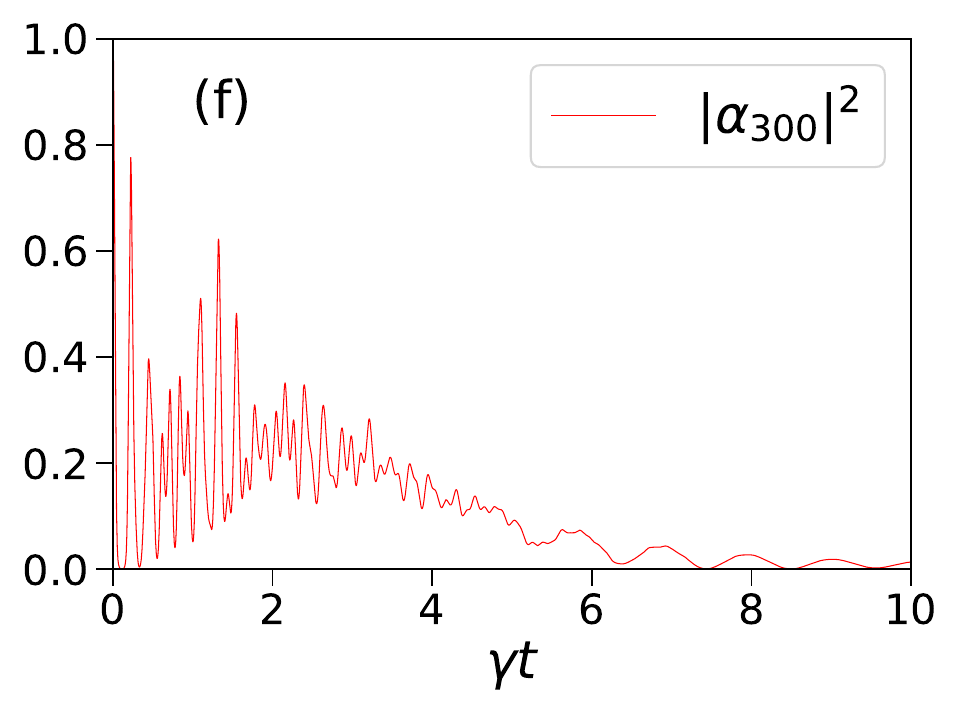}
    \end{subfigure}
    \begin{subfigure}[t]{50mm}
        \includegraphics[width=\linewidth]{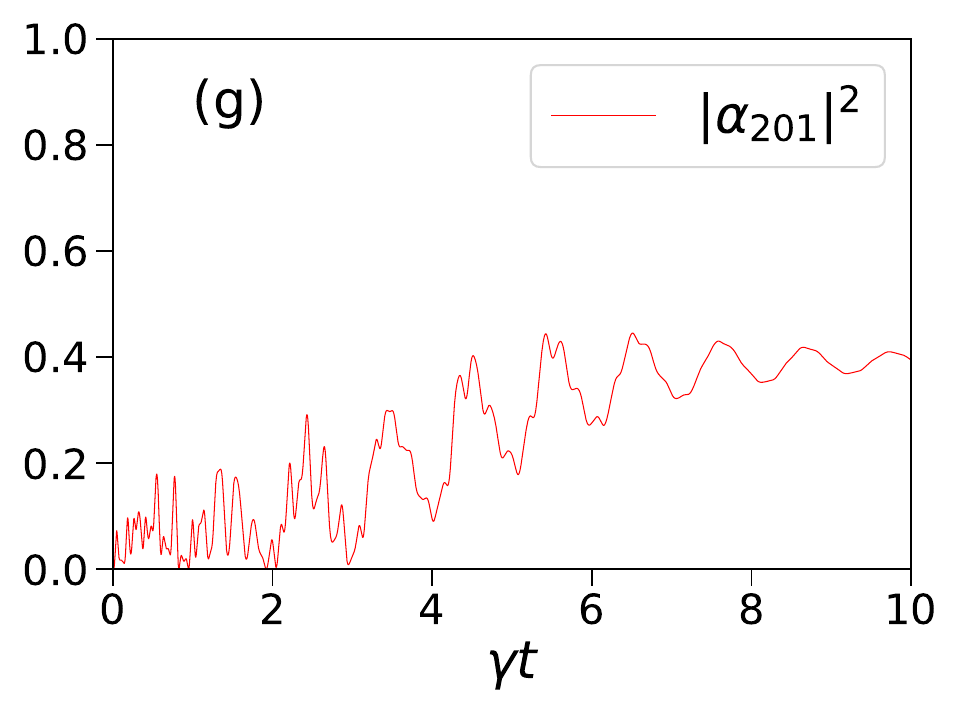}
    \end{subfigure}
    \begin{subfigure}[t]{50mm}
        \includegraphics[width=\linewidth]{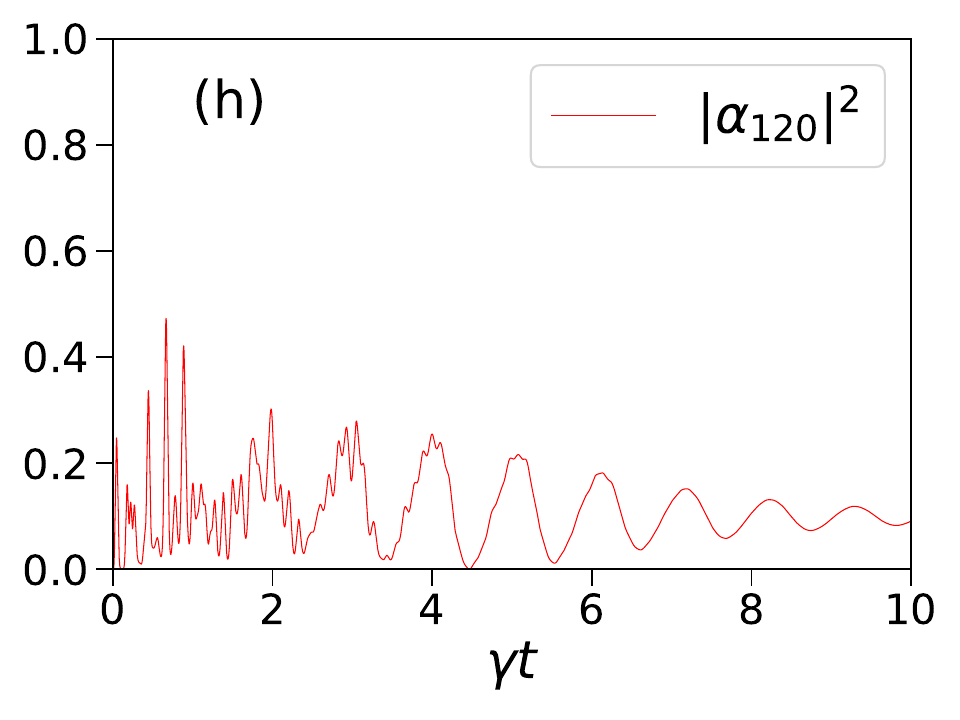}
    \end{subfigure}
    \caption{    Deterministic evolutions of the absolute squares of the amplitudes under non-local dephasing, generated by the non-Hermitian Hamiltonian (\ref{Eff-Hamiltonian-NLD}) are shown for $N=1$ (panels a–b), $N=2$ (panels c–e), and $N=3N$ (panels f–h). In all cases, the initial state is $\ket{N,0,0}$ and the parameter values are $J/\gamma = 20$, $U/\gamma = 5$. The rate-operator transformation employed is $C = k\, n_2$ with $k/\gamma = 6.0,\, 4.8$, and $k/\gamma = 4.0$ for particle numbers $N = 1,\, N = 2$, and $N = 3$, respectively. Components not shown either exhibit dynamics analogous to their symmetric counterparts or decay rapidly to zero. }
    \label{N=1,2,3-U05-kij01-Cn2}
\end{figure}

\begin{figure}[h]
    \begin{subfigure}[t]{50mm}
        \includegraphics[width=\linewidth]{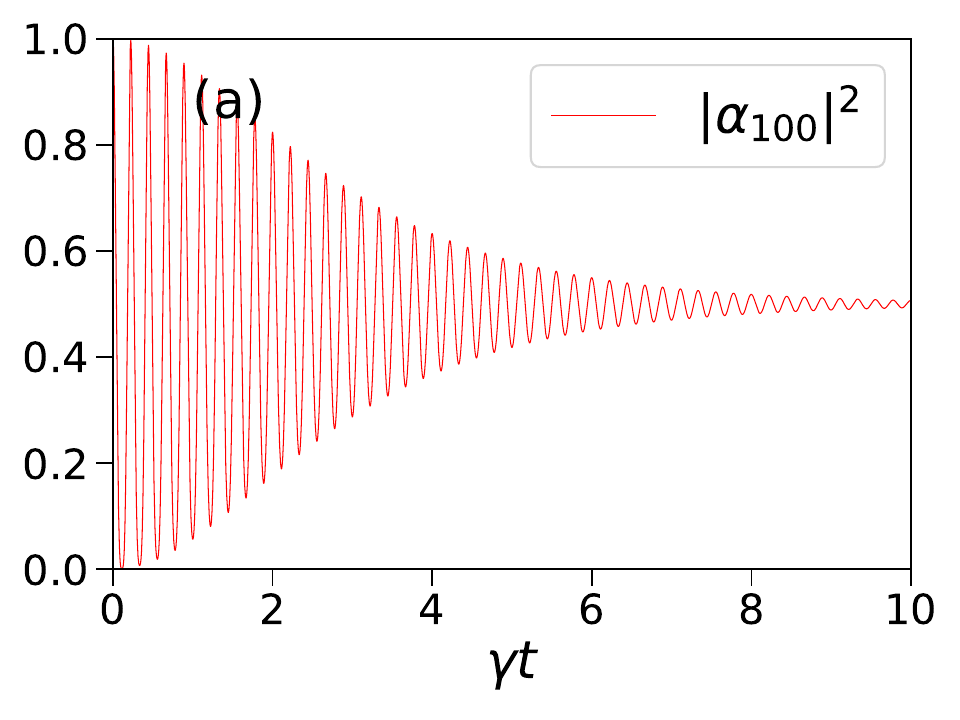}
    \end{subfigure}
    \begin{subfigure}[t]{50mm}
        \includegraphics[width=\linewidth]{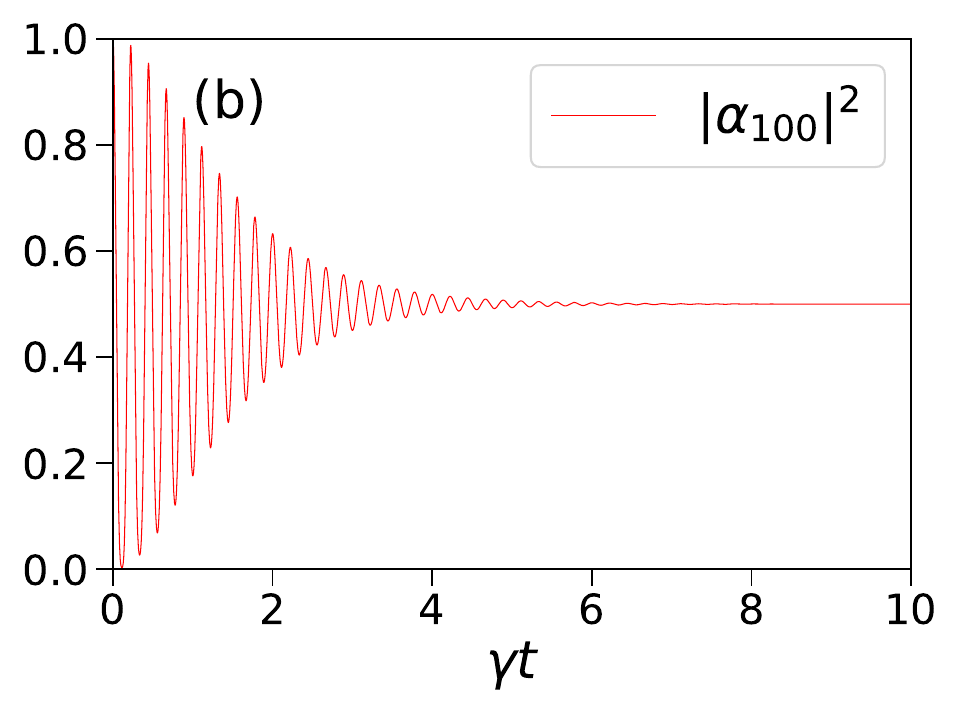}
    \end{subfigure}
    \begin{subfigure}[t]{50mm}
        \includegraphics[width=\linewidth]{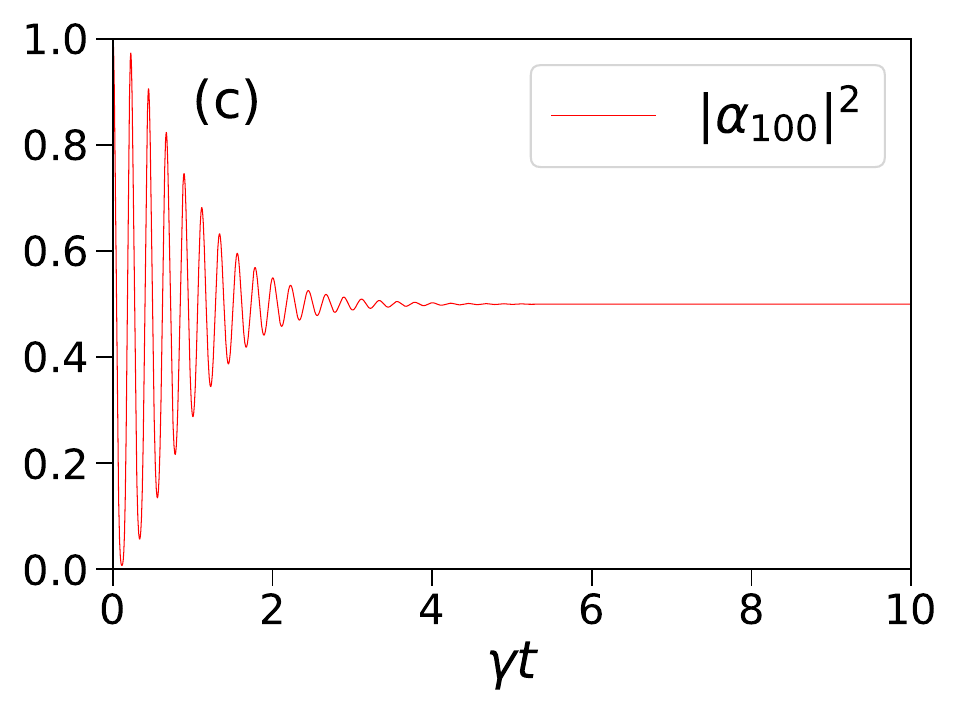}
    \end{subfigure}
    \caption{Evolution of $|\alpha_{100}|^2$ using non-local dephasing and rate-operator transformation $k\, n_2$. The initial $N=1$ state is $\ket{\psi_0}=\ket{1,0,0}$ and parameters are $J/\gamma=20$. Panels a-c correspond $k/\gamma=2.0$, 4.0 and 6.0, respectively.}
    \label{aikaskaala}
\end{figure}

Figure~\ref{aikaskaala} illustrates the influence of the parameter $k$ in the rate-operator transformation $C = k\, n_2$ under non-local dephasing. Varying $k$ affects exclusively the characteristic timescale of the evolution: larger values of $k$ lead to shorter evolution timescales. This behavior is advantageous for state-engineering applications based on non-Hermitian Hamiltonians. Analogous behavior is observed for $N = 2$ and $N = 3$. These observations motivate the use of the maximal admissible values of $k$.

\begin{figure}[h]
    \begin{subfigure}[t]{65mm}
        \includegraphics[width=\linewidth]{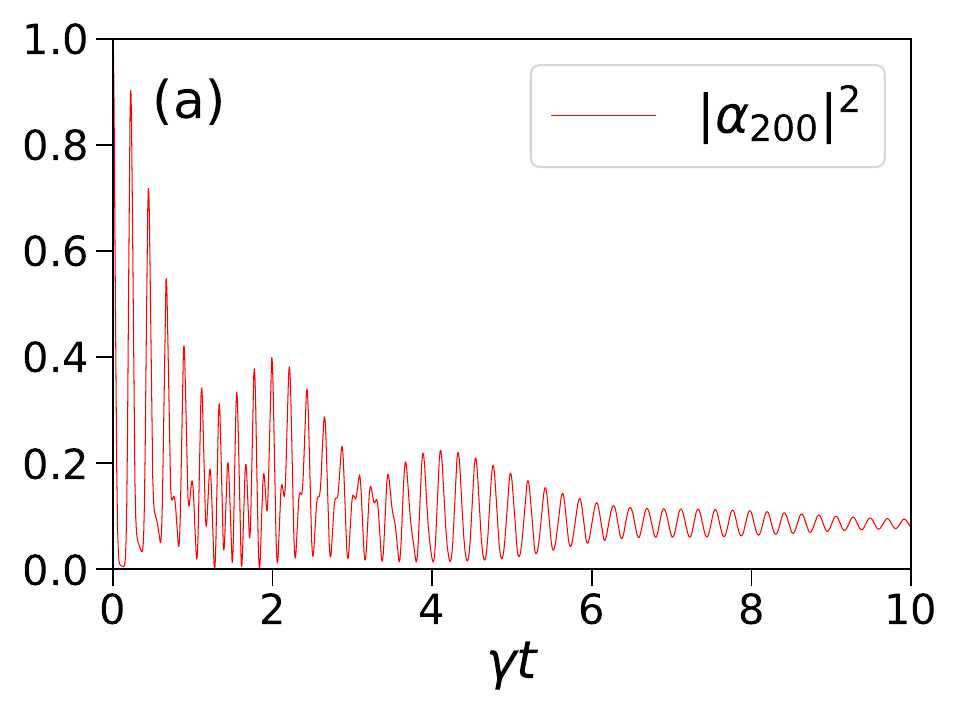}
    \end{subfigure}
    \begin{subfigure}[t]{65mm}
        \includegraphics[width=\linewidth]{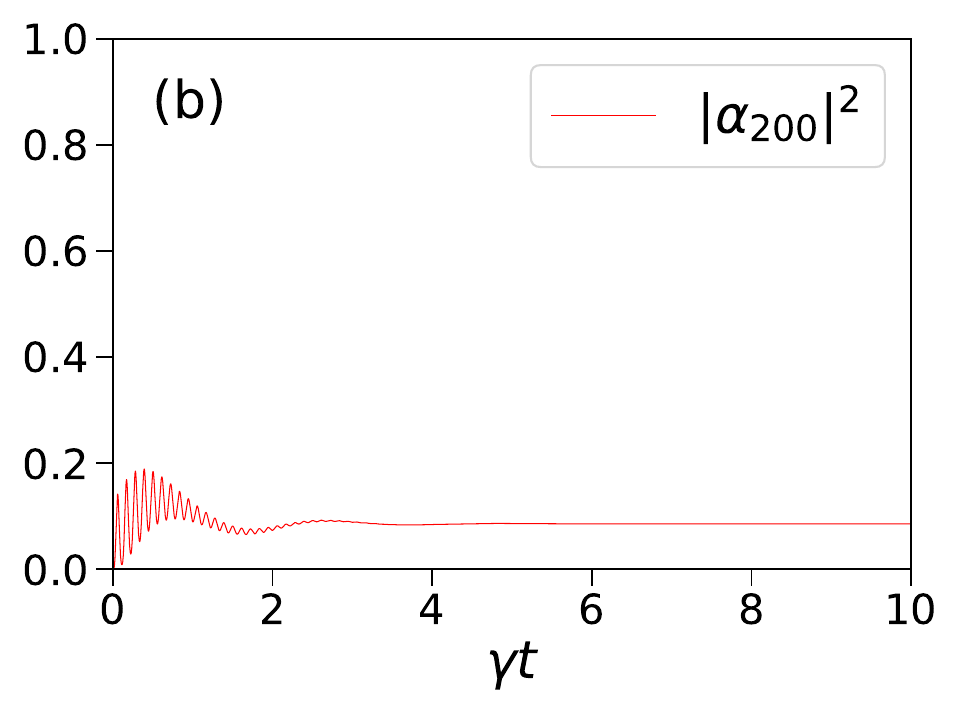}
    \end{subfigure}
    \caption{Difference in relaxation speed between different initial $N=2$ states using non-local dephasing. Figures show the evolution of absolute square of the amplitude 
    $\alpha_{200}$ with parameters $J/\gamma=20$, $U/\gamma=5$ and rate operator transformation $4.8/\gamma\, n_2$. The initial states are (a): $\ket{\psi_0}=\ket{2,0,0}$ and (b): $\ket{\psi_0}=\ket{1,0,1}$.}
    \label{relaxation-speed-N=2-U05-k048-200-101}
\end{figure}

Figure~\ref{relaxation-speed-N=2-U05-k048-200-101} illustrates the influence of the initial state on the relaxation speed under non‑local dephasing. The observed differences in relaxation rates between various initial states can be attributed to the magnitude of the overlap $\braket{\psi_{0}|\psi_{ss}}$, where $\ket{\psi_{0}}$ denotes the initial state and 
$\ket{\psi_{ss}}$ is the steady state of the effective Hamiltonian (\ref{Eff-Hamiltonian-NLD}). A larger overlap leads to a faster convergence toward the steady state, provided that such a steady state exists.

Figure~\ref{N=2-U0,05,10-k01-C048n2-200} shows the effect of varying the interaction strength $U/\gamma$ on the evolution of the state under non-local dephasing for particle number $N = 2$ and transformation $C = k\, n_2$, where $k=4.8\gamma$. Increasing the interaction strength $U/\gamma$ suppresses the rapid oscillations associated with the tunneling amplitude $J/\gamma$, leaving only oscillations occurring on longer time scales. The origin of these distinct time scales and their dependence on system parameters is discussed further in Section~\ref{sec-asymptotic}.

\begin{figure}[h]
    \begin{subfigure}[t]{50mm}
        \includegraphics[width=\linewidth]{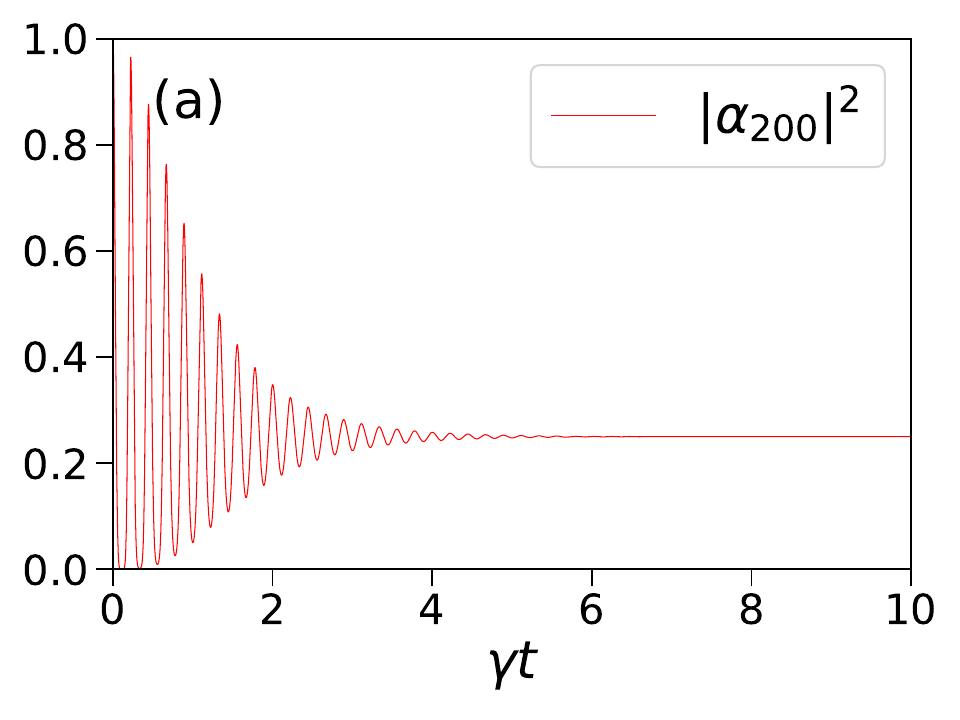}
    \end{subfigure}
    \begin{subfigure}[t]{50mm}
        \includegraphics[width=\linewidth]{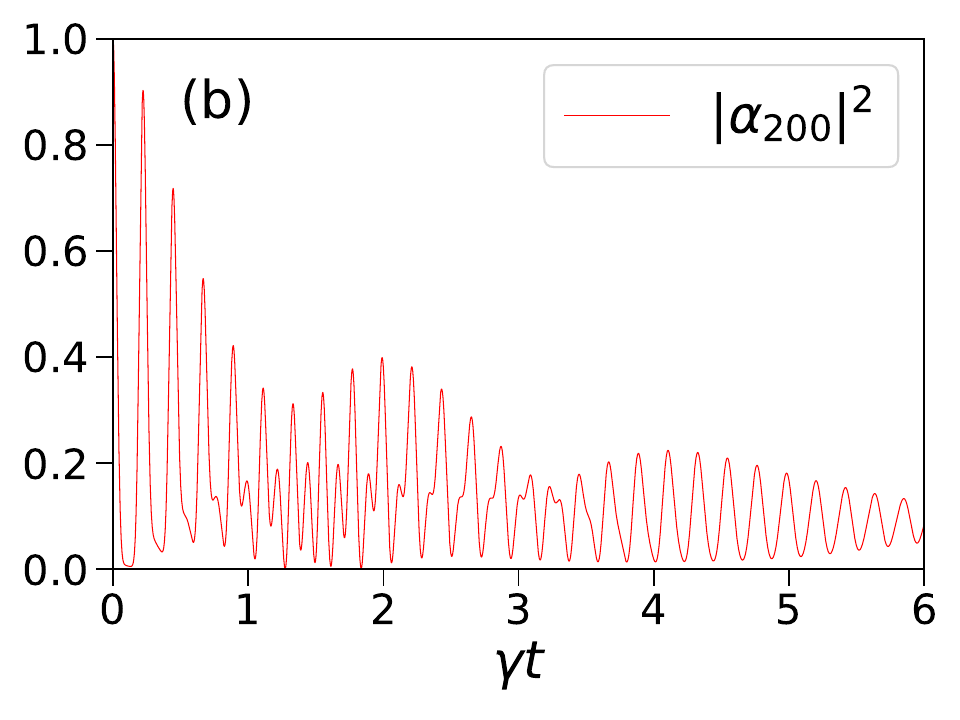}
    \end{subfigure}
    \begin{subfigure}[t]{50mm}
        \includegraphics[width=\linewidth]{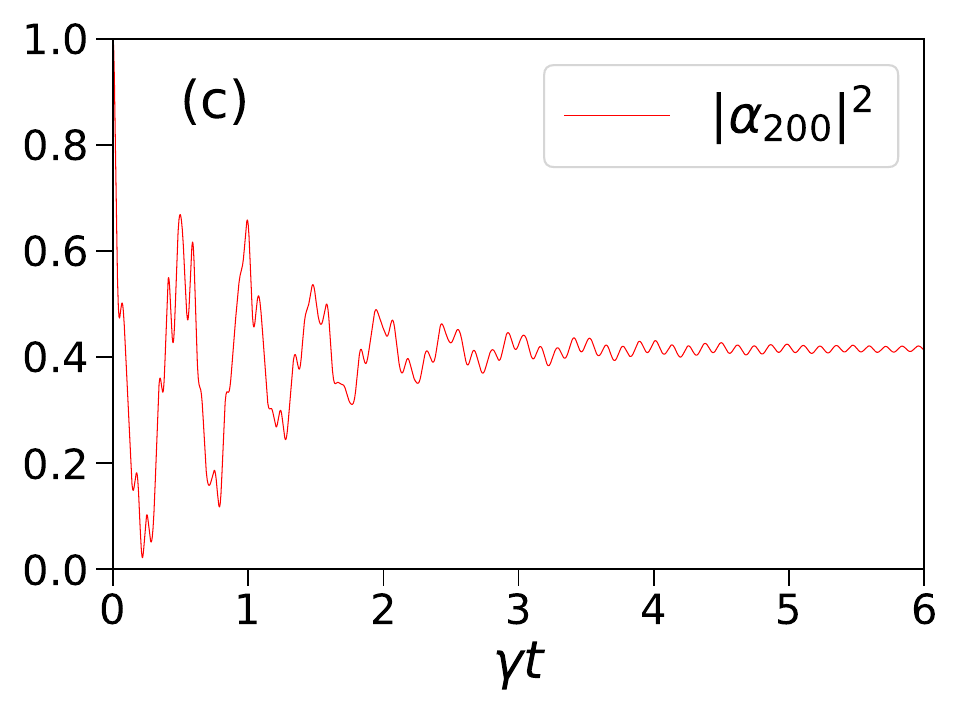}
    \end{subfigure} \\
    \begin{subfigure}[t]{50mm}
        \includegraphics[width=\linewidth]{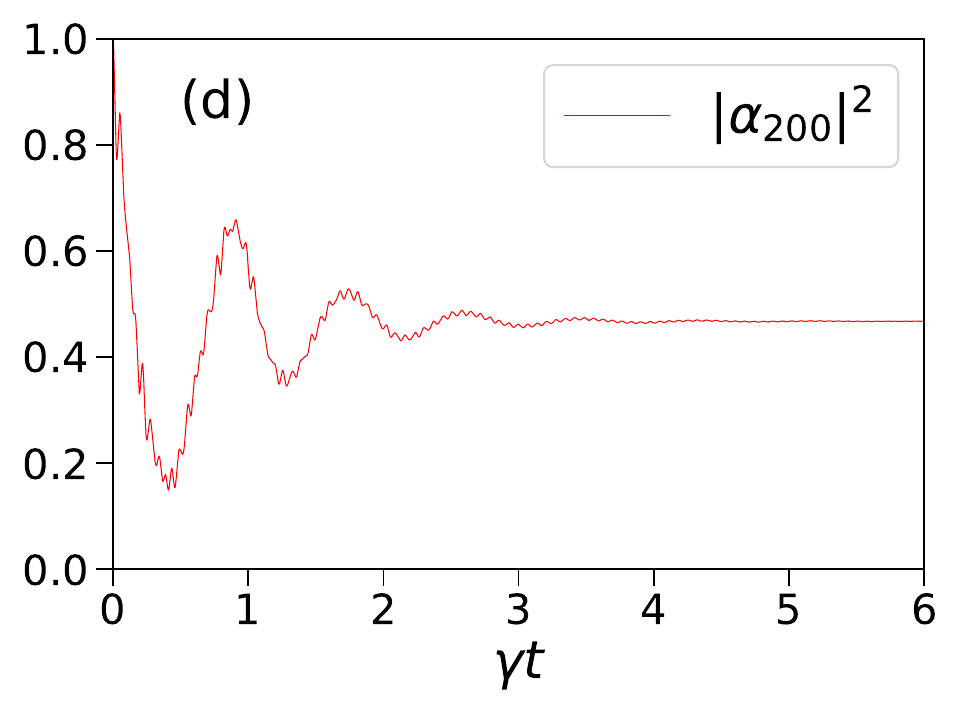}
    \end{subfigure}
    \begin{subfigure}[t]{50mm}
        \includegraphics[width=\linewidth]{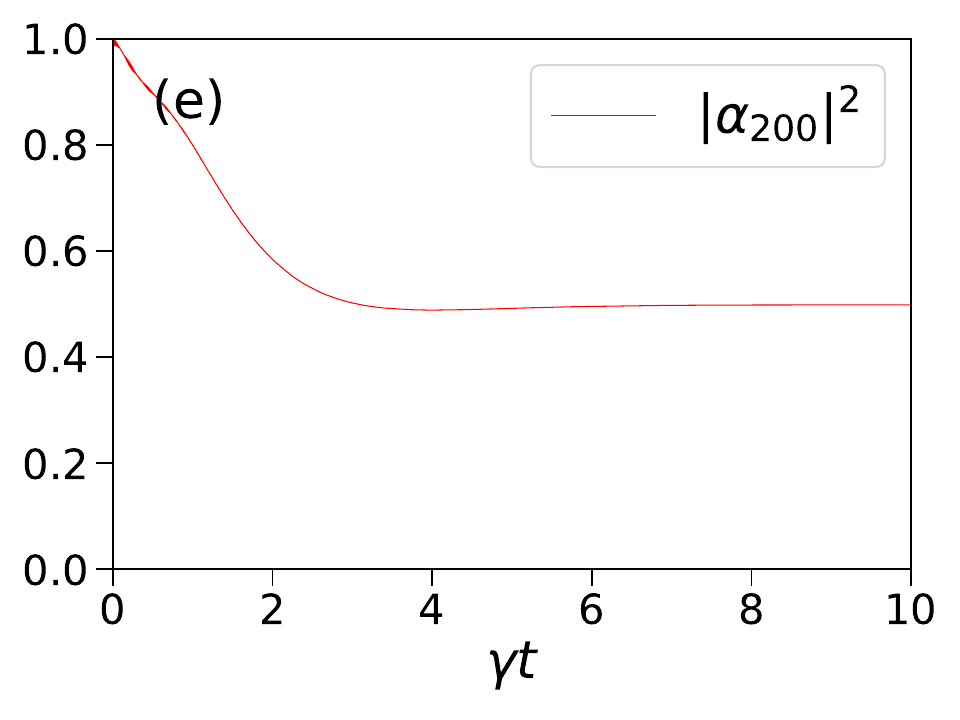}
    \end{subfigure}
    \caption{Deterministic evolution of $|\alpha_{200}|^2$ using non-local dephasing and rate operator transformation $C/\gamma=4.8\, n_2$, parameters $J/\gamma=20$ and the initial state $\ket{\psi_0}=\ket{2,0,0}$ for different values of interaction strength $U/\gamma$:  (a) $U=0$. (b) $U/\gamma=5$. (c) $U/\gamma=50$. (d) $U/\gamma=100$. (e) $U/\gamma=500$.}
    \label{N=2-U0,05,10-k01-C048n2-200}
\end{figure}

To compare the local and non-local dephasing schemes, we need the effective Hamiltonian associated with local dephasing. For $L_i = n_i$ and the transformation $C = k\, n_2$, it takes the form
\begin{align}
K' &= H - \frac{i}{2}\sum_{i=1}^3 \gamma_i n_i^2 - \frac{i}{2}k\, n_2. \label{Eff-Hamiltonian-LD}
\end{align}
If the decay rates are uniform, $\gamma_i \equiv \gamma \;\forall i$, the analytically solvable interval ensuring that the rate operator remains non-negative is
$k \in \bigl[\, N\gamma(1 - \sqrt{3}),\; N\gamma(1 + \sqrt{3}) \,\bigr]$.
For $N = 1$, the identity $n_i = n_i^2$ holds, implying that the effective Hamiltonian—and hence the deterministic evolution—are qualitatively identical for the local and non-local dephasing schemes. For $N > 1$, however, this equivalence no longer holds, and the two dephasing mechanisms yield different dynamical behavior.

\begin{figure}[h]
    \begin{subfigure}[t]{65mm}
        \includegraphics[width=\linewidth]{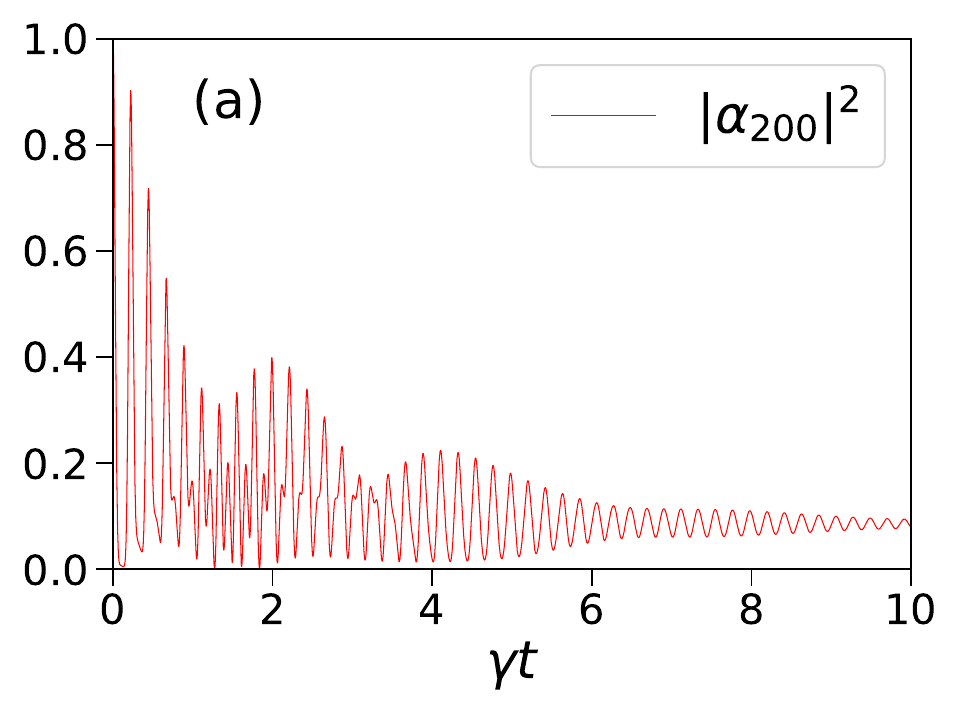}
    \end{subfigure}
    \begin{subfigure}[t]{65mm}
        \includegraphics[width=\linewidth]{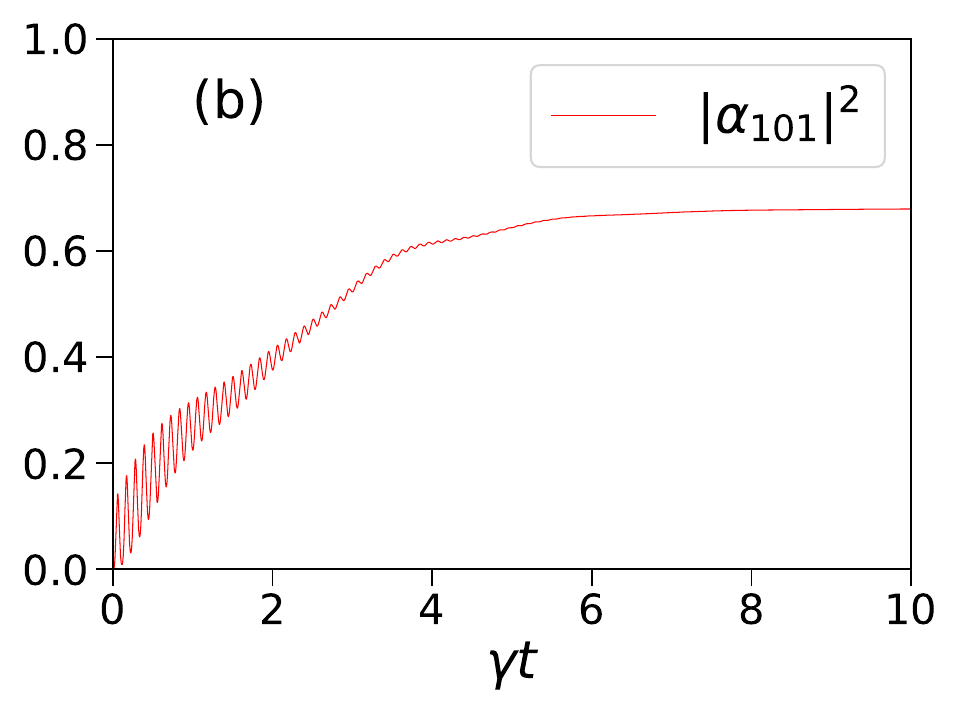}
    \end{subfigure} 
    \begin{subfigure}[t]{65mm}
        \includegraphics[width=\linewidth]{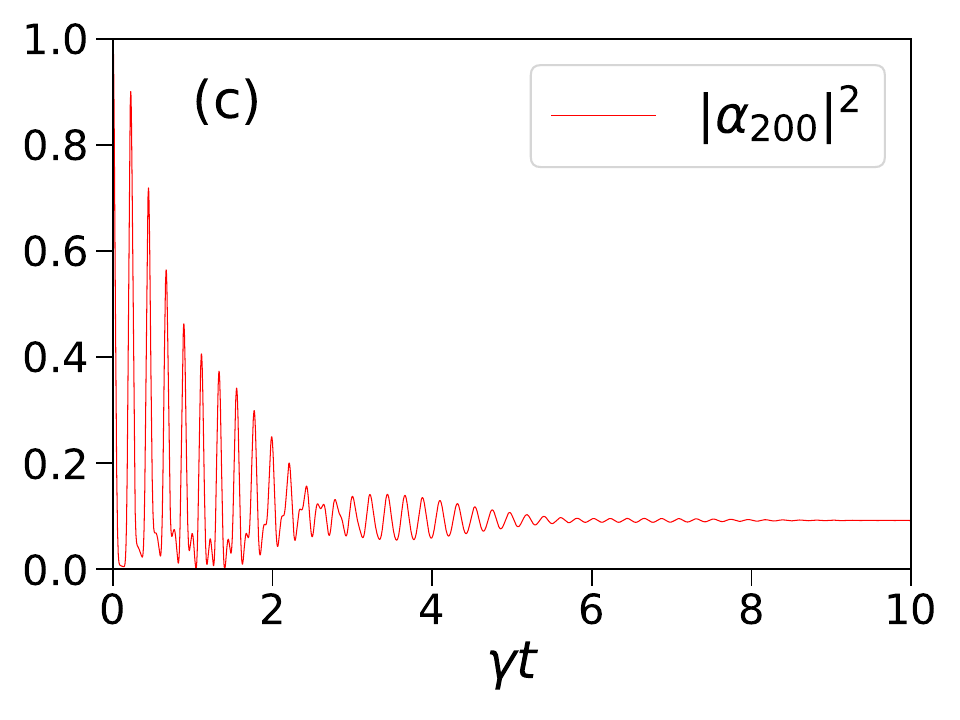}
    \end{subfigure}
    \begin{subfigure}[t]{65mm}
        \includegraphics[width=\linewidth]{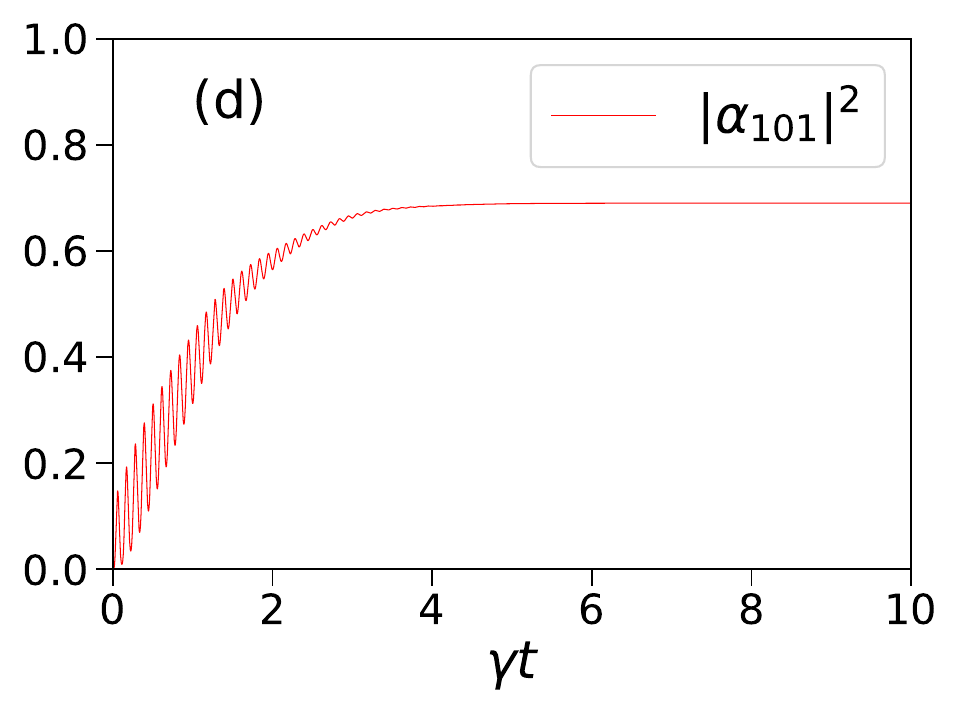}
    \end{subfigure}
    \caption{Deterministic evolutions given by non-Hermitian Hamiltonian $K'$ for non-local, (a-b), and local dephasing, (c-d). Parameters are $J/\gamma=2$ and $U/\gamma=0.5$. Rate operator transformation $C/\gamma=4.8\, n_2$ is used for both dephasing schemes. The initial state is $\ket{2,0,0}$ and amplitudes shown are $\alpha_{200}$ in figs. (a) and (c) and $\alpha_{101}$ in figs. (b) and (d).}
    \label{LvNL-N=2}
\end{figure}

Figure~\ref{LvNL-N=2} compares the dynamics generated by the non‑local and local dephasing schemes for the case $N = 2$. For both schemes we set $k/\gamma = 4.8$, chosen such that the corresponding rate operator remains non‑negative, enabling a fair comparison.
The remaining parameters are fixed to $J/\gamma = 20$ and  $U/\gamma = 5$. The figure displays the evolution of the absolute squares of the amplitudes associated with the components $\ket{2,0,0}$ and $\ket{1,0,1}$.
The most pronounced difference between the two dephasing schemes lies in the characteristic timescale: the relaxation is noticeably faster under local dephasing. The components not shown exhibit no additional qualitative differences. The disparity in timescales arises from the fact that the rates in the two schemes are not fully comparable. In the NLD scheme, the Lindblad operators contribute a term proportional to the identity, which does not affect the evolution of the absolute values of the amplitudes.
In contrast, the LD scheme introduces a term proportional to $\sum_i n_i^2$, which directly influences the amplitudes. Indeed, the shorter timescale observed for local dephasing appears to result from the presence of the term $-\frac{i}{2}\gamma \sum_i n_i^2$.

\begin{figure}[h!]
    \begin{subfigure}[t]{65mm}
        \includegraphics[width=\linewidth]{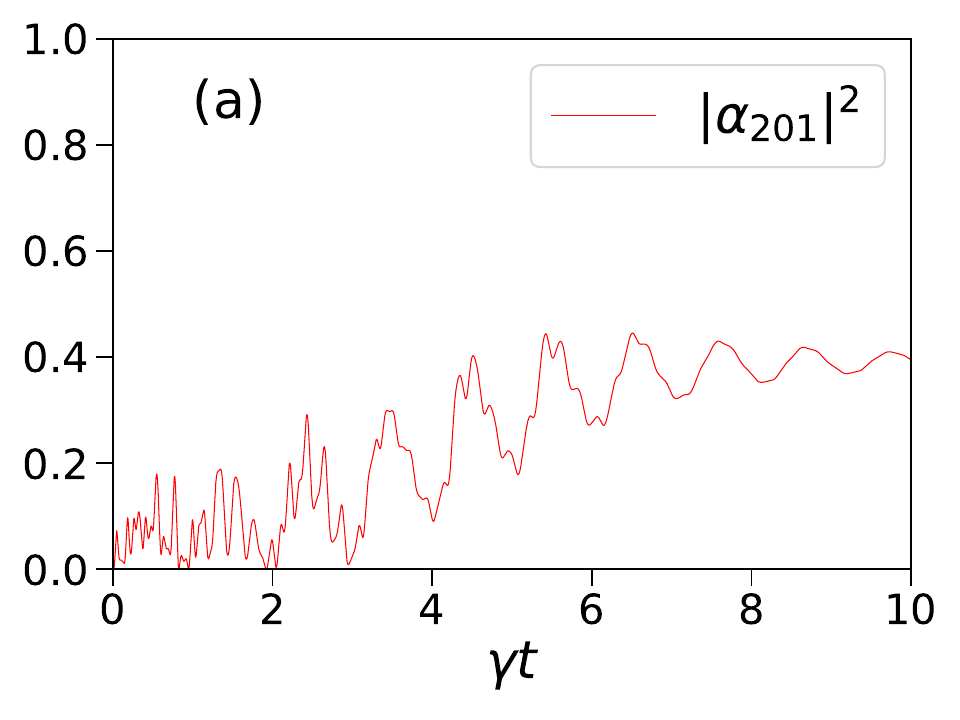}
    \end{subfigure}
    \begin{subfigure}[t]{65mm}
        \includegraphics[width=\linewidth]{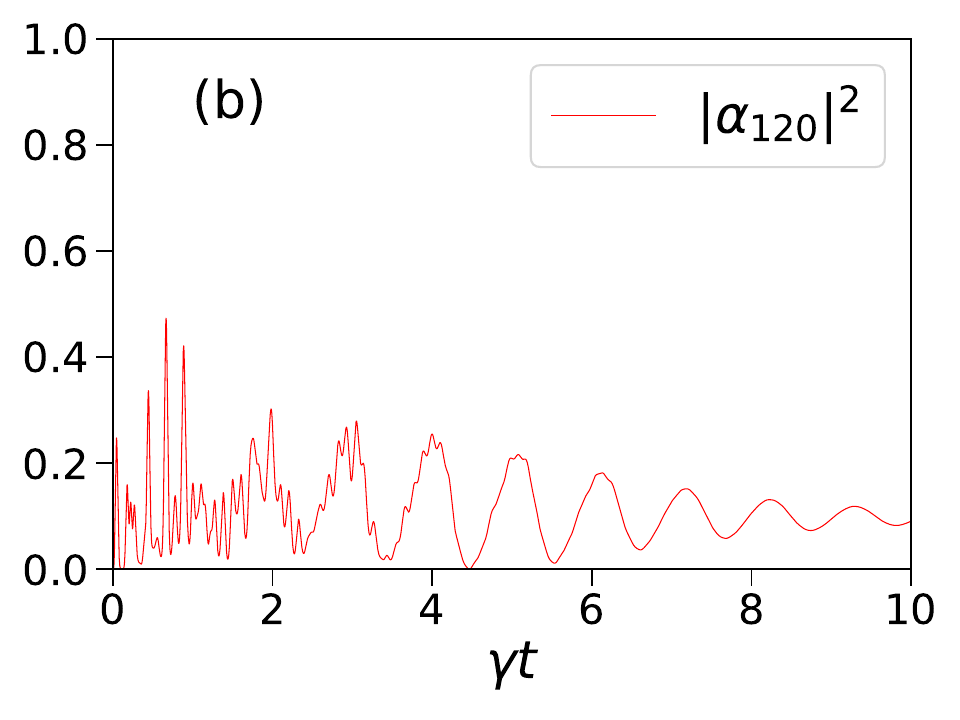}
    \end{subfigure} 
    \begin{subfigure}[t]{65mm}
        \includegraphics[width=\linewidth]{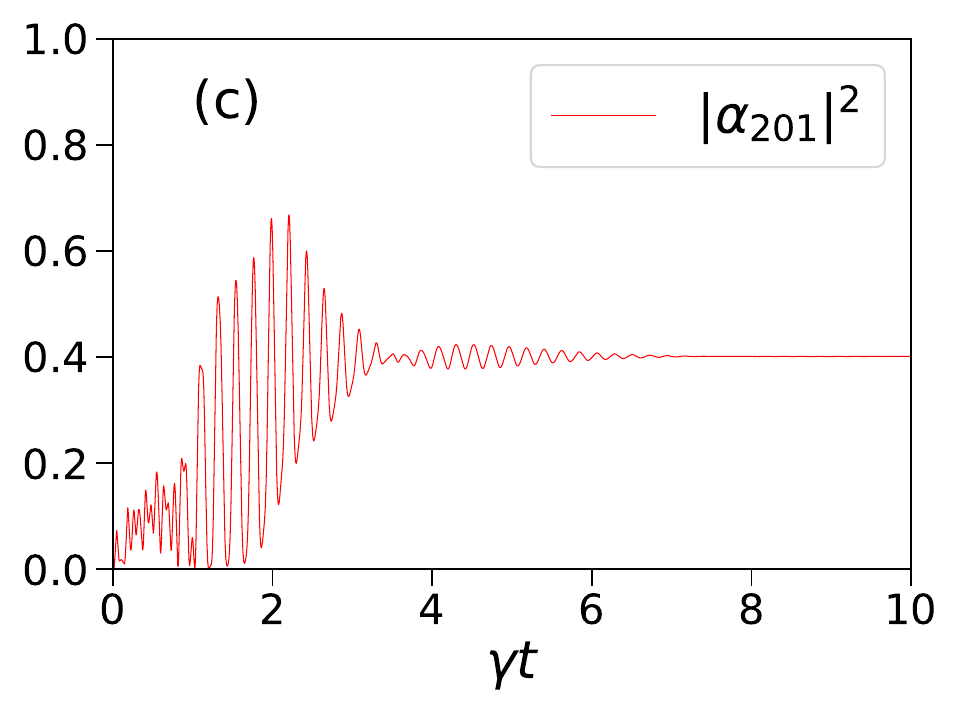}
    \end{subfigure}
    \begin{subfigure}[t]{65mm}
        \includegraphics[width=\linewidth]{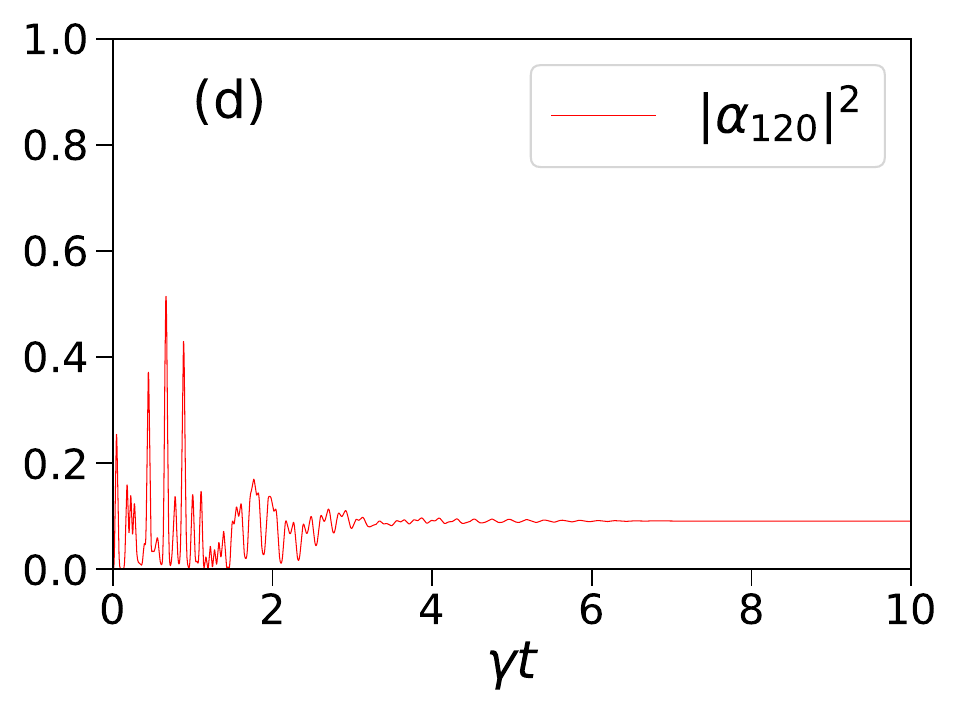}
    \end{subfigure}
    \caption{Deterministic evolutions given by non-Hermitian Hamiltonian $K'$ for non-local, (a-b), and local dephasing, (c-d). Parameters are $J/\gamma=20$ and $U/\gamma=5$. Rate operator transformation $C=k\, n_2$, where $k=0.4\gamma$, is used for both dephasing schemes. The initial state is $\ket{3,0,0}$ and amplitudes shown are $\alpha_{201}$ in figs. (a) and (c), and $\alpha_{120}$ in figs. (b) and (d).}
    \label{LvNL-N=3}
\end{figure}

Figure \ref{LvNL-N=3} shows a similar comparison for the case $N=3$. The parameters are the same as before, but $k=4.0$ to keep the rate operator non-negative. We can again see a quicker relaxation in the case of local dephasing without other major differences.

We have also performed simulations with non‑homogeneous rates $\kappa_{ij}$ and $\gamma_i$. For the non‑local dephasing scheme, interpreting the physical implications of modifying individual rates is more challenging, and several representative cases are discussed in Appendix~\ref{sec-non-hom-rates}. In contrast, for local dephasing, individual rate control is experimentally feasible. The earlier results in this work rely on enhancing the decay of components with large occupation of the middle site through the term 
$-\tfrac{i}{2}k\, n_2$. It is therefore evident that by adjusting the coefficients of the analogous local terms $-\tfrac{i}{2}\gamma_i n_i^2$, one can induce similar modifications to the deterministic evolution.
This behavior is illustrated in Fig.~\ref{local-C=0-0,01,0} in Appendix~\ref{sec-non-hom-rates}, where $\gamma_1 = \gamma_3 = 0$, $\gamma_2 \neq 0$, and $C = 0$. In this configuration, the relaxation time is controlled directly by the value of $\gamma_2$.

\subsection{Asymptotic behavior of deterministic evolution \label{sec-asymptotic}}

In this section, we analyze the asymptotic state of the system and the reduction of the norm of the unnormalized state vector. The flexibility provided by rate‑operator transformations allows one to construct effective Hamiltonians that possess a unique steady state, enabling their use in quantum‑state engineering. In this setting, the squared norm of the unnormalized state represents the probability that the deterministic trajectory is realized under the measurement scheme defined by the chosen rate operator. It is therefore important to determine how rapidly the system can be driven toward its steady state while still retaining a sufficiently large norm to ensure an appreciable probability of success.

\subsubsection{Evolution to steady state and reduction in norm \label{sec-asymptotic-A}}

\begin{figure}[h!]
    \begin{subfigure}[t]{50mm}
        \includegraphics[width=\linewidth]{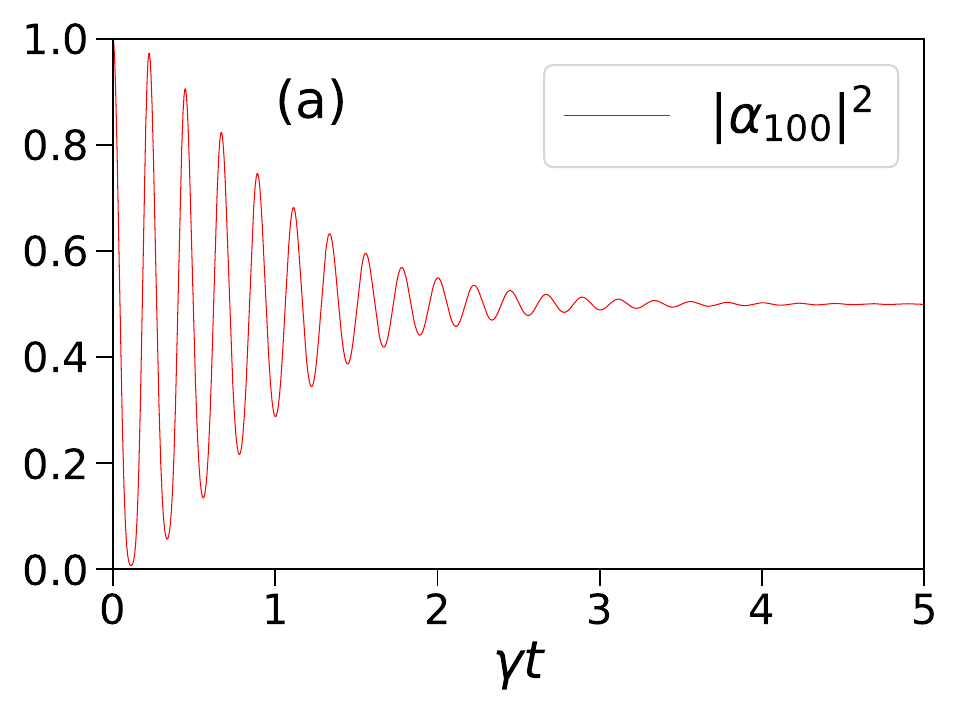}
    \end{subfigure}
    \begin{subfigure}[t]{50mm}
        \includegraphics[width=\linewidth]{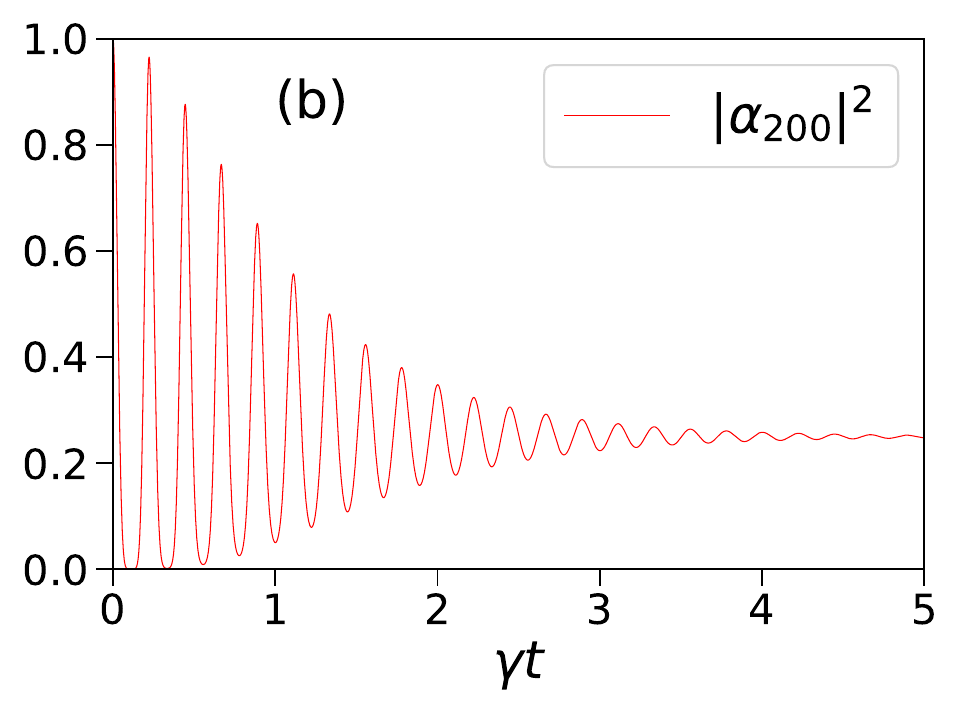}
    \end{subfigure}
    \begin{subfigure}[t]{50mm}
        \includegraphics[width=\linewidth]{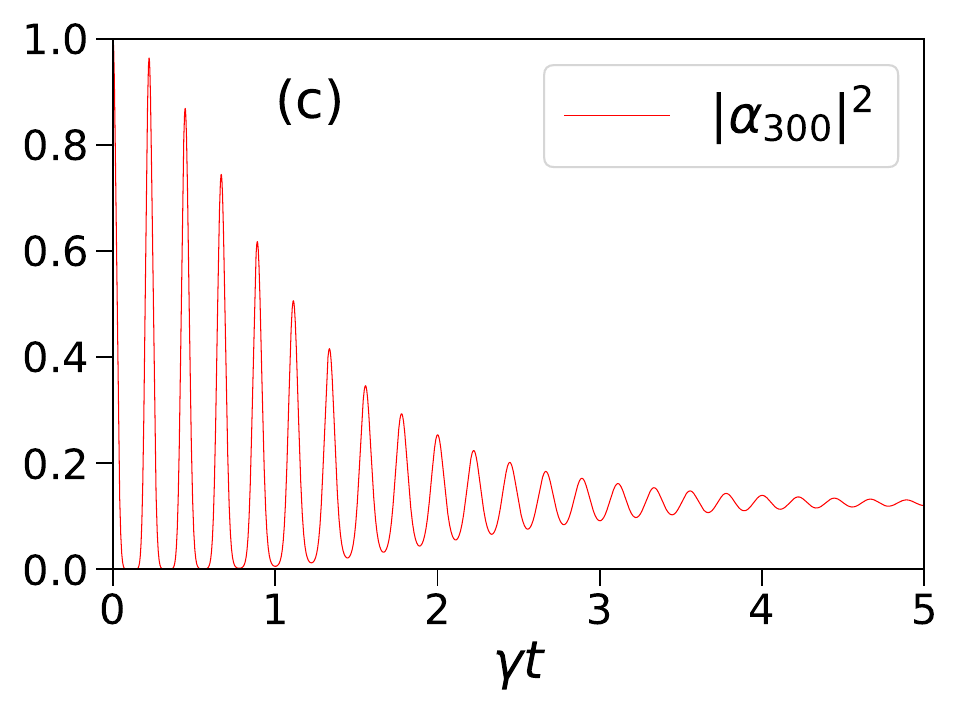}
    \end{subfigure}\\
    \begin{subfigure}[t]{50mm}
        \includegraphics[width=\linewidth]{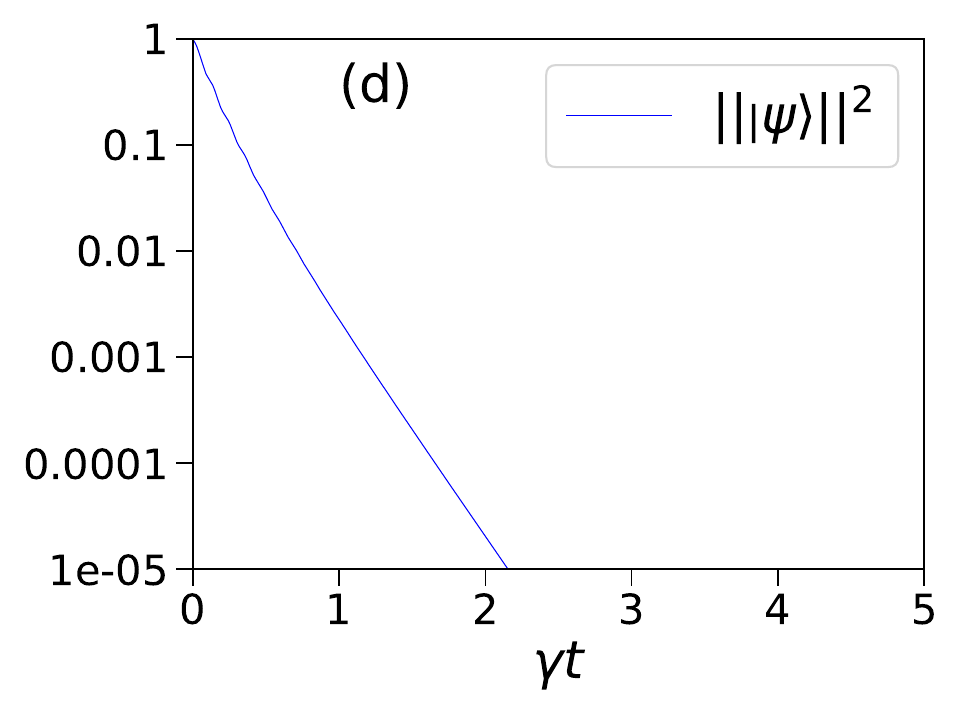}
    \end{subfigure}
    \begin{subfigure}[t]{50mm}
        \includegraphics[width=\linewidth]{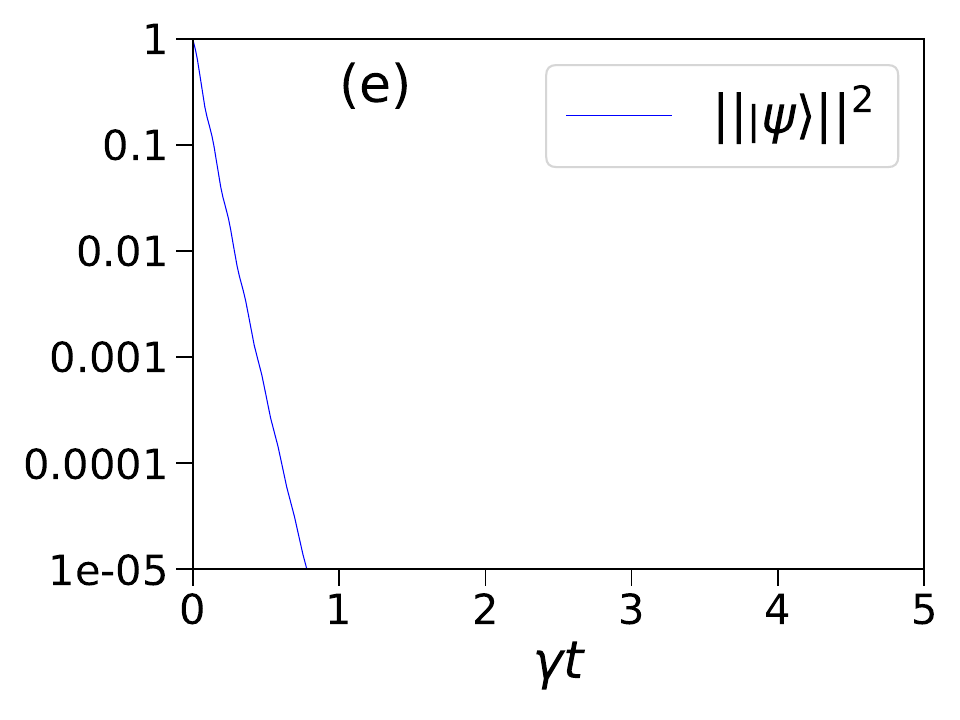}
    \end{subfigure}
    \begin{subfigure}[t]{50mm}
        \includegraphics[width=\linewidth]{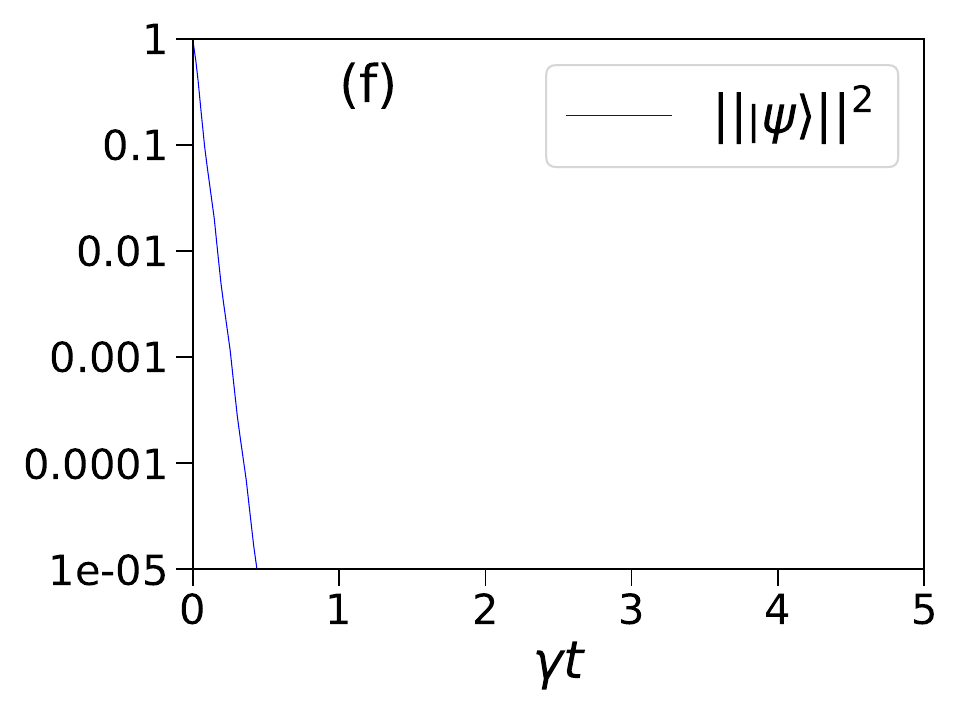}
    \end{subfigure}
    \caption{Evolution of $|\alpha_{N00}|^2$ and $\|\ket{\psi(t)}\|^2$ using non-local dephasing with $J/\gamma=20$ and $U=0$. The initial states are 
    $\ket{N,0,0}$ and panels (a),(d): $N=1$, $C/\gamma=6.0\,n_2$; (b),(e): $N=2$, $C=k\,n_2$, where $k=4.8\gamma$; 
    (c),(f): $N=3$, $C=k\,n_2$, where $k=4.0\gamma$.}
    \label{norm-U=0-N=1,2,3}
\end{figure}

The asymptotic behavior of the deterministic evolution can be inferred from the (right) eigenvalues and eigenstates of the effective Hamiltonian $K'$. The dynamics converge to the direct sum of eigenspaces associated with the eigenvalues whose imaginary parts are maximal. If this eigenvalue is unique and non‑degenerate, its corresponding eigenspace is one‑dimensional, and the associated eigenstate constitutes the unique steady state 
$\ket{\psi_{ss}}$ of the evolution.

As shown above, different choices of rates and transformations lead to distinct steady states. The reduction of $\|\ket{\psi(t)}\|^{2}$ during the deterministic evolution corresponds to the cumulative probability that a quantum jump has occurred since the beginning of the trajectory. Consequently, for state‑engineering purposes, it is desirable that the norm remains sufficiently large at the time when the deterministic evolution has converged close to the steady state. This ensures that the engineered state is obtained with an appreciable probability under the specified measurement scheme.

In the context of state engineering, the most favorable scenario was observed in the non‑interacting limit $U = 0$. Figures~\ref{norm-U=0-N=1,2,3} and \ref{norm-U=0-N=1,2,3-local} present representative deterministic evolutions for both non‑local and local dephasing schemes: the absolute squares of the amplitudes associated with the states $\ket{N,0,0}$ for particle numbers $N = 1, 2, 3$ obtained using rate‑operator transformations $C = k\, n_2$, together with the evolution of the squared norm of the unnormalized state $\ket{\psi(t)}$.
In all cases, the squared norm decreases to very small values before the deterministic evolution has fully converged, and this decay becomes more rapid as the particle number increases. Furthermore, we observe that for non‑local dephasing the relaxation speed increases with particle number, whereas for local dephasing the trend is reversed. These findings suggest that state engineering with larger particle numbers is more feasible under local dephasing than under non‑local dephasing. Nevertheless, for both schemes the probability of realizing a deterministic trajectory without jumps during the relaxation time remains far below $10^{-3}$.
Despite this limitation, the results clearly demonstrate that the rate‑operator formalism enables the controlled engineering of state trajectories and steady states.

\begin{figure}[h!]
    \begin{subfigure}[t]{50mm}
        \includegraphics[width=\linewidth]{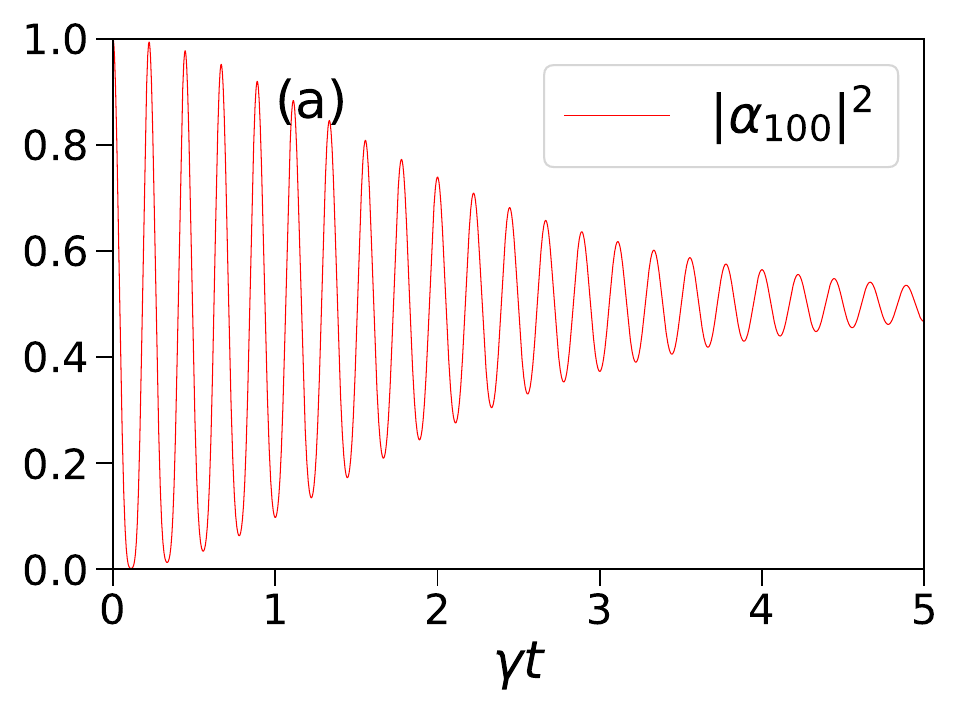}
    \end{subfigure}
    \begin{subfigure}[t]{50mm}
        \includegraphics[width=\linewidth]{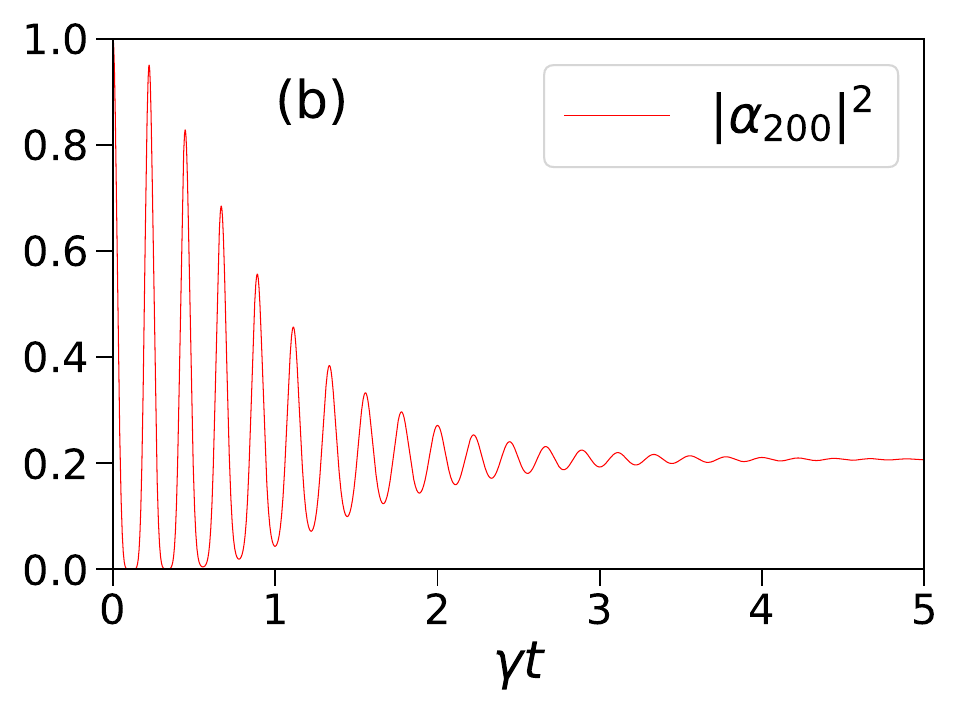}
    \end{subfigure}
    \begin{subfigure}[t]{50mm}
        \includegraphics[width=\linewidth]{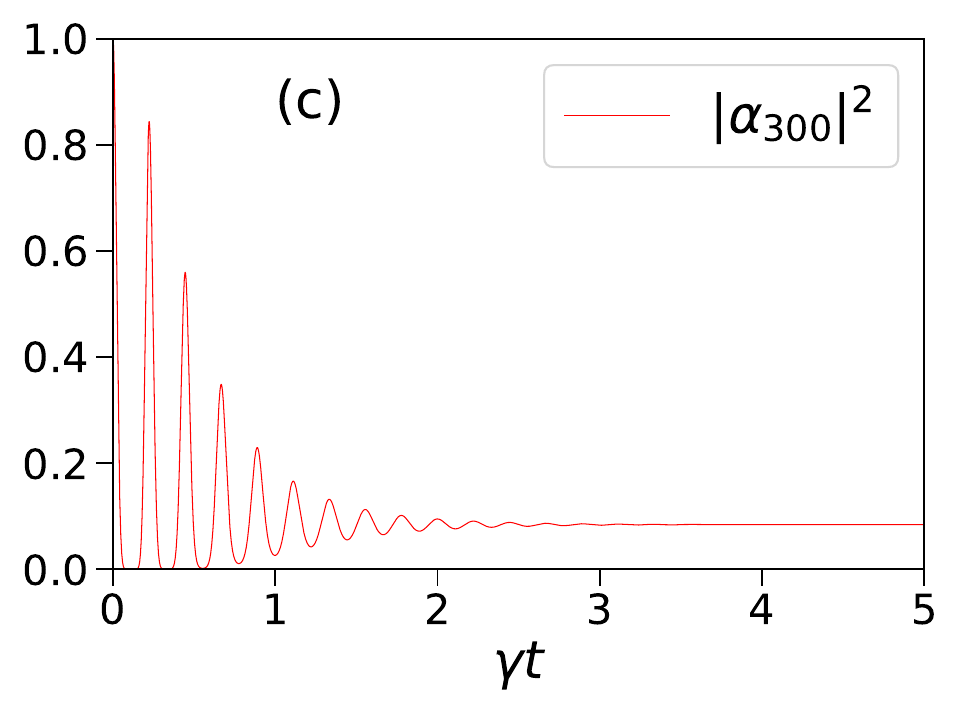}
    \end{subfigure}\\
    \begin{subfigure}[t]{50mm}
        \includegraphics[width=\linewidth]{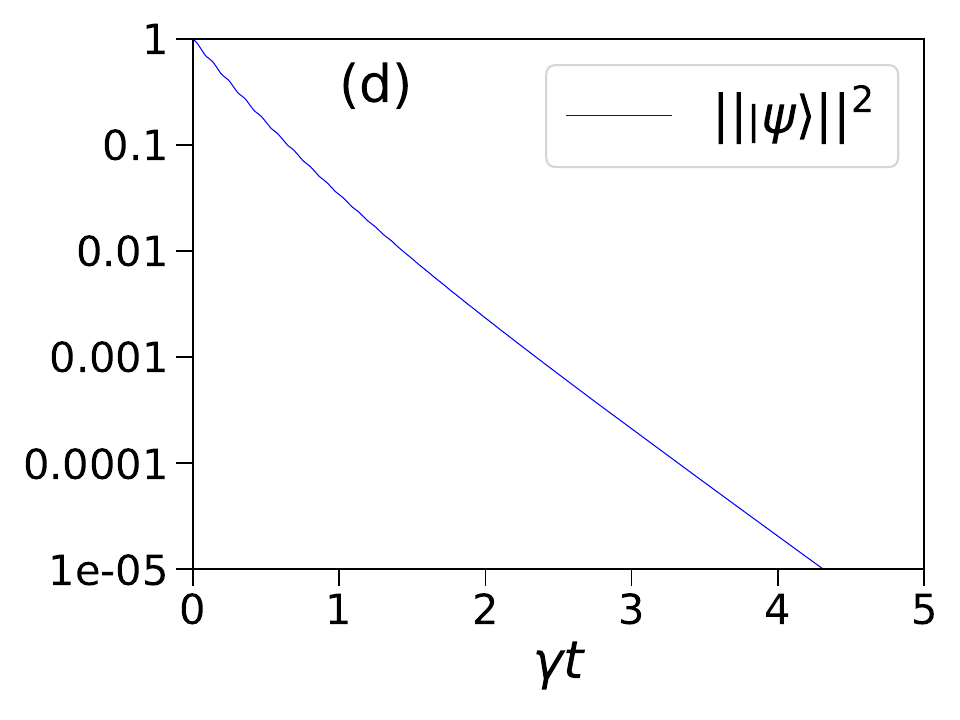}
    \end{subfigure}
    \begin{subfigure}[t]{50mm}
        \includegraphics[width=\linewidth]{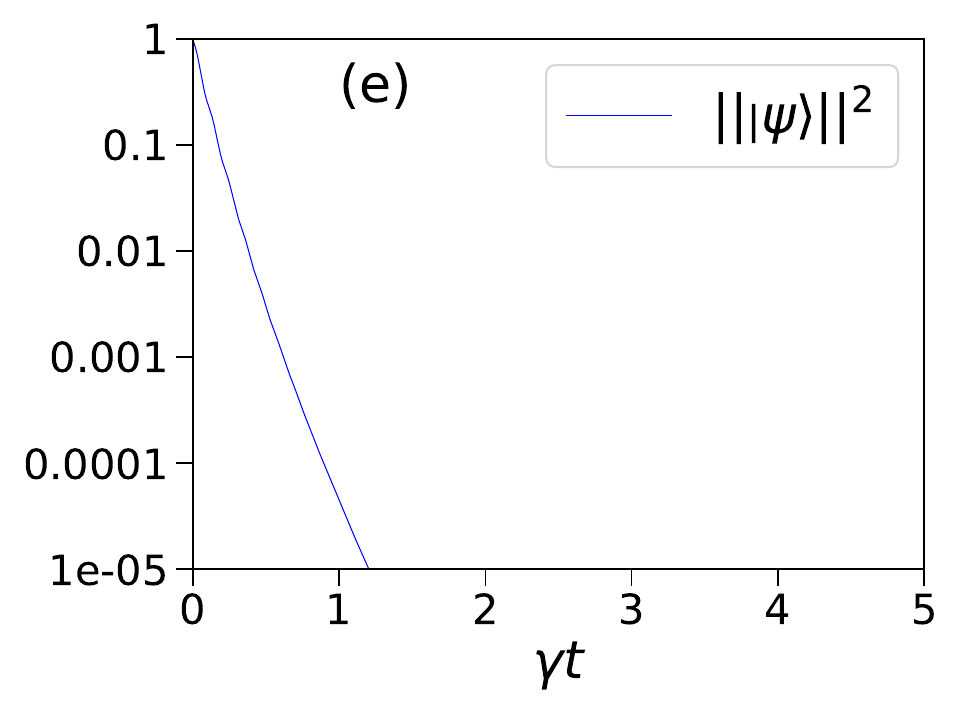}
    \end{subfigure}
    \begin{subfigure}[t]{50mm}
        \includegraphics[width=\linewidth]{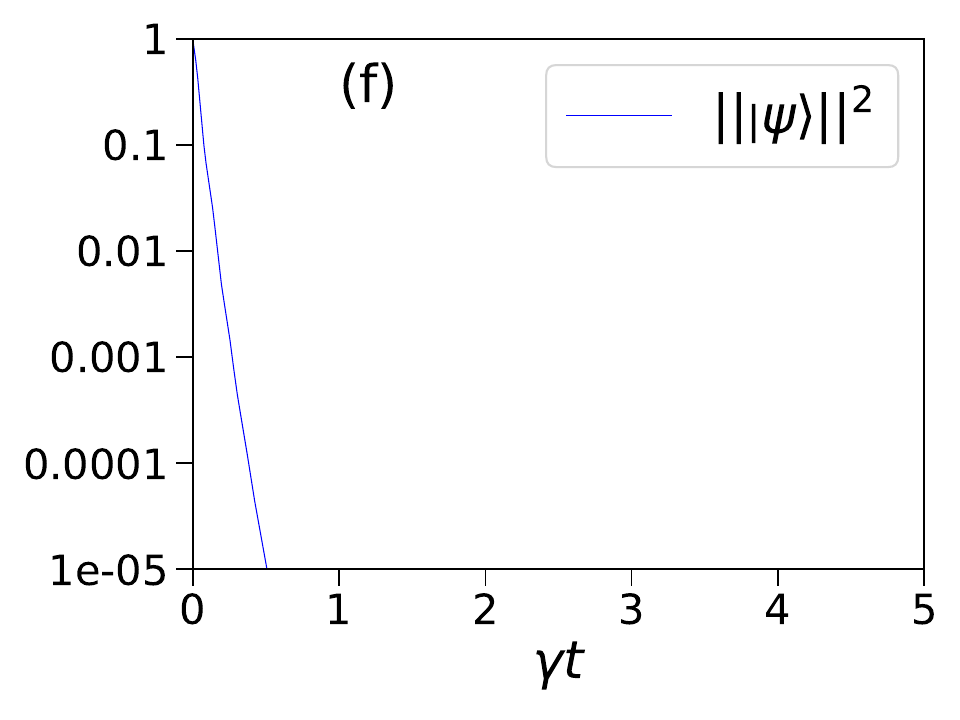}
    \end{subfigure}
    \caption{Evolution of $|\ket{N,0,0}|^2$ and $\|\ket{\psi}\|^2$ in local dephasing scheme with $J/\gamma=20$ and $U=0$. The initial states are $\ket{N,0,0}$,  the transformations $C=N\gamma(1+\sqrt{3})n_2$, and panels (a),(d): $N=1$. (b),(e): $N=2$.
    (c),(f): $N=3$.}
    \label{norm-U=0-N=1,2,3-local}
\end{figure}

Larger relaxation times can be achieved in non‑homogeneous systems. Figure~\ref{local-C=0-0,01,0} in Appendix~\ref{sec-non-hom-rates} provides an example of local dephasing with 
$\gamma_1 = \gamma_3 = 0$ and $\gamma_2 \neq 0$, using $C = 0$. In this configuration, the relaxation time is longer (with $\gamma_2=\gamma$); however, the norm approaches a non‑zero constant asymptotically due to the absence of terms that deplete probability from states of the form $\ket{n,0,m}$. From an experimental state‑engineering perspective, an optimal strategy would therefore be to minimize dephasing on the edge sites while maintaining stronger dephasing on the central site. 

In conclusion, the rate‑operator transformations can be employed to construct effective Hamiltonians that possess a unique steady state and that relax toward this steady state more rapidly than in the absence of such a transformation. Our results demonstrate that the flexibility enabled by rate‑operator transformations provides a powerful framework for exploring how different continuous‑measurement schemes influence quantum‑state engineering.

\section{Results: Stationary properties \label{sec-results_b}}  

In Section~\ref{sec-asymptotic-B}, we analyze stationary properties of the steady states; how the steady state of the effective Hamiltonian depends on the interaction strength $U$ and on the particular choice of rate-operator transformation. There appears to be rich phase structure depending crucially on the dephasing scheme and particle number: the $T=0$ quantum phases and corresponding phase transitions are also dependent on the rate-operator transformation.  

\subsection{The phase structure of the steady states \label{sec-asymptotic-B}}

Next, we investigate the steady state of the effective Hamiltonian for varying values of the parameters $U$ and $k$. We find that the steady state undergoes abrupt changes as one crosses a boundary in the $(k/\gamma,\,U/\gamma)$ parameter space. In practice, these transitions occur when the two largest imaginary parts of the eigenvalues of the effective Hamiltonian exchange order. The $(k/\gamma,\,U/\gamma)$ plane is thus partitioned into distinct regions separated by phase boundaries corresponding to sudden changes in the steady‑state structure.
These transitions constitute quantum phase transitions \cite{Vojta_2003} at $T = 0$, analogous to those defined by the ground state of a Hermitian Hamiltonian \cite{Kuhner1998,Kollath2007,Chen2016,Shimizu2018,Dengis2025,Wang2025,Mal2025} or by the steady state of a Lindblad master equation \cite{Kessler2012,Minganti2018,Debecker2023,Debecker2024}, both of which have been extensively studied. There also exists a growing body of work on measurement‑induced phase transitions (MIPTs) \cite{Skinner2019,Biella2021,Turkeshi2021,DiFresco2024}, in which the phase of a quantum system changes as a function of the measurement rate or strength.
Our findings exhibit clear similarities to these results in the no‑click limit, in which the system follows a trajectory with no quantum jumps and evolves solely under the effective Hamiltonian $K'$. In particular, we observe phase transitions induced by varying a parameter associated with the measurement process. In other works, this parameter typically represents the measurement strength, whereas in our setting the parameter $k$ modifies the (state‑dependent) measurement basis defined by the rate operator $R_{\psi(t)}$, without altering the parameters of the underlying master equation. Changing of the measurement basis leading to MIPTs has been noticed before in the context of transformations of Kraus operators in \cite{Vovk2022,Vovk2024}.

The interpretation is that the deterministic evolution $\ket{\psi(t)}$ exhibits an instability with respect to the measurement basis determined by the rate operator $R_{\psi(t)}$. This measurement basis is defined for each state $\ket{\psi} \in \mathcal{H}$ given a fixed value of $k \in \mathbbm{R}$ in the rate‑operator transformation $C = k\, n_2$. More general choices of transformation would presumably lead to an even richer and more intricate phase structure. Since the phases are determined solely by the effective Hamiltonian, they may be characterized using state‑independent rate‑operator transformations, whereas the corresponding measurement bases associated with these phases are, in general, state dependent. In what follows, we restrict our analysis to the structure of the steady‑state phase space only.

\begin{figure}[h!]
    \begin{subfigure}[t]{75mm}
        \includegraphics[width=\linewidth]{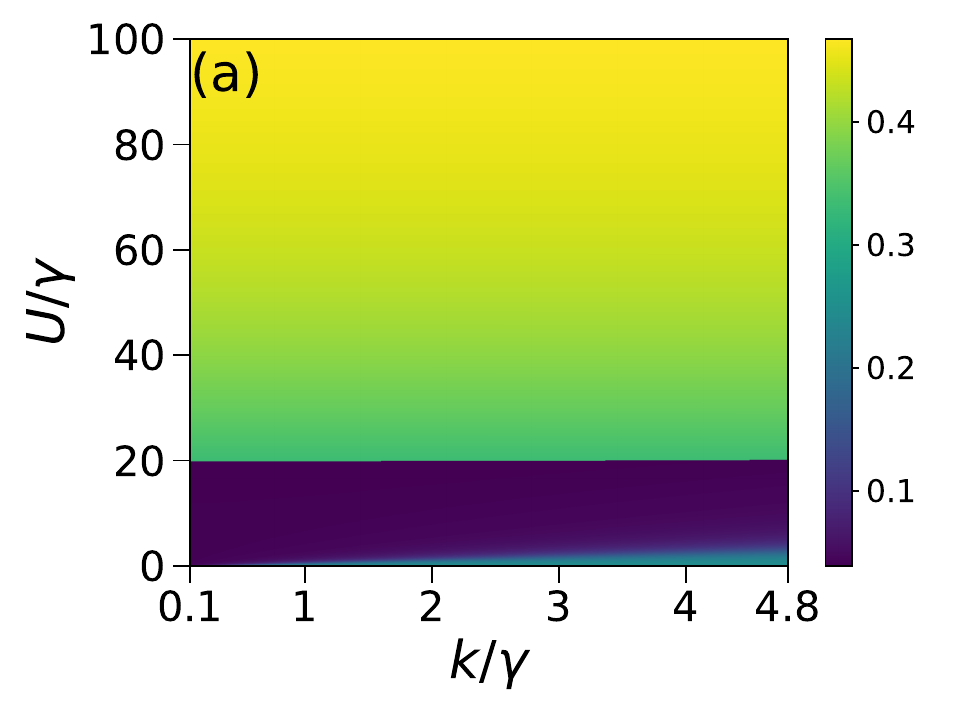}
    \end{subfigure}
    \begin{subfigure}[t]{75mm}
        \includegraphics[width=\linewidth]{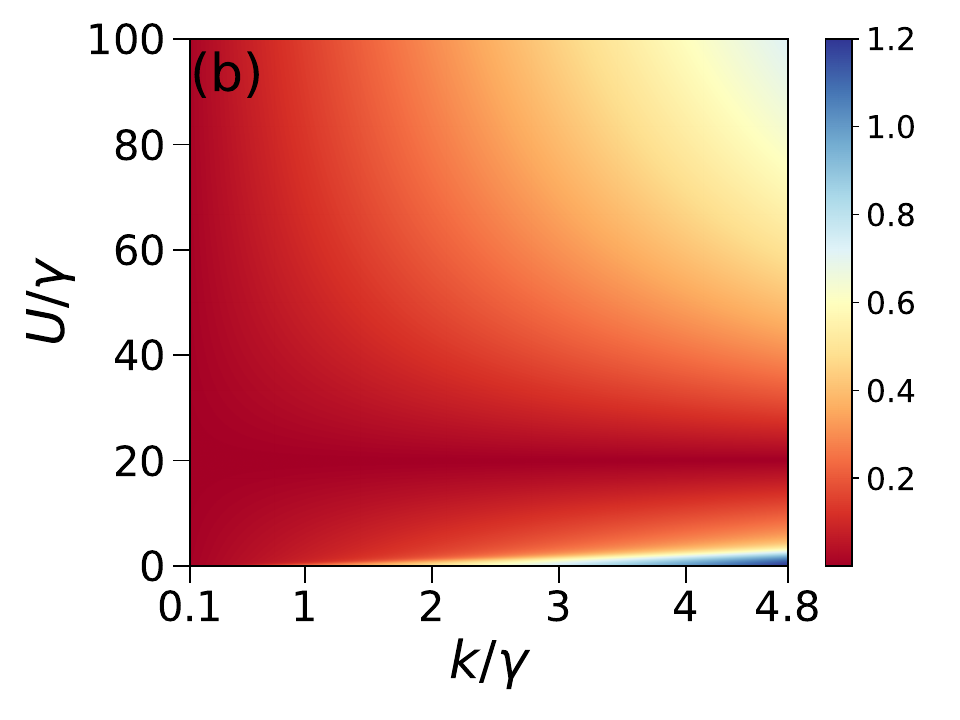}
    \end{subfigure}
    \begin{subfigure}[t]{75mm}
        \includegraphics[width=\linewidth]{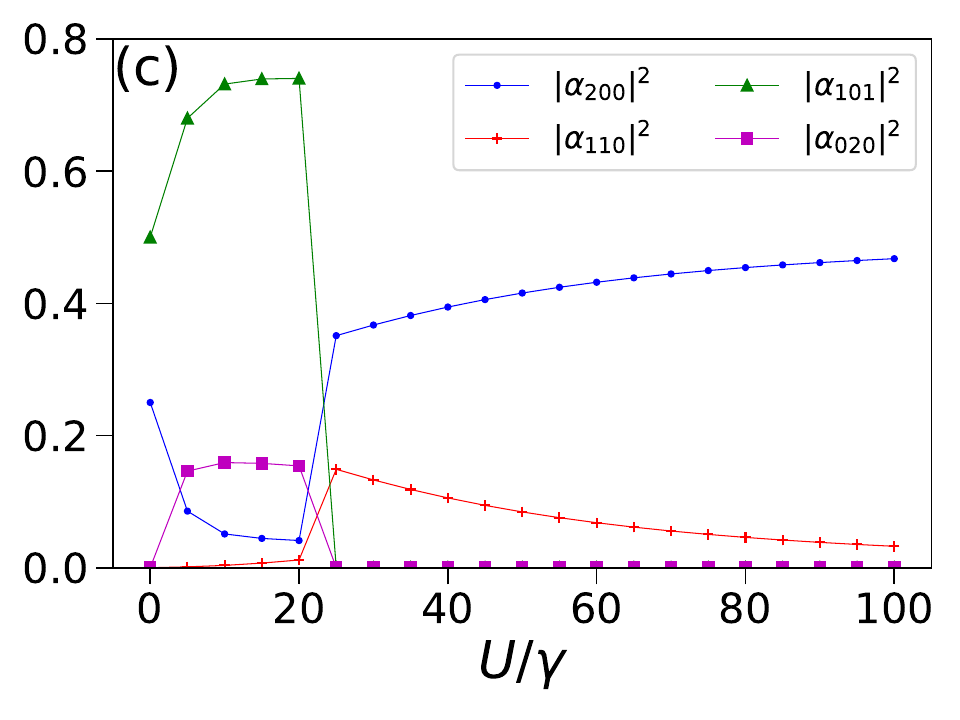}
    \end{subfigure}
    \caption{Phase transitions of the steady state $\ket{\psi_{ss}}$ in non-local dephasing with $C=kn_2$. Parameters are $N=2$ and $J/\gamma=20$, and panels are (a): 
    $|\alpha_{200}|^2$ of the steady state $\ket{\psi_{ss}}$ in $(k/\gamma,\,U/\gamma)$ plane; (b): the gap 
    $\Delta$ of the effective Hamiltonian $K'/\gamma$ in $(k/\gamma,\,U/\gamma)$ plane; (c): an example of absolute squares of the amplitudes of the steady state as a function of $U/\gamma$ with $k/\gamma=4.8$. Some coefficients are not shown as they have the same amplitudes as their symmetric counterparts.}
    \label{steadystate-heatmap-N=2}
\end{figure}

\begin{figure}[htp]
    \begin{subfigure}[t]{75mm}
        \includegraphics[width=\linewidth]{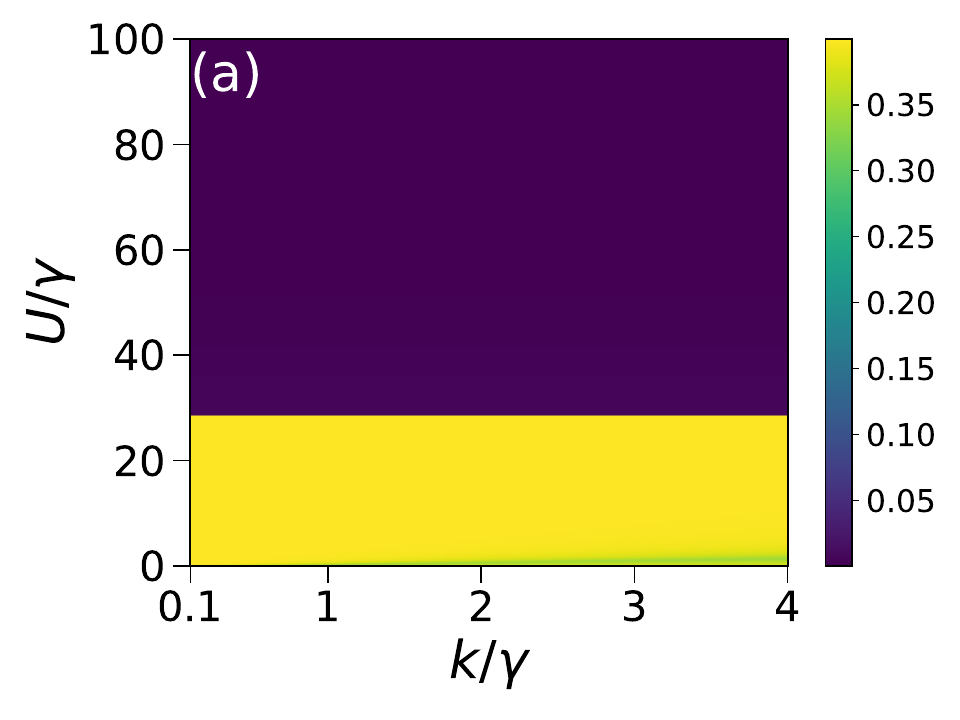}
    \end{subfigure}
    \begin{subfigure}[t]{75mm}
        \includegraphics[width=\linewidth]{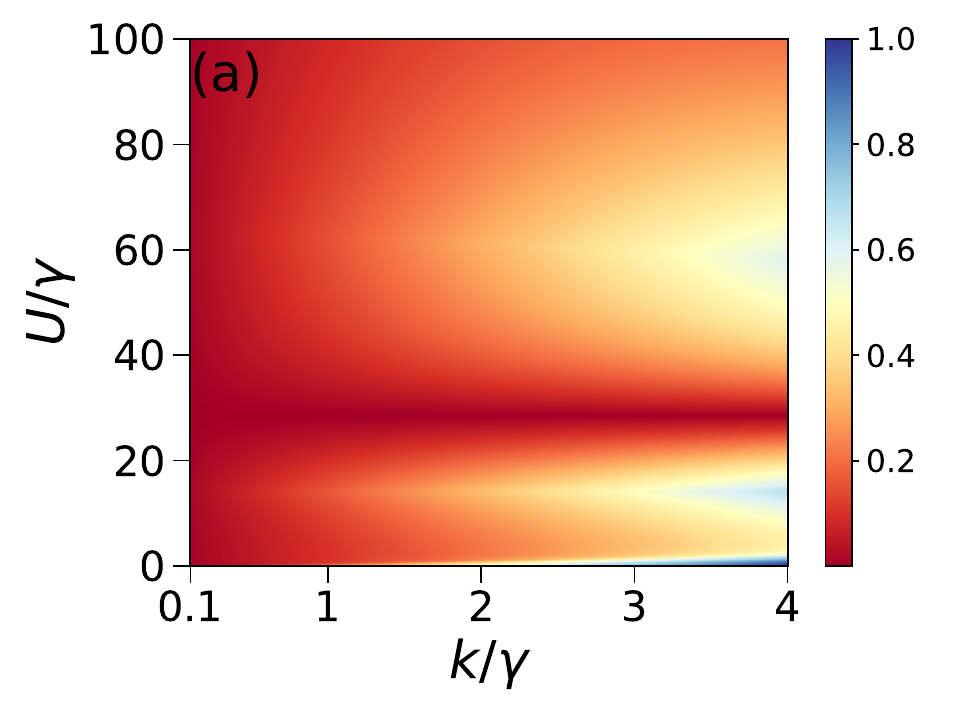}
    \end{subfigure}
    \begin{subfigure}[t]{75mm}
        \includegraphics[width=\linewidth]{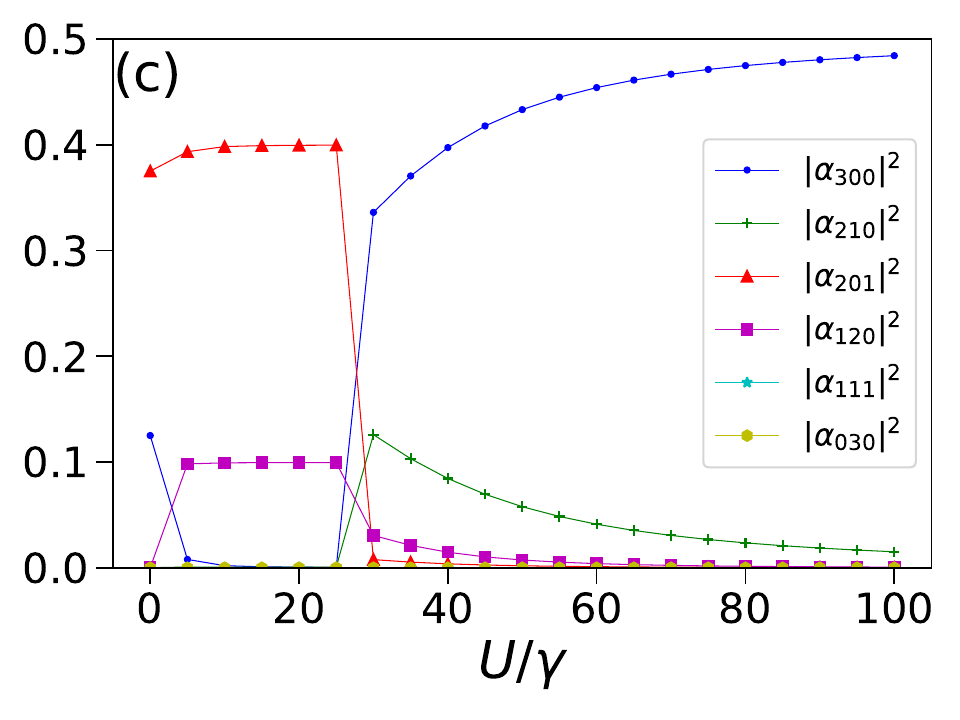}
    \end{subfigure}
    \caption{Phase transitions of the steady state $\ket{\psi_{ss}}$ in non-local dephasing with $C=k\,n_2$. Parameters are $N=3$ and $J/\gamma=20$, and panels are (a): 
    $|\alpha_{201}|^2$ of the steady state $\ket{\psi_{ss}}$ in $(k/\gamma,\,U/\gamma)$ plane; (b): the gap 
    $\Delta$ of the effective Hamiltonian $K'/\gamma$ in $(k/\gamma,\,U/\gamma)$ plane; (c): an example of absolute squares of the amplitudes of the steady state as a function of $U/\gamma$ with $k/\gamma=4.0$. Some components are not shown as they have the same amplitudes as their symmetric counterparts.}
    \label{steadystate-heatmap-N=3}
\end{figure}

\begin{figure}[htp]
    \begin{subfigure}[t]{75mm}
        \includegraphics[width=\linewidth]{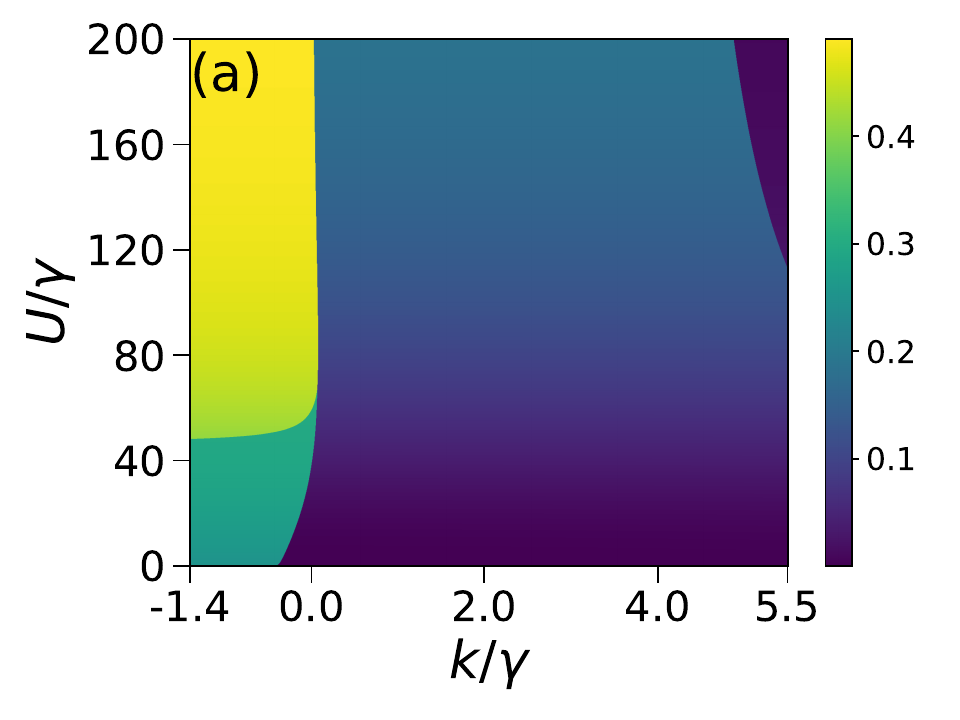}
    \end{subfigure}
    \begin{subfigure}[t]{75mm}
        \includegraphics[width=\linewidth]{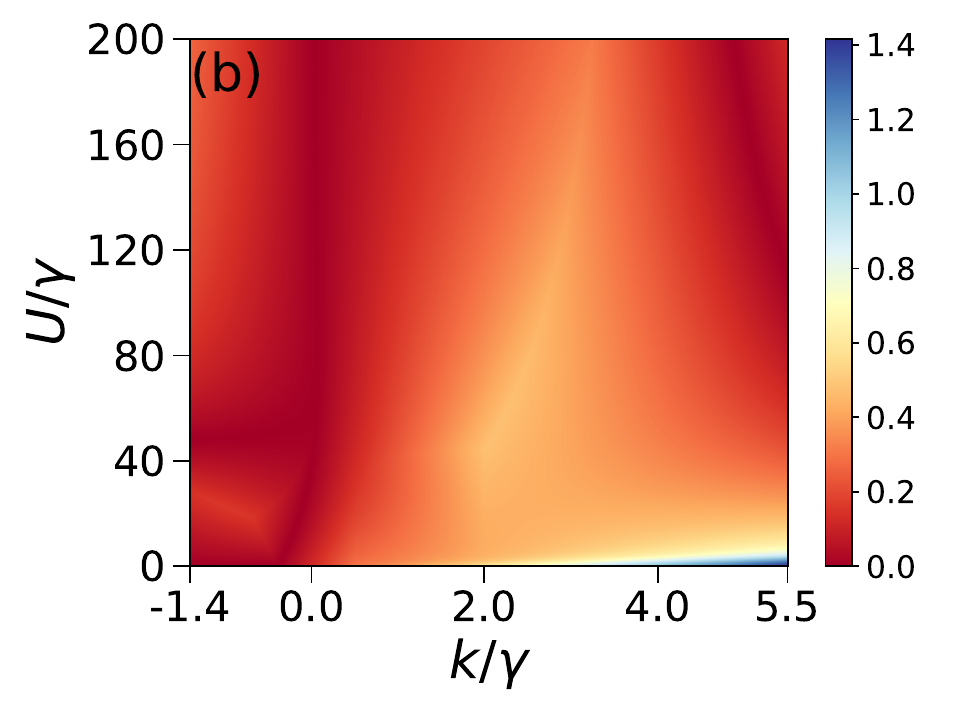}
    \end{subfigure} \\
    \begin{subfigure}[t]{75mm}
        \includegraphics[width=\linewidth]{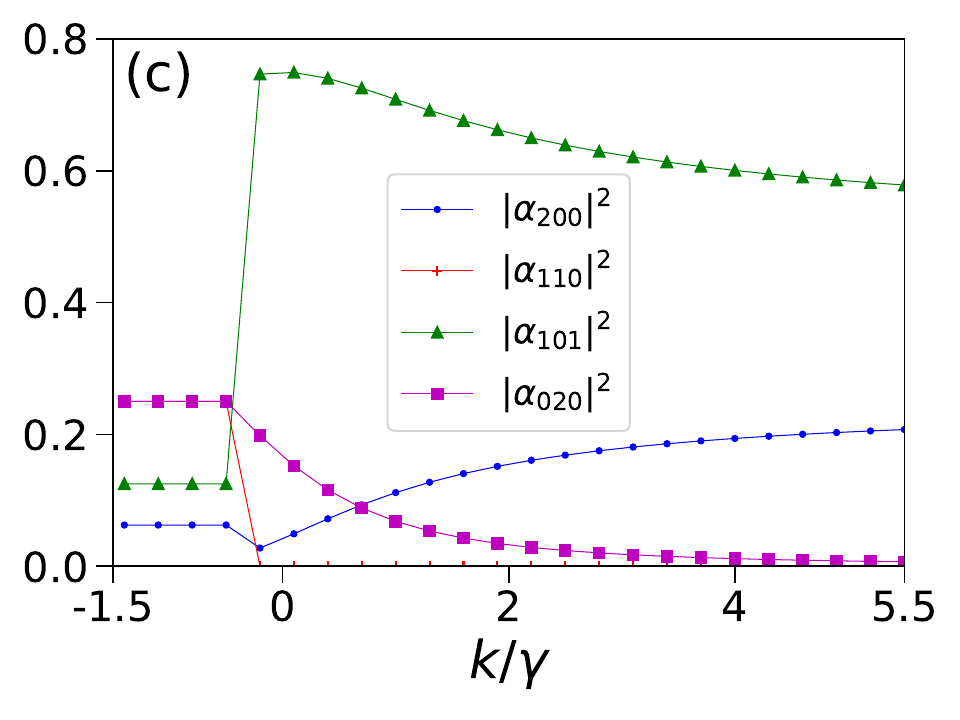}
    \end{subfigure}
    \begin{subfigure}[t]{75mm}
        \includegraphics[width=\linewidth]{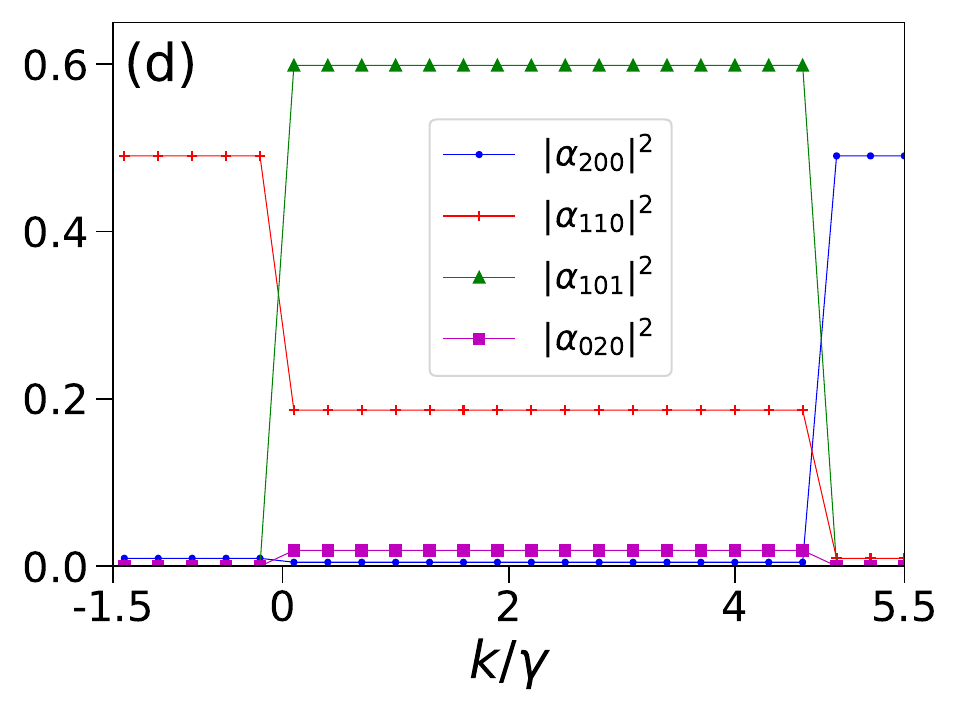}
    \end{subfigure} \\
    \begin{subfigure}[t]{75mm}
        \includegraphics[width=\linewidth]{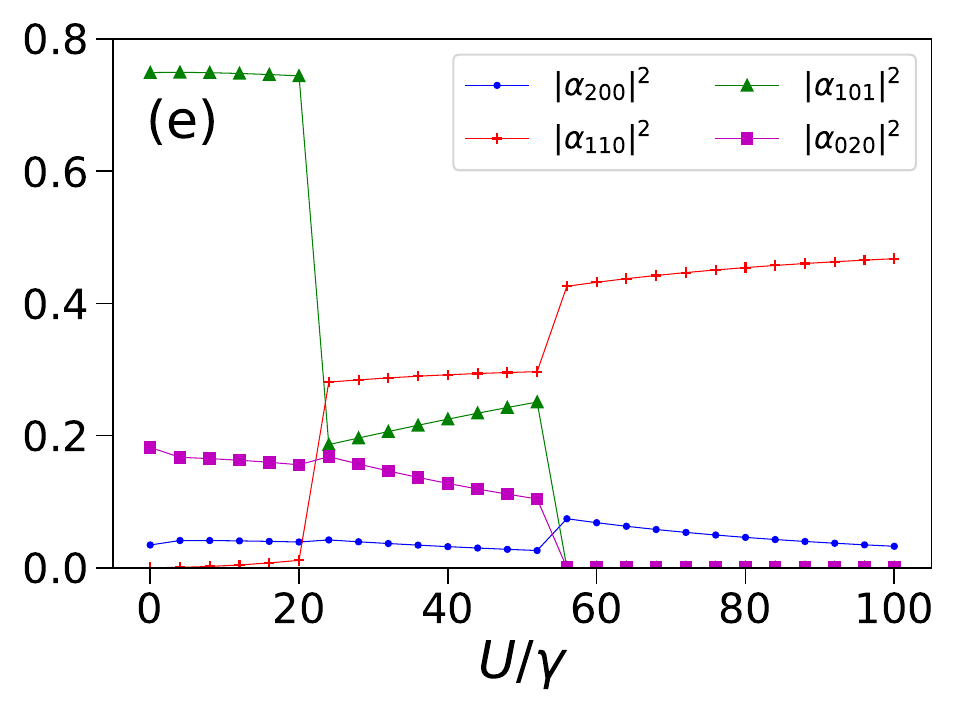}
    \end{subfigure}
    \begin{subfigure}[t]{75mm}
        \includegraphics[width=\linewidth]{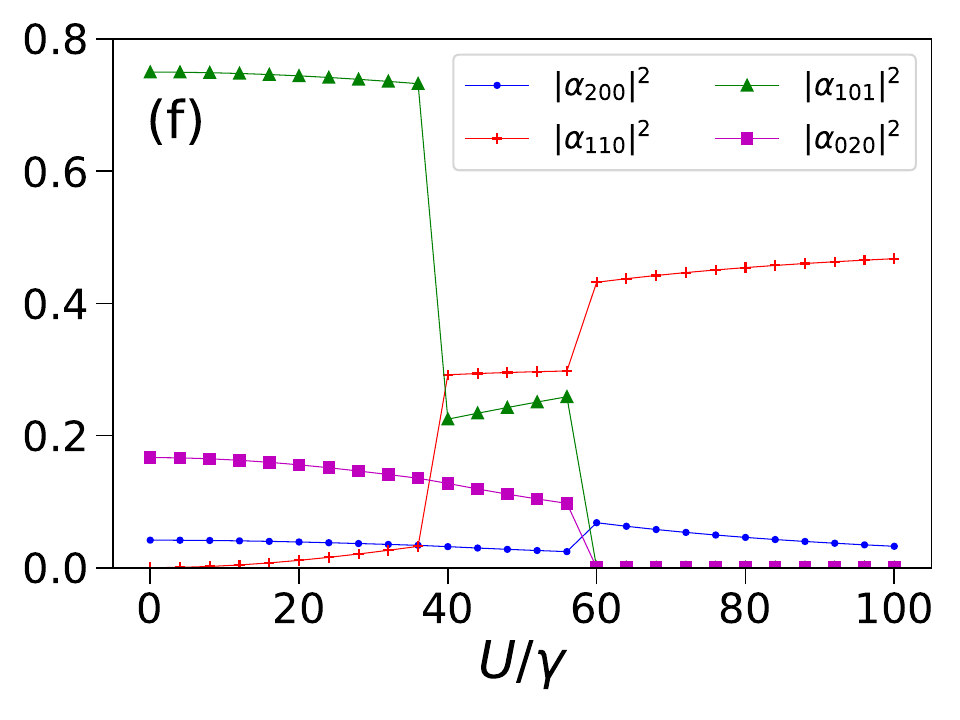}
    \end{subfigure}
    \caption{Phase transitions of the steady state $\ket{\psi_{ss}}$ in local dephasing with $N=2$ and $J/\gamma=20$. (a): $|\alpha_{200}|^2$ in $(k/\gamma,\,U/\gamma)$ plane; (b): the gap $\Delta$ of the effective Hamiltonian $K'/\gamma$ in $(k/\gamma,\,U/\gamma)$ plane; (c-d): examples of absolute squares of the amplitudes of the steady state as a function of $k/\gamma$ with $U=0$ and $U/\gamma=200$, respectively; (e-f): examples of absolute squares of the amplitudes of the steady state as a function of $U/\gamma$ with $k/\gamma=-0.1$ and $k/\gamma=0$, respectively. Some components are not shown as they have the same amplitudes as their symmetric counterparts.}
    \label{steadystate-heatmap-N=2-local}
\end{figure}

\begin{figure}[htp]
    \begin{subfigure}[t]{75mm}
        \includegraphics[width=\linewidth]{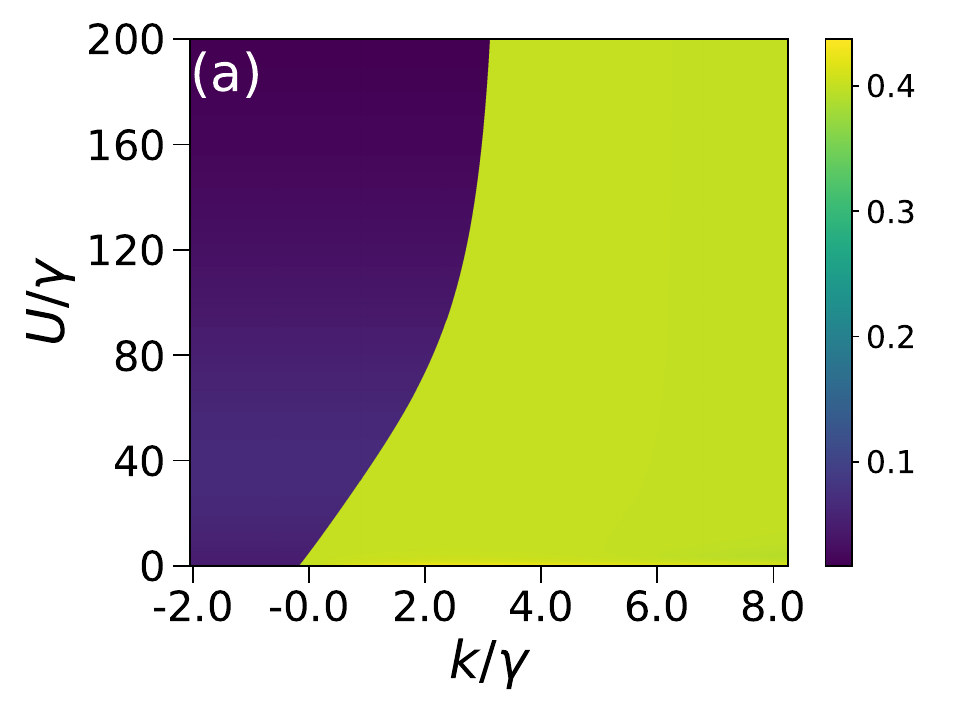}
    \end{subfigure}
    \begin{subfigure}[t]{75mm}
        \includegraphics[width=\linewidth]{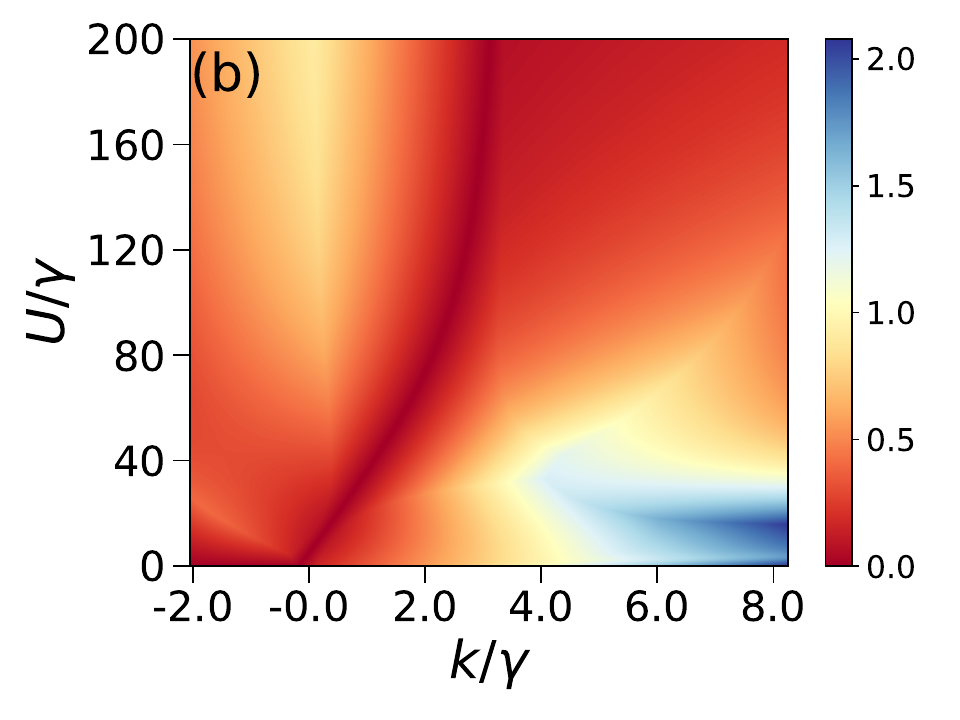}
    \end{subfigure} \\
    \begin{subfigure}[t]{75mm}
        \includegraphics[width=\linewidth]{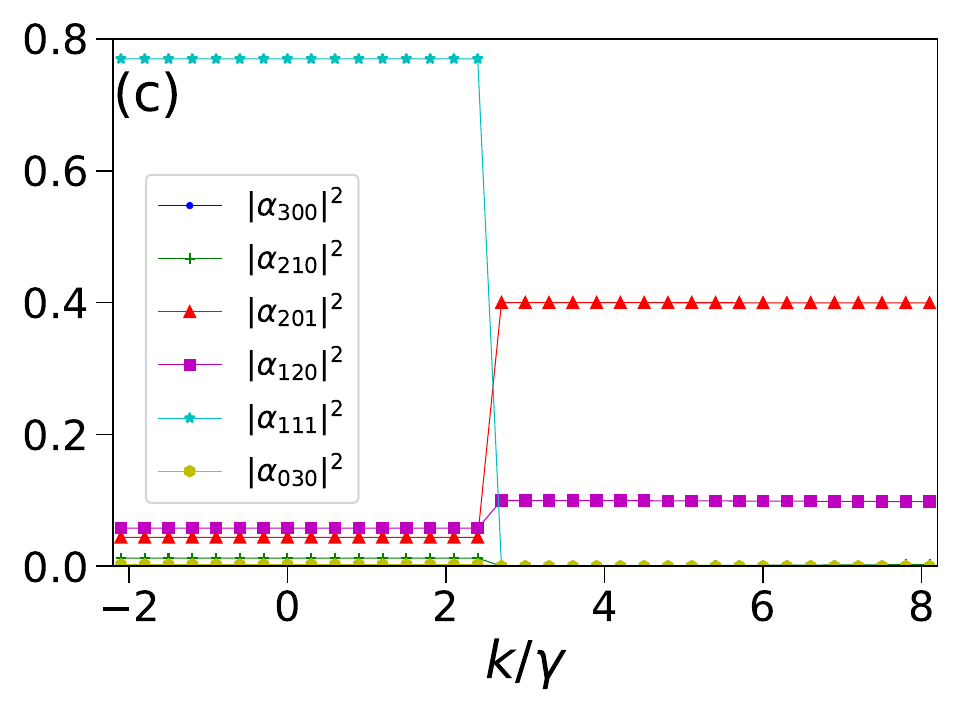}
    \end{subfigure}
    \begin{subfigure}[t]{75mm}
        \includegraphics[width=\linewidth]{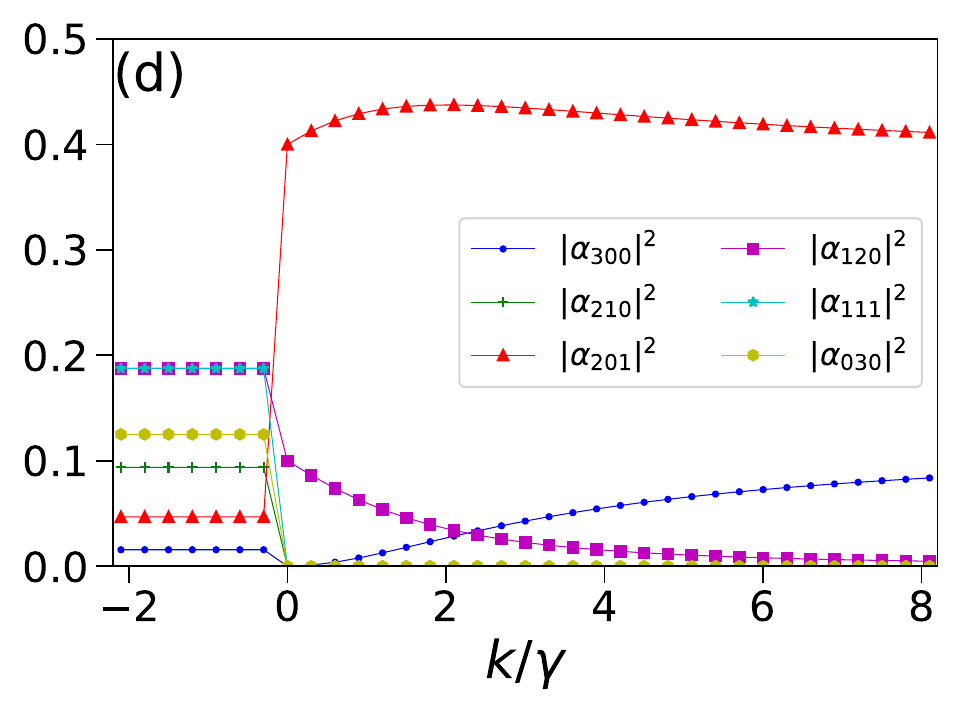}
    \end{subfigure}
    \caption{Phase transitions of the steady state $\ket{\psi_{ss}}$ in local dephasing with $N=3$ and $J/\gamma=20$. (a): $|\alpha_{201}|^2$ in $(k/\gamma,\,U/\gamma)$ plane; (b): The gap $\Delta$ of the effective Hamiltonian $K'/\gamma$ in $(k/\gamma,\,U/\gamma)$ plane; (c-d): Examples of absolute squares of the amplitudes of the steady state as a function of $k/\gamma$ with $U/\gamma=100$ and $U=0$, respectively.}
    \label{steadystate-heatmap-N=3-local}
\end{figure}

 We compare the results for non‑local and local dephasing in the cases $N = 2$ and $N = 3$ Figures~\ref{steadystate-heatmap-N=2}–\ref{steadystate-heatmap-N=3-local} present data obtained from exact diagonalization of the effective Hamiltonian $K'/\gamma$ with fixed tunneling strength $J/\gamma = 20$. In subfigures (a), we display the absolute square of a representative component of the steady state to highlight the distinct regions in the $(k/\gamma,\,U/\gamma)$ parameter space corresponding to different phases. Again, the parameter $k/\gamma$ is chosen within the interval that ensures the non-negativity of the rate operator. For the non‑local dephasing scheme, the case $k/\gamma = 0$ is omitted, as it does not yield a unique steady state.
Subfigures (b) show the gap $\Delta = \mathrm{Im}(\lambda_1) - \mathrm{Im}(\lambda_2)$, defined as the difference between the two largest imaginary parts of the eigenvalues of $K'/\gamma$, where $\lambda_1$ and $\lambda_2$ denote the eigenvalues with the largest and second‑largest imaginary parts, respectively. The quantity $\Delta$ characterizes the relaxation speed toward the eigenstate $\ket{\psi_{ss}}$ associated with $\lambda_1$.
Finally, subfigures (c)–(f) show selected absolute squares of the steady‑state amplitudes as functions of either $k/\gamma$ or $U/\gamma$, revealing sharp transitions that signal phase boundaries in the effective‑Hamiltonian steady‑state landscape.

 Another important quantity is the imaginary part $\mathrm{Im}(\lambda_1)$ of the eigenvalue associated with the steady state. Its absolute value provides an approximation of the asymptotic decay rate of the norm of the state. Since $\mathrm{Im}(\lambda_1)$ is always negative, larger values of it correspond to slower decay. Maps of it as a function of the parameter $k/\gamma$ and the interaction strength $U/\gamma$ are presented in Appendix~\ref{sec-decay-rates}. In Appendix~\ref{sec-energies}, we also show the expectation value of the energy 
 $\langle H/\gamma \rangle$ in the steady state. At zero temperature, changes in $\langle H/\gamma \rangle$ coincide with changes in the free energy across a phase boundary.

In Figs.~\ref{steadystate-heatmap-N=2}.a and \ref{steadystate-heatmap-N=3}.a, we present the results for non‑local dephasing with $N = 2$ and $N = 3$, respectively. In both cases, the two steady‑state phases are separated by a boundary that is approximately constant in $U/\gamma$, forming a coexistence curve: for $N = 2$ the transition occurs near $U/\gamma \approx 20$, while for $N = 3$ it appears near $U/\gamma \approx 30$. As shown in Figs.~\ref{steadystate-heatmap-N=2}.b and \ref{steadystate-heatmap-N=3}.b, these coexistence curves coincide with $\Delta = 0$, as expected from the exchange of the two largest imaginary parts of the eigenvalues of the effective Hamiltonian. Furthermore, the steady‑state phase transition exhibits discontinuities in the energy across the boundary, as evident in Figs.~\ref{energies}.a–b. For both cases, the fastest relaxation is achieved at $U/\gamma = 0$ and with $k/\gamma$ chosen as large as possible while maintaining the non‑negativity of the rate operator. The figures also identify the regions in the parameter space where the evolution toward the steady state is most rapid.

In Fig. \ref{steadystate-heatmap-N=2}.c, we illustrate how the steady state for $N = 2$ changes as the interaction strength $U/\gamma$ is varied, using the transformation $C/\gamma = 4.8\, n_{2}$. For $U/\gamma = 0$, only the amplitudes $\alpha_{200},\alpha_{002}$, and $\alpha_{101}$ are non-zero, with
$|\alpha_{200}|^{2} =|\alpha_{002}|^{2} = \frac{1}{4},\ |\alpha_{101}|^{2} = \frac{1}{2}$.
As $U/\gamma$ increases, the amplitude $\alpha_{101}$ rapidly becomes dominant, while the other amplitudes remain small but non-zero. Once $U/\gamma$ reaches approximately $U/\gamma \approx 20$, the steady state undergoes a discontinuous phase transition to a regime in which $\alpha_{200}$ and 
$\alpha_{002}$ dominate. For larger interaction strengths, the steady state approaches
$\ket{\psi_{ss}} \approx \frac{1}{\sqrt{2}}\left(\ket{2,0,0} - \ket{0,0,2}\right)$.
From Fig. \ref{steadystate-heatmap-N=2}.b, we observe that at higher values of $U/\gamma$, the gap 
$\Delta$ remains relatively large compared with most other points in the $(k/\gamma,\,U/\gamma)$ parameter space, indicating a robust relaxation toward the steady state.
In summary, for $N = 2$, the phase at small interaction strengths $U/\gamma$ is characterized by a dominant amplitude $\alpha_{101}$, whereas at larger values of $U/\gamma$, the dominant contributions stem from $\alpha_{200}$ and $\alpha_{002}$.

In Fig. \ref{steadystate-heatmap-N=3}.c, we show the corresponding steady‑state amplitudes for the case $N = 3$ as functions of the interaction strength $U/\gamma$, with the transformation parameter fixed to $k/\gamma = 4.0$. The qualitative behavior closely mirrors the $N = 2$ case. For large values of $U/\gamma$, the steady state is dominated by the amplitudes $\alpha_{300}$ and 
$\alpha_{003}$, while the remaining components approach zero. For smaller interaction strengths, $U/\gamma \lesssim 30$, the amplitudes with the largest magnitudes are $\alpha_{201}$ and 
$\alpha_{102}$.

Figures \ref{steadystate-heatmap-N=2-local} and \ref{steadystate-heatmap-N=3-local} present the phase diagrams for the local dephasing scheme. A notable difference emerges between the cases $N = 2$ and $N = 3$: the two‑particle system exhibits a substantially more intricate phase structure than either the three‑particle system or the non‑local dephasing model.
In Fig.~\ref{steadystate-heatmap-N=2-local}.a ($N = 2$), a vertical phase boundary appears near $k/\gamma \approx 0$. For $k/\gamma \lesssim 0$, three distinct phases are observed, while for $k/\gamma \gtrsim 0$, two phases appear—yielding a total of four steady‑state phases. In contrast, for $N = 3$ (Fig.~\ref{steadystate-heatmap-N=3-local}), only two phases are present, separated by a steep coexistence curve around $k/\gamma \approx 3.0$. It remains open if additional phases emerge at larger values of $U/\gamma$; however, computations performed up to $U/\gamma = 100$ do not indicate the presence of further phase boundaries.

Figures \ref{steadystate-heatmap-N=2-local}.b and \ref{steadystate-heatmap-N=3-local}.b show that, in both cases, the fastest relaxation occurs in the lower‑right region of the $(k/\gamma,\,U/\gamma)$ plane, where $k/\gamma$ is large and $U/\gamma$ is small. For $N = 2$, only the non‑interacting case $U/\gamma = 0$ exhibits a clearly maximal relaxation rate. In contrast, for $N = 3$, all interaction strengths in the interval $U/\gamma \in [0,30]$ yield relatively rapid relaxation, with local maxima at $U/\gamma = 0$ and around $U/\gamma \approx 15$.

Figures \ref{steadystate-heatmap-N=2-local}.c–d display the steady-state components as functions of the transformation parameter $k/\gamma$ for fixed interaction strengths $U/\gamma = 0$ and $U/\gamma = 200$, respectively. For $U/\gamma = 0$, two distinct phases are observed. In the low-$k/\gamma$ phase, the amplitudes $\alpha_{110},\ \alpha_{011}$, and $\alpha_{020}$ have comparable absolute values that are larger than those of the remaining components. In contrast, in the high-$k/\gamma$ regime, $k/\gamma > -0.5$, the amplitude $\alpha_{101}$ becomes dominant, while $\alpha_{110}$ and 
$\alpha_{011}$ tend toward zero. This indicates that configurations in which the two particles occupy adjacent sites are suppressed.
For $U/\gamma = 200$, three distinct phases emerge. At small values of $k/\gamma$, the amplitudes 
$\alpha_{110}$ and $\alpha_{011}$ dominate, with all other components remaining close to zero, such that the steady state can be approximated as $\ket{\psi_{ss}} \approx (\ket{1,1,0} - \ket{0,1,1})/\sqrt{2}$. In the large-$k/\gamma$ regime, the system enters a phase in which the steady state approaches $\ket{\psi_{ss}} \approx (\ket{2,0,0} - \ket{0,0,2})/\sqrt{2}$, analogous to the behavior observed for non‑local dephasing. In the intermediate range of $k/\gamma$, the state 
$\ket{1,0,1}$ constitutes the dominant component of the steady state.

Figures~\ref{steadystate-heatmap-N=2-local}.e–f illustrate the dependence of the steady state on the interaction strength $U/\gamma$ for fixed values of the parameter $k/\gamma$. For $k/\gamma = 0$, and interaction strengths in the range $0 \le U/\gamma \lesssim 40$, nearly the entire population resides in the amplitude $\alpha_{101}$. In the interval $40 \lesssim U/\gamma \lesssim 60$, the amplitudes $\alpha_{101}$, $\alpha_{110}$, and $\alpha_{011}$ attain nearly equal magnitudes, while the remaining components remain close to zero. For $U/\gamma \gtrsim 60$, the amplitudes $\alpha_{110}$ and $\alpha_{011}$ become dominant, whereas $\alpha_{101}$ drops to zero.
Notably, at the first transition around $U/\gamma \approx 40$, only the amplitudes corresponding to states with particles distributed over different lattice sites exhibit discontinuous changes. By contrast, at the second transition near $U/\gamma \approx 60$, all amplitudes undergo a discontinuous jump. For a small non-zero value $k/\gamma = -0.1$, the location of the first transition shifts from $U/\gamma \approx 40$ to $U/\gamma \approx 20$. As shown in Fig.~\ref{steadystate-heatmap-N=2-local}.a, when $k/\gamma \lesssim -0.4$, the first phase no longer appears. The position of the second transition, however, shows a weaker dependence on the value of $k/\gamma$ for $k/\gamma \le 0$.
In summary, for $N = 2$, the phase located on the right-hand side of the heat maps in Figs. \ref{steadystate-heatmap-N=2-local}.a–b is characterized by a large magnitude of the amplitude $\alpha_{101}$. In contrast, the phases on the left-hand side are defined by large magnitudes of the amplitudes $\alpha_{110}$ and $\alpha_{011}$.

Figure~\ref{steadystate-heatmap-N=3-local}.c illustrates the dependence of the steady state on the parameter $k/\gamma$ for a fixed interaction strength $U/\gamma = 100$. As discussed previously, only two distinct phases are present, with a phase transition occurring at approximately $k/\gamma \approx 2.5$. For values of $k/\gamma$ below this threshold, the amplitude $\alpha_{111}$ has the largest magnitude, with $|\alpha_{111}|^2 \approx 0.75$. As the interaction strength $U/\gamma$ is reduced, this magnitude decreases continuously, reaching $|\alpha_{111}|^2 \approx 0.2$ at $U/\gamma = 0$. At this point, the amplitudes satisfy $|\alpha_{120}|^2 = |\alpha_{021}|^2 = |\alpha_{111}|^2$.
In the phase located on the right-hand side of Figs.~\ref{steadystate-heatmap-N=3-local}.a–b, the dominant contributions arise from the amplitudes $\alpha_{201}$ and $\alpha_{102}$, with magnitudes $|\alpha_{201}|^2 = |\alpha_{102}|^2 \approx 0.4$. The remaining probability weight is primarily distributed among the amplitudes $\alpha_{120}$ and $\alpha_{021}$, each having magnitudes $|\alpha_{120}|^2 = |\alpha_{021}|^2 \approx 0.1$. These amplitudes, together with the remaining components of the steady state, are found to be approximately independent of both the interaction strength $U/\gamma$ and the parameter $k/\gamma$ over the parameter ranges considered here. An exception occurs for vanishing interaction strength $U/\gamma\approx 0$, where the magnitudes of the amplitudes $\alpha_{120}$ and $\alpha_{021}$ vary continuously together with those of $\alpha_{300}$ and $\alpha_{003}$, as shown in Fig.~\ref{steadystate-heatmap-N=3-local}.d. 

It is evident that local and non‑local dephasing schemes exhibit both significant similarities and pronounced differences in their steady‑state structures. In particular, the resulting phase diagrams differ substantially between the two dephasing mechanisms.
For non‑local dephasing with $N = 2$, the phase occurring at larger interaction strengths, 
$U/\gamma \gtrsim 20$, is characterized by the probability being predominantly shared between the states $\ket{2,0,0}$ and $\ket{0,0,2}$, indicating strong localization of particles at the edge sites. In the local dephasing scheme, a comparable localization behavior is observed only for sufficiently large interaction strengths, $U/\gamma \gtrsim 110$, in combination with large values of the transformation parameter $k/\gamma$. The non‑local dephasing scheme also exhibits localization for the case $N = 3$ when the interaction strength exceeds $U/\gamma \gtrsim 30$.
In contrast, the local dephasing scheme does not display analogous localization for $N = 3$. Instead, at large interaction strengths and sufficiently small values of $k/\gamma$, a phase emerges that is dominated by a large magnitude of the amplitude $\alpha_{111}$. For smaller values of both $U/\gamma$ and $k/\gamma$, no single amplitude dominates the steady state. At sufficiently large values of $k/\gamma$, the amplitudes $\alpha_{201}$ and $\alpha_{102}$ become dominant. Unlike the non‑local dephasing scheme, the phase structure in the local dephasing case exhibits a strong dependence on the transformation parameter $k/\gamma$.

The expectation value of the energy $\braket{H/\gamma}$, shown in Appendix \ref{sec-energies} in Figs. \ref{energies}.a–b, indicate that increasing the interaction parameter $U/\gamma$, which corresponds to stronger on-site interaction energy, drives the system toward a higher-energy steady state. As illustrated in Figs.~\ref{steadystate-heatmap-N=2}.c and \ref{steadystate-heatmap-N=3}.c, these transitions are accompanied by a change in the structure of the steady state: the system evolves into a configuration in which particles are more likely to occupy the same lattice site. This behavior is counterintuitive when compared to the properties of the ground state of the corresponding Hermitian Hamiltonian. In the latter case, increasing the on-site interaction strength $U/\gamma$ suppresses multiple occupancy and favors spatial delocalization of particles. In contrast, the steady state of the non-Hermitian effective Hamiltonian exhibits enhanced localization with increasing $U/\gamma$, highlighting a fundamental qualitative difference between dissipative steady states and equilibrium ground-state physics.

The maps of the quantity $\mathrm{Im} \lambda_1$ presented in Fig.~\ref{prob-decay-heatmaps} in Appendix~\ref{sec-decay-rates} indicate that, in all cases considered, vanishing interaction strength $U/\gamma$ yields an optimal or near-optimal regime in terms of minimizing the decay rate of the state norm. As expected, increasing the parameter $k/\gamma$ generally leads to more negative values of $\mathrm{Im} \lambda_1$, as can be inferred directly from the contribution $-\frac{i}{2} k \,n_2$ in the effective Hamiltonian $K'$.
In the case of local dephasing, where negative values of $k/\gamma$ are admissible, the decay of the norm can be significantly slowed for $k/\gamma < 0$. However, this improvement comes at the expense of a longer relaxation timescale toward the steady state, as evidenced in Figs.~\ref{steadystate-heatmap-N=2-local}.b and \ref{steadystate-heatmap-N=2-local}.b. Furthermore, Figs.~\ref{prob-decay-heatmaps}.a–b reveal a region of enhanced norm decay when $k/\gamma$ is large and the interaction strength $U/\gamma$ is sufficiently small but nonzero. This behavior likely arises because, in this parameter regime, the steady state exhibits larger amplitudes associated with configurations in which particles occupy the central site, thereby increasing the contribution to norm decay through the operator $n_2$ appearing in the rate-operator transformation.

\section{Summary and discussion \label{summary}}

In this work, we have conducted a detailed investigation of the Bose–Hubbard model as an engineered open quantum system subject to both local and non-local dephasing, incorporating rate-operator transformations. Our results demonstrate that the dynamical behavior of the system can be significantly controlled through an appropriate choice of dephasing mechanisms and transformation parameters. Moreover, we observe the emergence of quantum phase transitions in the steady-state structure. The resulting phase landscape is highly non-trivial, revealing a rich interplay between coherent dynamics, dissipation, and measurement-induced effects.

We have described the phases of the steady state of the effective Hamiltonian for both dephasing schemes with $N=2$ and $N=3$ using rate operator transformation of the form $C=k\,n_2$. The phases are essentially independent of the parameter $k$ when using non-local dephasing, while for local dephasing there is heavy dependence on the parameter $k$. In any case there are more than one quantum phase separated by clear phase boundary and different free energies. Some of the steady state phases are characterized by localization of particles while in others the particles delocalize. This shows that the different measurement schemes defined by the rate operator can lead to significant differences in the steady states. We also saw how the rate operator transformations has a large impact, together with the interaction strength $U$, on the relaxation speed of the quantum system.

We have characterized the steady‑state phases of the effective Hamiltonian for both local and non‑local dephasing schemes with $N = 2$ and $N = 3$, employing rate‑operator transformations of the form $C = k\, n_2$. For non‑local dephasing, the phase structure is largely independent of the parameter $k$, whereas for local dephasing the steady states exhibit a pronounced dependence on $k$. In all cases, multiple quantum phases are observed, separated by well‑defined phase boundaries and associated with distinct free‑energy values. Some of these steady‑state phases are characterized by particle localization, while others correspond to delocalized particle distributions. These findings demonstrate that different measurement schemes, as encoded by the choice of rate operator, can lead to substantial qualitative differences in the resulting steady states. Furthermore, we have shown that rate‑operator transformations, together with the interaction strength $U$, have a significant impact on the relaxation speed of the system toward its steady state.

A significant part of recent literature on non-Hermitian dynamics studies the effects of exceptional points and $PT$-symmetry to non-Hermitian dynamics and the phase structures of non-Hermitian Hamiltonians. Although it is beyond the scope of this paper, it is possible to use the rate operator formalism to engineer non-Hermitian Hamiltonians with these properties.

\section*{Acknowledgements}
We thank Federico Settimo for useful discussions. J.L. thanks University of Turku Graduate School (UTUGS) Doctoral Programme in Exact Sciences (EXACTUS) for funding. K.L. and J.L.  gratefully acknowledge financial support from the Emil
Aaltonen Foundation. J.L. thanks Eva and Niilo Pakkala Fund for travel grants.

\bibliography{RO-BH-bib}

\appendix

 \section{Non-homogenous rates \label{sec-non-hom-rates}}

\begin{figure}[htp]
    \begin{subfigure}[t]{75mm}
        \includegraphics[width=\linewidth]{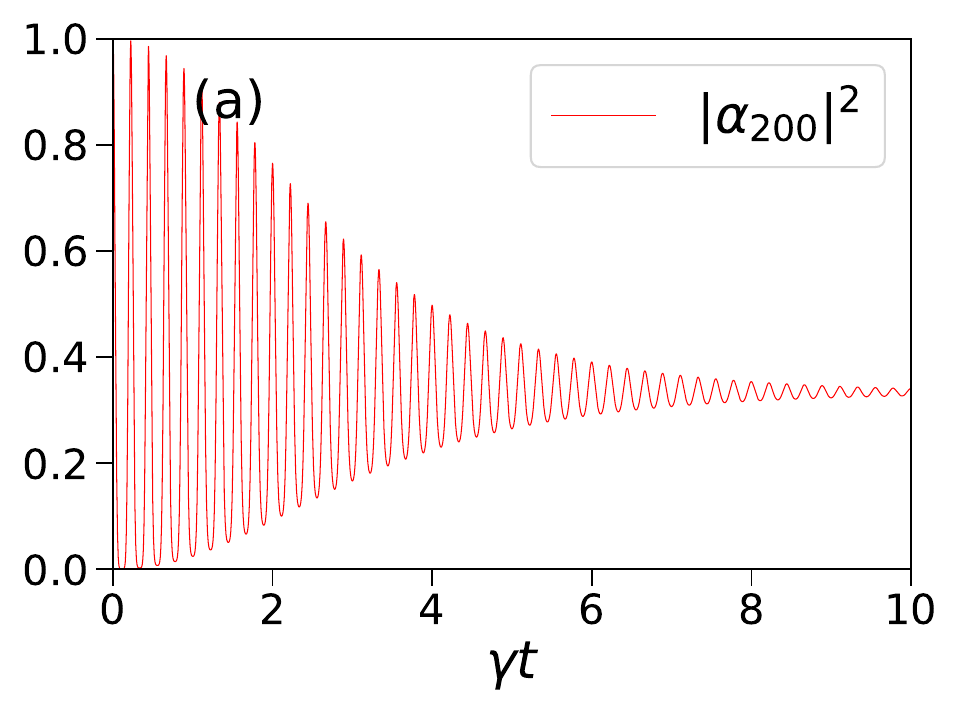}
    \end{subfigure}
    \begin{subfigure}[t]{75mm}
        \includegraphics[width=\linewidth]{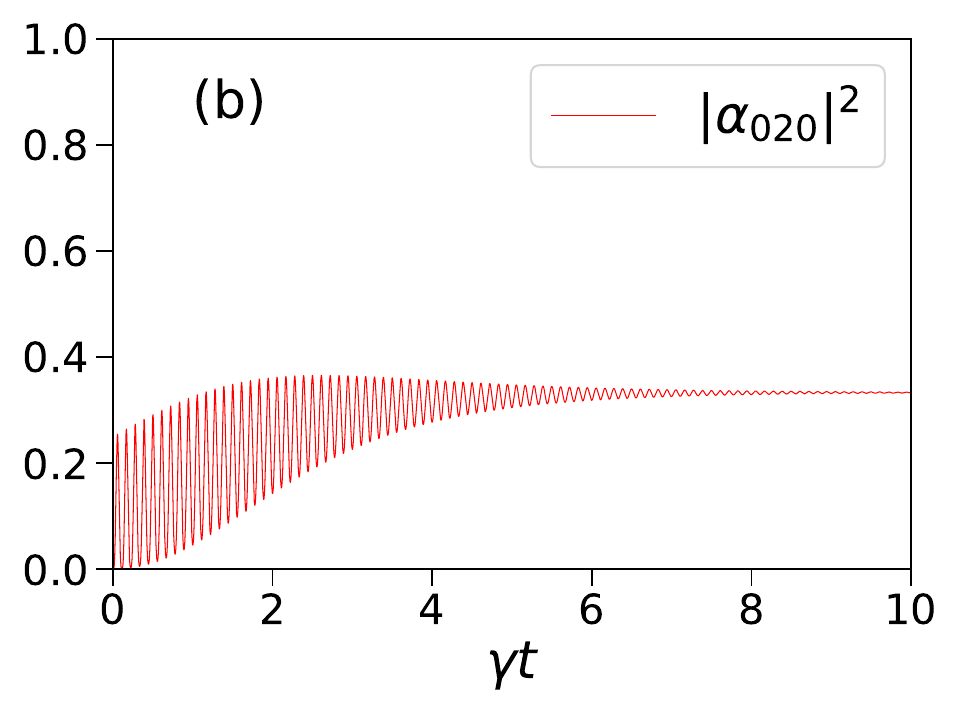}
    \end{subfigure} \\
    \begin{subfigure}[t]{75mm}
        \includegraphics[width=\linewidth]{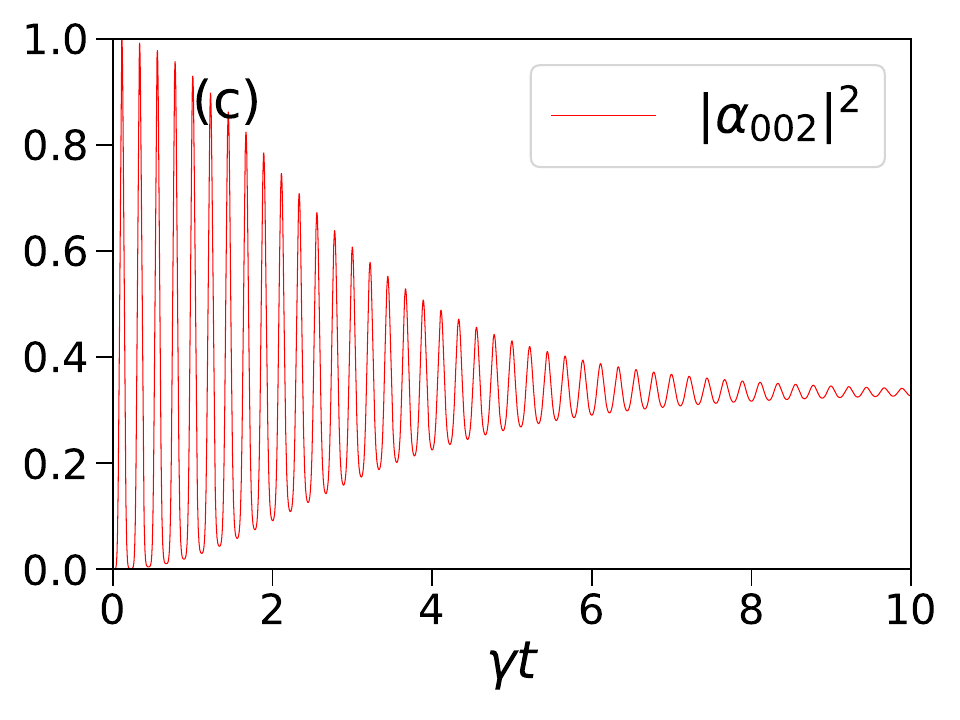}
    \end{subfigure}
    \begin{subfigure}[t]{75mm}
        \includegraphics[width=\linewidth]{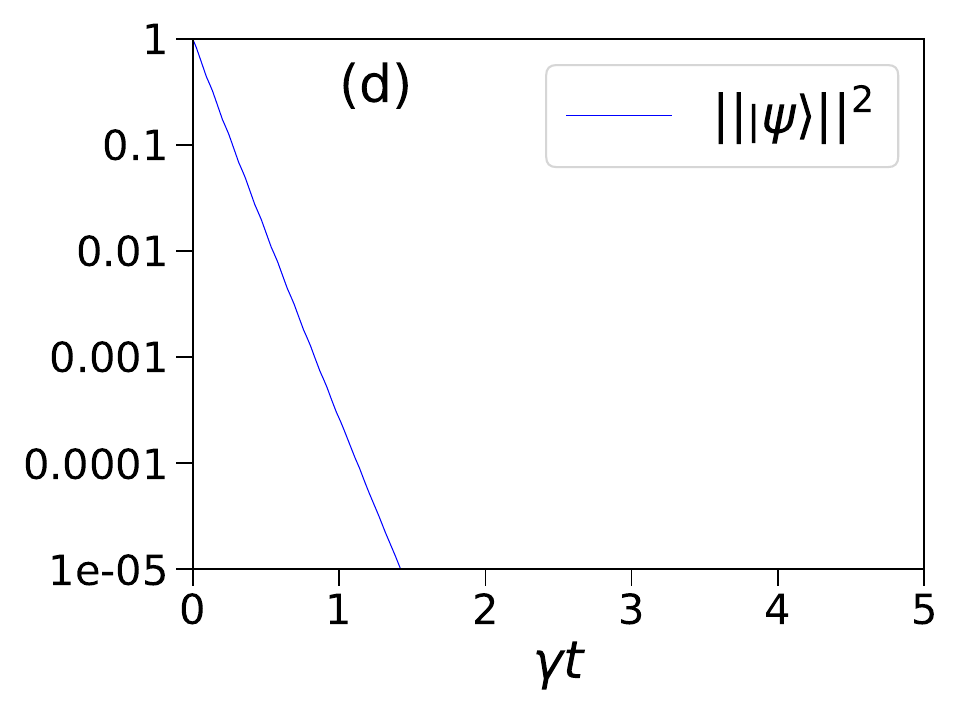}
    \end{subfigure}
    \caption{Deterministic evolution of the absolute squares of asymptotically non-zero amplitudes without rate operator transformation using non-local dephasing. Parameters are $J/\gamma=20$, $U=0$ and the inhomogeneous rates are $\kappa_{ij}=\gamma\neq 0$ except $\kappa_{200,020}=\kappa_{200,002}=\kappa_{020,002}=0$. The initial state is $\ket{\psi_0}=\ket{2,0,0}$ and the panels are (a): $|\alpha_{200}|^2$, (b): $|\alpha_{020}|^2$, (c): $|\alpha_{002}|^2$, (d): $\|\ket{\psi}\|^2$ on a logarithmic scale.}
    \label{N=2-C=0-k14k16k46=0-200}
\end{figure}

This section examines the effects of unbalanced decay rates in the master equation on the deterministic evolution. Figure~\ref{N=2-C=0-k14k16k46=0-200} presents the case of non-local dephasing in which the rates $\kappa_{200,020}=\kappa_{200,002}=\kappa_{020,002}$ are set to zero, while all remaining rates are fixed to be equal but non-zero. In this example, no rate-operator transformation is applied.
Under these conditions, the effective Hamiltonian $K$ reduces the norm of the states $\ket{2,0,0}$, $\ket{0,2,0}$, and $\ket{0,0,2}$ at a slower rate than that of the remaining basis states. As a result, the deterministic evolution asymptotically converges to a steady state that is a linear combination of these three states. The chosen set of decay rates leads to a steady state in which the components $\ket{2,0,0}$, $\ket{0,2,0}$, and $\ket{0,0,2}$ appear with equal magnitude of their amplitudes.

Figure~\ref{local-C=0-0,01,0} shows the deterministic evolution under local dephasing with $\gamma_1=\gamma_3=0$ and $\gamma_2\neq 0$. The norms of the states $\ket{n,0,m}$ are not reduced by the Lindblad operator $n_2$, since these states are eigenvectors of $n_2$ with zero eigenvalue. If a linear combination of such states is simultaneously an eigenvector of the Hamiltonian $H$, it constitutes a steady state of the evolution generated by the effective Hamiltonian $K$. In this case, the imaginary part of the corresponding eigenvalue vanishes, implying that the asymptotic value of the norm remains non-zero.
As a consequence, the system asymptotically approaches a state with non-zero amplitudes only for basis states of the form $\ket{n,0,m}$. Such an eigenstate is referred to as a dark state and is also a steady state of the master equation~(\ref{me-local-dephasing}). In the present example, numerical diagonalization indicates that this dark state is unique for $N=1$, $N=2$, and $N=3$.

\begin{figure}[htp]
    \begin{subfigure}[t]{50mm}
        \includegraphics[width=\linewidth]{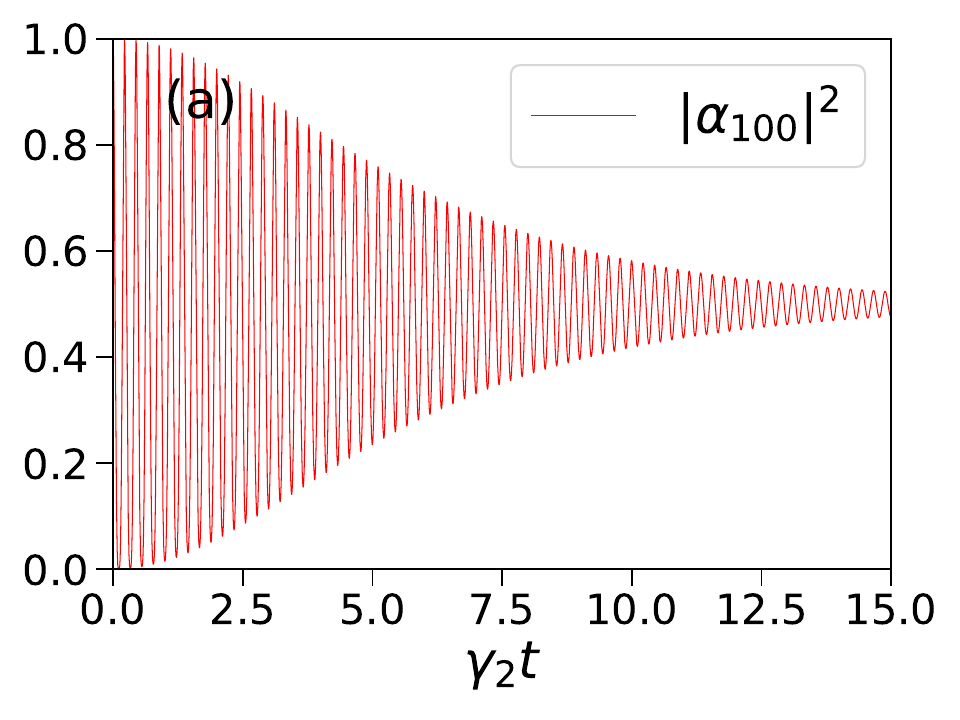}
    \end{subfigure}
    \begin{subfigure}[t]{50mm}
        \includegraphics[width=\linewidth]{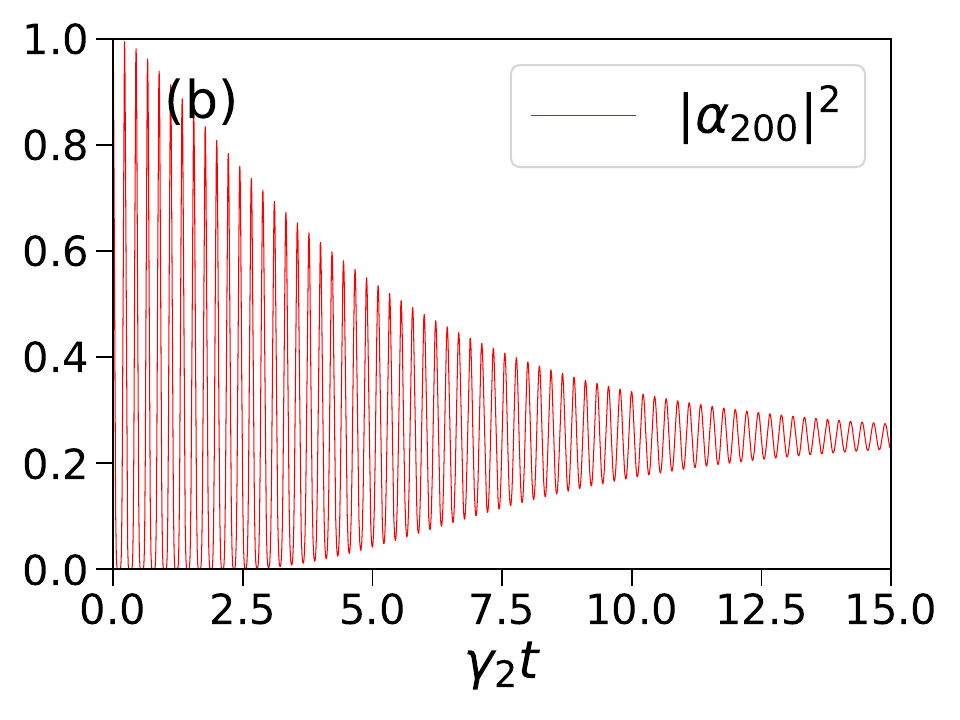}
    \end{subfigure} 
    \begin{subfigure}[t]{50mm}
        \includegraphics[width=\linewidth]{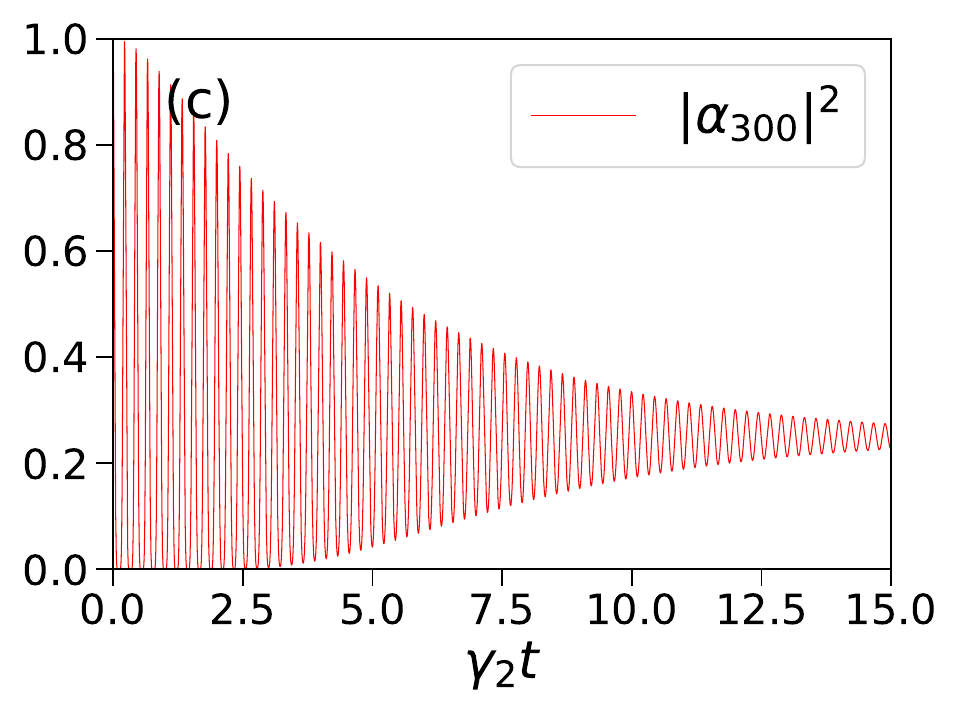}
    \end{subfigure} \\
    \begin{subfigure}[t]{50mm}
        \includegraphics[width=\linewidth]{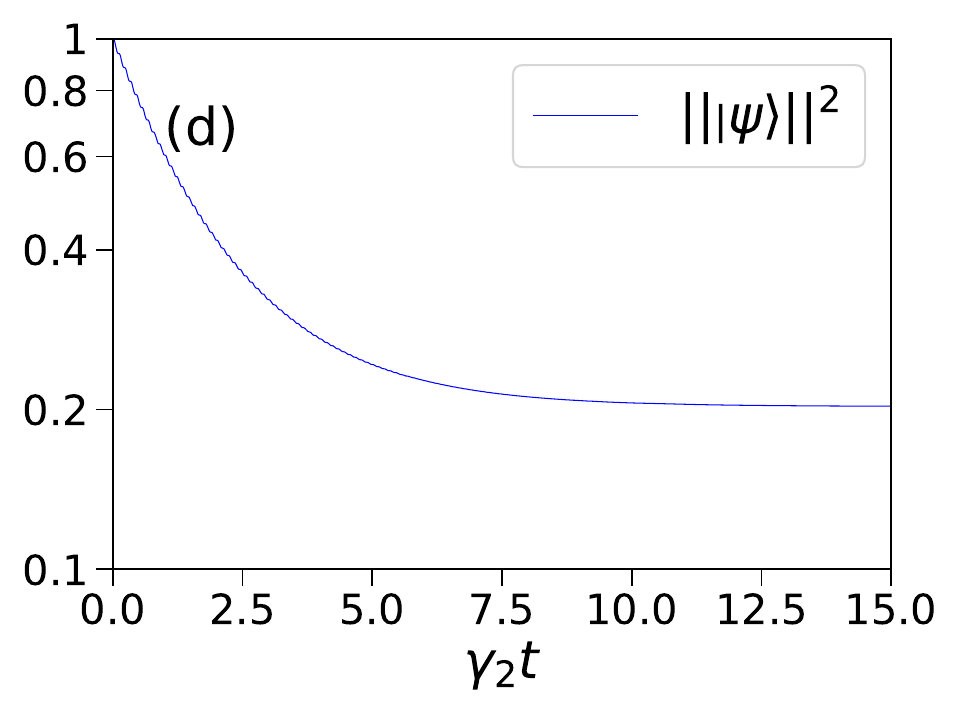}
    \end{subfigure}
    \begin{subfigure}[t]{50mm}
        \includegraphics[width=\linewidth]{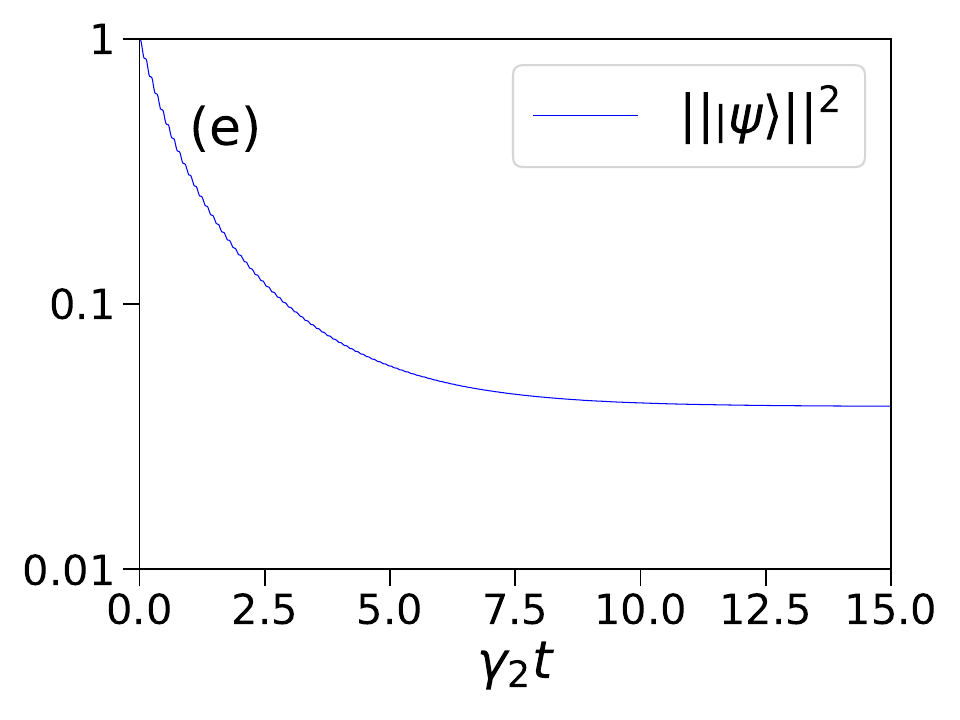}
    \end{subfigure} 
    \begin{subfigure}[t]{50mm}
        \includegraphics[width=\linewidth]{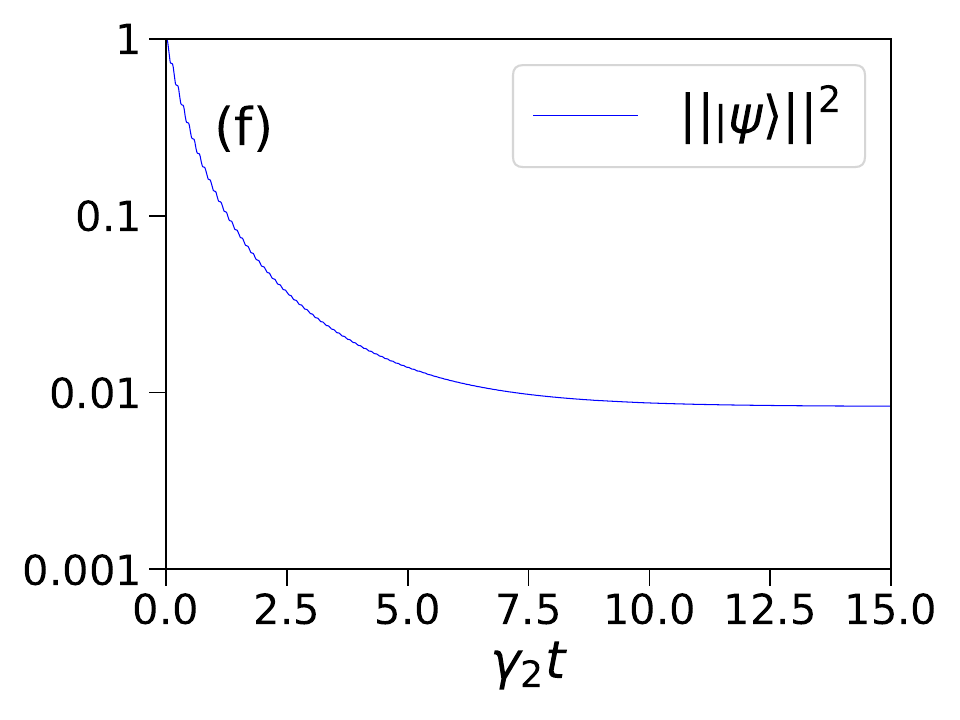}
    \end{subfigure}
    \caption{Deterministic evolution of $|\alpha_{N00}|^2$ and the norm of the state with $C=0$ for $\gamma_1=\gamma_3=0$ and $\gamma_2\neq 0$, $J/\gamma_2=20$ and $U=0$, for local dephasing. The initial states are $\ket{\psi_0}=\ket{N,0,0}$, and the panels are (a): $|\alpha_{100}|^2$, (b): $|\alpha_{200}|^2$, (c): $|\alpha_{300}|^2$, (d-f): $\|\ket{\psi}\|^2$ on a logarithmic scale for $N=1$, 2 and 3, respectively.}
    \label{local-C=0-0,01,0}
\end{figure}

\section{Decay rates of the state norm} \label{sec-decay-rates}

\begin{figure}[htp]
    \begin{subfigure}[t]{75mm}
        \includegraphics[width=\linewidth]{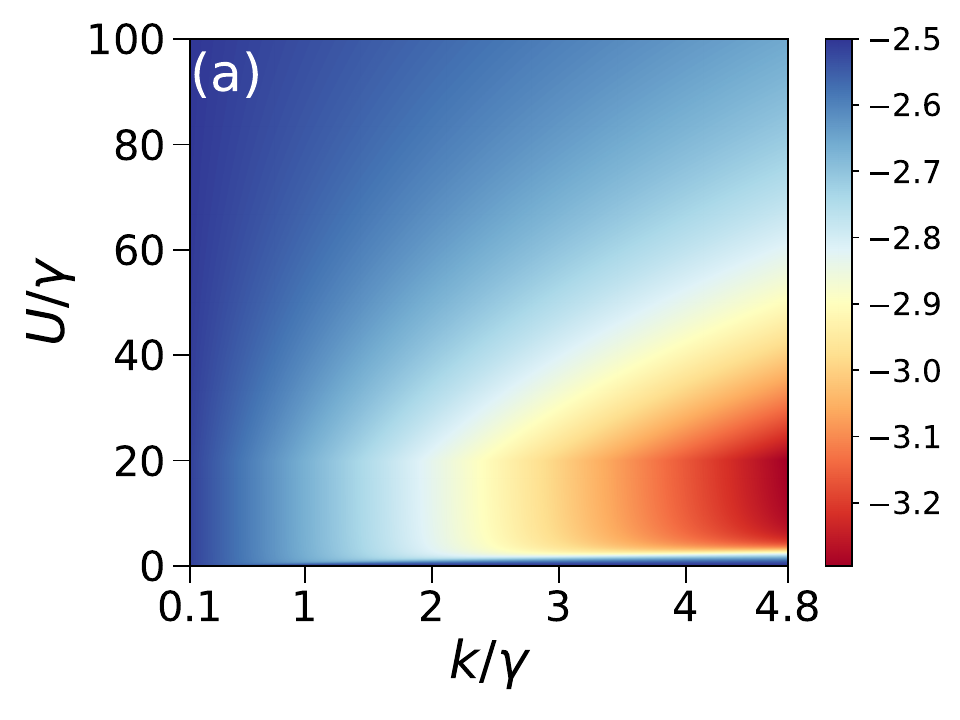}
    \end{subfigure}
    \begin{subfigure}[t]{75mm}
        \includegraphics[width=\linewidth]{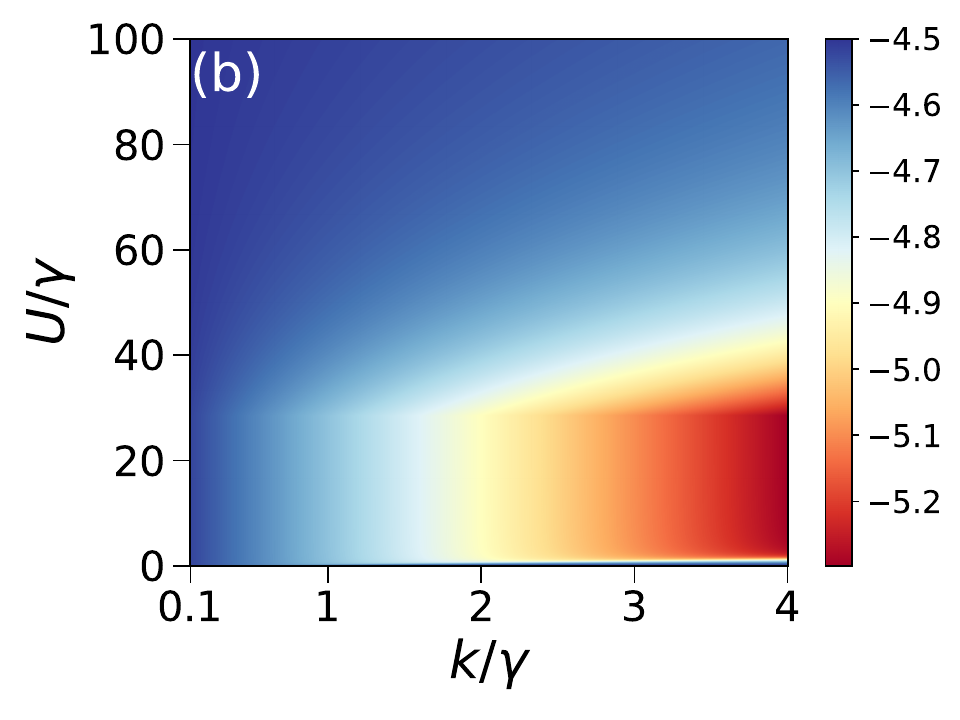}
    \end{subfigure} \\
    \begin{subfigure}[t]{75mm}
        \includegraphics[width=\linewidth]{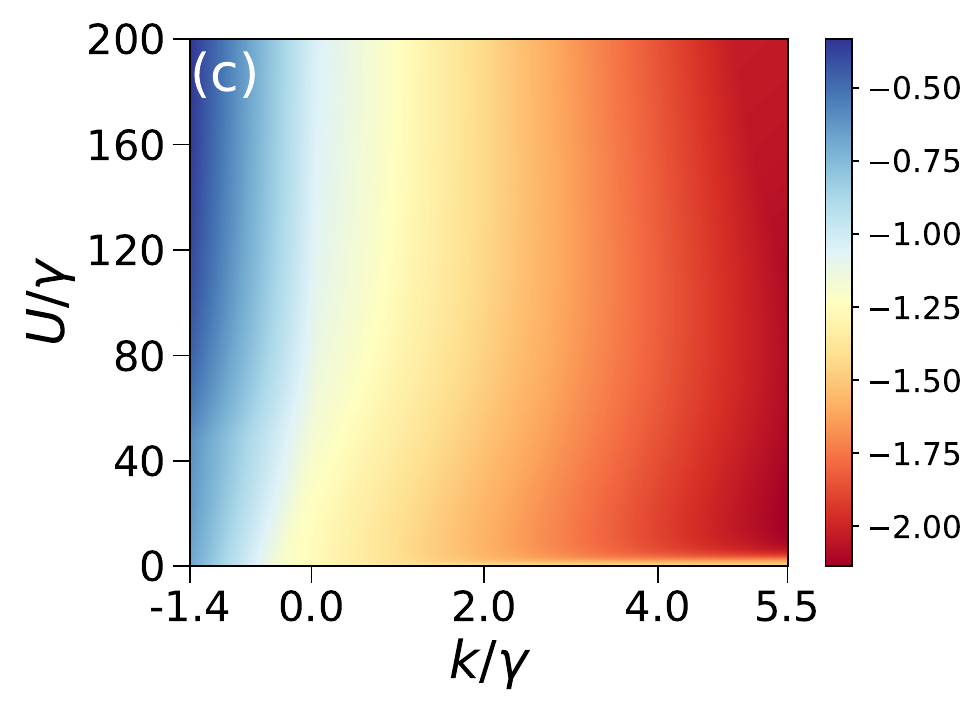}
    \end{subfigure}
    \begin{subfigure}[t]{75mm}
        \includegraphics[width=\linewidth]{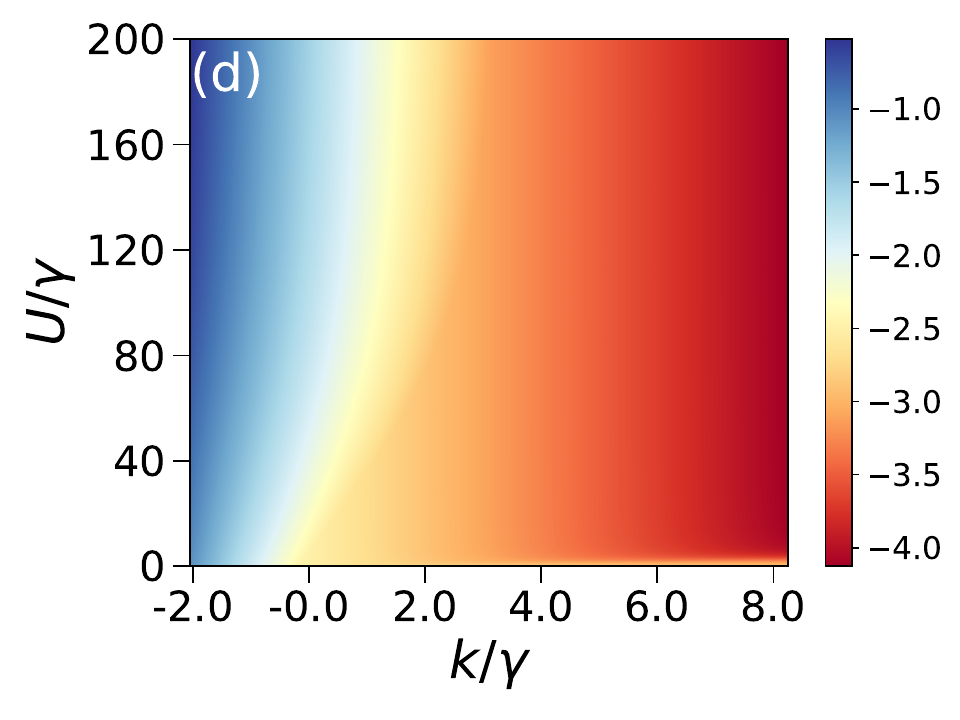}
    \end{subfigure}
    \caption{Imaginary part $\text{Im}(\lambda_1)$ of the eigenvalue of the steady state in $(k/\gamma,\,U/\gamma)$ plane. The panels are (a): non-local dephasing, $N=2$; (b): non-local dephasing, $N=3$; (c): local dephasing, $N=2$; (d): local dephasing, $N=3$.}
    \label{prob-decay-heatmaps}
\end{figure}

Figure \ref{prob-decay-heatmaps} presents the decay rates $\mathrm{Im}(\lambda_1)$, i.e. the largest imaginary part of the eigenvalues, of the state norm in $(k/\gamma,\,U/\gamma)$ plane for both non‑local and local dephasing schemes with particle numbers $N = 2$ and $N = 3$. 

\section{Free energies $\braket{H/\gamma}$ of the steady states} \label{sec-energies}

\begin{figure}[htp]
    \begin{subfigure}[t]{75mm}
        \includegraphics[width=\linewidth]{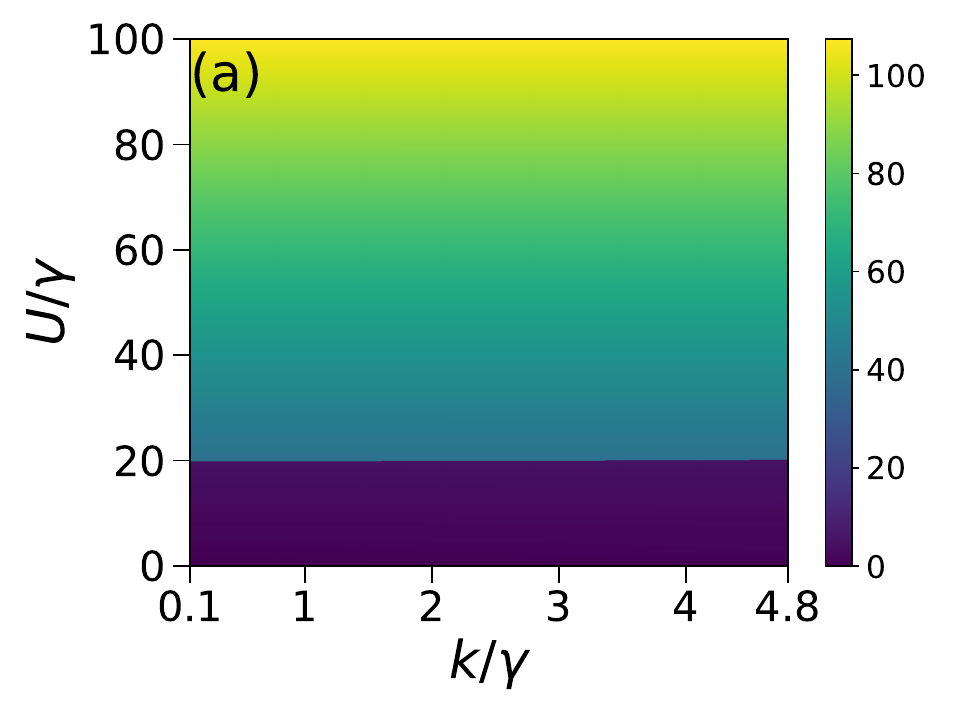}
    \end{subfigure}
    \begin{subfigure}[t]{75mm}
        \includegraphics[width=\linewidth]{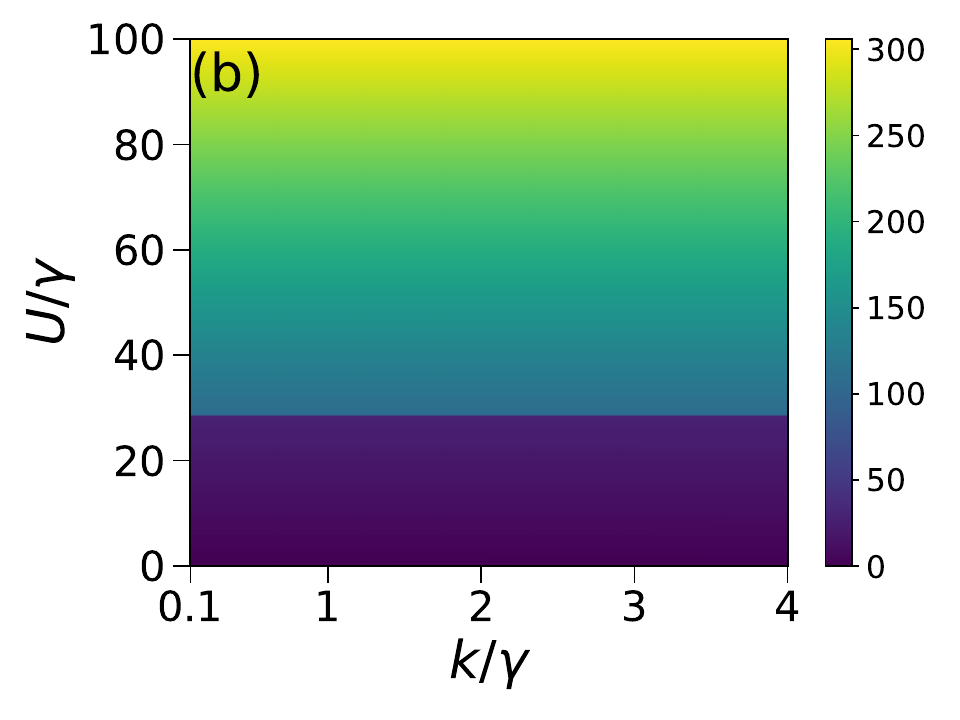}
    \end{subfigure} \\
    \begin{subfigure}[t]{75mm}
        \includegraphics[width=\linewidth]{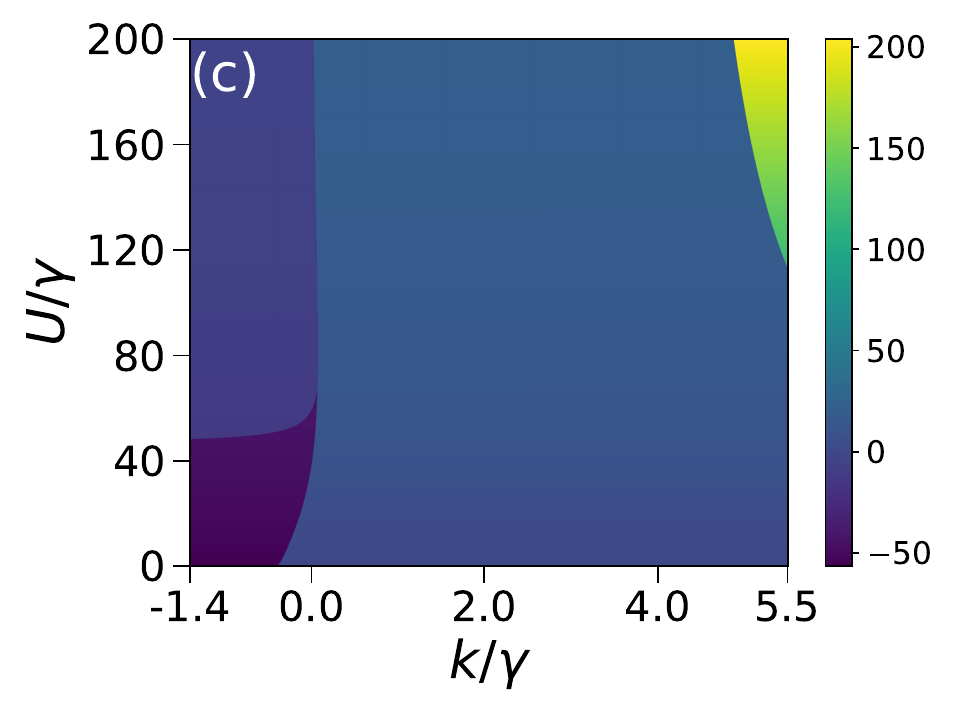}
    \end{subfigure}
    \begin{subfigure}[t]{75mm}
        \includegraphics[width=\linewidth]{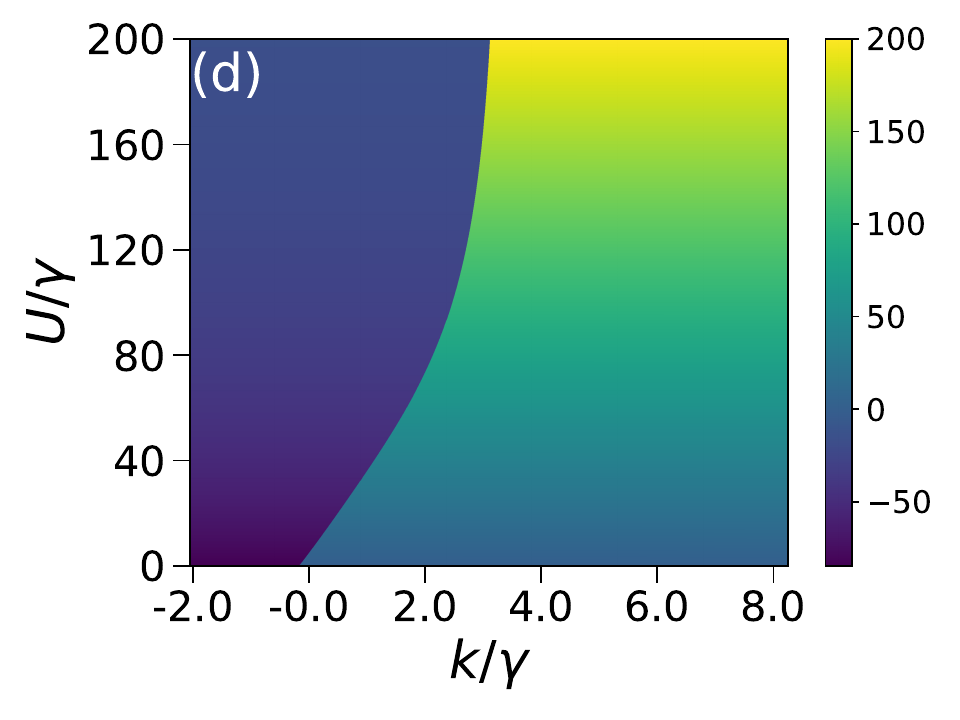}
    \end{subfigure}
    \caption{Expectation value $\braket{H/\gamma}$ of the energy of the steady state as a function of parameter $k/\gamma$ and interaction strength $U/\gamma$. The panels are (a): Non-local dephasing, $N=2$; (b): Non-local dephasing, $N=3$; (c): Local dephasing, $N=2$; (d): Local dephasing, $N=3$.}
    \label{energies}
\end{figure}

Figure~\ref{energies} presents the expectation values $\braket{H/\gamma}$ of the energy associated with the steady states for both non‑local and local dephasing schemes with particle numbers $N = 2$ and $N = 3$. In the context of quantum phase transitions at $T = 0$, the Hermitian Hamiltonian $H$ of the Bose–Hubbard model plays the role of the (Helmholtz) free energy of the system, encapsulating the characteristic signatures of the phase transition.

\end{document}